\documentclass[twoside,twocolumn,english,prc,amsmath]{revtex4}
\usepackage[T1]{fontenc}
\usepackage{geometry}
\usepackage{amsmath}
\usepackage{graphicx}
\usepackage{amssymb}

\makeatletter
\def\fnum@table{\tablename~{\bf\thetable}}
\def\fnum@figure{\figurename~{\bf\thefigure}}
\def\tablename{\footnotesize{\bf Table}}
\def\figurename{\footnotesize{\bf Figure}}

\def\VYP#1#2#3{{\bf #1}, #3 (#2)}  % Volume, page (Year)
\newcommand{\etal}{\mbox{\textit et al.}}                       %
\usepackage{dcolumn}

\usepackage{babel}
\makeatother
\begin{document}

\title{Macroscopic Treatment of Radio Emission from Cosmic Ray Air Showers
based on Shower Simulations}

\author{\textbf{Klaus WERNER }}

\affiliation{SUBATECH, Université de Nantes -- IN2P3/CNRS -- EMN, Nantes, France}

\author{\textbf{Olaf SCHOLTEN }}

\affiliation{Kernfysisch Versneller Instituut, University of Groningen,9747 AA,
Groningen, The Netherlands}

\begin{abstract}
We present a macroscopic calculation of coherent electro-magnetic
radiation from air showers initiated by ultra-high energy cosmic rays,
based on currents obtained from Monte Carlo simulations of air showers
in a realistic geo-magnetic field. We can clearly relate the time
signal to the time dependence of the currents. We find that the the
most important contribution to the pulse is related to the time variation
of the currents. For showers forming a sufficiently large angle with
the magnetic field, the contribution due to the currents induced by
the geo-magnetic field is dominant, but neither the charge excess
nor the dipole contribution can be neglected. We find a characteristic
bipolar signal. In our calculations, we take into account a realistic
index of refraction, whose importance depends on the impact parameter
and the inclination. Also very important is the role of the positive
ions. 
\end{abstract}
\maketitle

\section{Introduction}

The main motivation for this work is the apparent need of radio detection
experiments ({\small LOPES}~\cite{Fal05,Ape06}, {\small CODALEMA}~\cite{Ard06})
for reliable calculations. Already in the earliest works on radio
emission from air showers~\cite{Jel65,Por65,Kah66,All71}, a macroscopic
treatment of the radio emission was proposed, but at the time the
assumptions about the currents were rather crude. In more recent work~\cite{Fal03,Sup03,Hue03,Hue05,Hue07},
a microscopic picture of coherent synchrotron radiation from secondary
shower electrons and positrons gyrating in the Earth's magnetic field
was employed. 

Recently, we performed macroscopic calculations, which allow one under
simplifying conditions to obtain a simple analytic expression for
the pulse shape, showing a clear relation between the pulse shape
and the shower profile \cite{olaf}. The picture used was very similar
to the one in Ref.~\cite{Kah66}, which has been refined by using
a more realistic shower profile and where  we calculate the time-dependence
of the pulse.  

In the present paper, we advance further by computing first the four-current
from a realistic Monte Carlo simulation (in the presence of a geo-magnetic
field), and then solve the Maxwell equations to obtain the electric
field, while considering a realistic (variable) index of refraction.
Although this index varies only between 1 and 1.0003, this variation
has quite interesting consequences. 

For the moment, we neglect the finite extension of the shower (pancake
thickness and lateral extension), at a given time. As shown in \cite{olaf},
this is a good approximation for large impact parameters; for smaller
ones the finite extension has to be considered. There are no conceptual
problems in extending the present approach to finite size sources,
which is simply a three-dimensional integral over the pointlike expressions.
This will be discussed in a future publication.

The shower moves with almost the vacuum velocity of light $c$. There
is a constant creation of electrons and positrons at the shower front,
with somewhat more electrons than positrons (electron excess). This
is compensated by positive ions in the air, essentially at rest. The
electrons and positrons of the shower scatter and lose energy, and
therefore they move slower than the shower front, falling behind,
and finally drop out as {}``slow electrons / positrons''. Close
to the shower maximum, the charge excess of the {}``dropping out''
particles is compensated by the positive ions, since there is no current
before or behind the shower. Taking all together we have the situation
of a moving charge, moving with almost the vacuum velocity of light,
even though the electrons and positrons are moving slower. For a consistent
picture one must not forget the positive ions, which are usually not
considered when talking about the shower. They are very slow, but
their position of creation moves with the velocity of light (shower
front), as well as the position of the {}``charge neutralization''
at the {}``shower tail''. The above picture is true with or without
geo-magnetic field, in the former case the electrons and positrons
move on curved trajectories due to the Lorentz force. Neglecting the
finite dimension of the shower, one has a four-current \begin{equation}
j(t',\vec{x})=J(t')\,\delta^{3}(\vec{x}-\vec{\xi}(t')),\label{eq:current}\end{equation}
with a longitudinal component due to charge excess, and a transverse
component due to the geo-magnetic field. The precise form of $J(t')$
will be obtained from realistic shower simulations, using {\small CONEX}
\cite{conex1,conex2}. Solving Maxwell's equations, we can express
the electric field in terms of the four-current $J$ and its time
derivative $K=dJ/dt$, evaluated at the retarded time. 

In this paper, we will develop explicitly the formalism of electric
fields created from general pointlike currents of the form shown in
eq. (\ref{eq:current}), in particular investigating carefully the
role of a position dependent index of refraction. We analyze the electric
fields for some typical high energy showers, their dependence on the
observer position and on the orientation of the showers with respect
to the magnetic field. We restrict ourself of sufficinently large
impact parameters, so that singularities (related to Chrenkov radiation)
do not matter. In all cases, we can clearly relate the electric field
signal to the sources. The dominant contribution comes always from
the time variation of the currents. For finite angles between the
shower axis and the magnetic field, the biggest contribution is due
to the time derivative of the transverse currents induced  by the
geo-magnetic field; but there is also a contribution (three times
smaller, opposite polarization) due to the transverse current itself.
There is also a sizable contribution from the time derivative of the
charge excess, and from the dipole moment. The strength of these two
contributions is roughly 1/4 each compared to the dominant contribution
from the transverse current. The fact that the time variation of the
currents is the dominant mechanism leads to a characteristic bipolar
signal.

These results actually confirm our earlier findings \cite{olaf} of
a bipolar signal due to the time derivatives of the sources. This
is in qualitative contrast to the findings of \cite{Fal03,Sup03,Hue03,Hue05,Hue07},
where a microscopic picture has been employed, leading to a single
pulse. The fundamental difference is that the time variation of the
currents in the macroscopic picture treats in an effective way the
creation and disappearance of charges, which contributes significantly.
In a microscopic picture this has to be treated explicitly.

\section{Geometry}

We characterize the trajectory of a moving point-like current (referred
to as the shower trajectory) in the following way %
\begin{figure}[htb]
\begin{center}~\end{center}

\begin{center}\includegraphics[%
  scale=0.4]{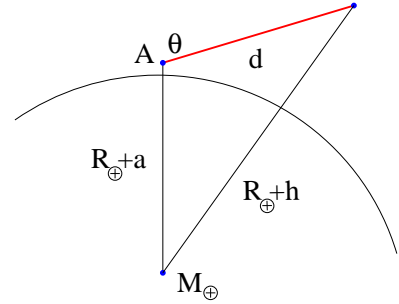}\end{center}

\caption{Characterizing the trajectory: $A$ is the intersection of the trajectory
with a sphere of radius $R_{\oplus}+a$ around the center $M_{\oplus}$
of the Earth , with $R_{\oplus}$ being the radius of the Earth; the
angle $\theta$ measures the inclination. \label{cap:charac}}
\end{figure}
(see fig.\ref{cap:charac}): we imagine a sphere around the center
$M_{\oplus}$ of the Earth with radius $R_{\oplus}+a$, with $R_{\oplus}$
being the radius of the Earth. The point $A$ is by definition the
intersection of the trajectory with the sphere (supposed the intersection
exists). The angle between the trajectory and the axis $M_{\oplus}A$
(referred to as vertical) is called $\theta$. The trajectory is characterized
by the point $A$, the angle $\theta$ (and an azimuthal angle, which
is not important for the following discussion). A position on the
trajectory may be characterized by the distance $d$ from the point
$A$. Using these definitions, the height along the trajectory is
given as\begin{equation}
h(d)=\sqrt{(R_{\oplus}+a)^{2}+d^{2}+2(R_{\oplus}+a)d\,\cos\theta}-R_{\oplus}.\end{equation}

\begin{figure}[htb]
\begin{center}\includegraphics[%
  scale=0.4]{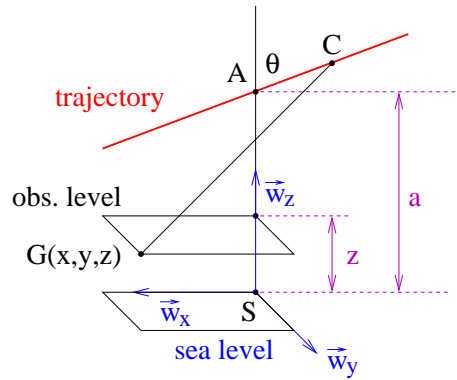}\end{center}

\caption{The trajectory as seen by an observer $G$ situated on the observation
level. We use the Earth frame $\{ S,\vec{w}_{x},\vec{w}_{y},\vec{w}_{z}\}$
to represent points and vectors.\label{cap:charac2}}
\end{figure}

Let $S$ be the point on the axis $M_{\oplus}A$ representing the
sea level. We define the {}``Earth frame'' $\{ S$,$\vec{w}_{x}$,
$\vec{w}_{y}$, $\vec{w}_{z}\}$ (see fig. \ref{cap:charac2}) such
that $\vec{w}_{z}$ represents the vertical upward unit vector, and
$\vec{w}_{y}$ is orthogonal to the {}``shower plane'' defined by
the shower trajectory and the point $S$. In other words, $\{ S,\vec{w}_{x},\vec{w}_{z}\}$
is the shower plane. The shower velocity vector is given as\begin{equation}
\vec{V}=\left(\begin{array}{c}
v\,\sin\theta\\
0\\
-v\,\cos\theta\end{array}\right)\end{equation}
with respect to the Earth frame basis.

 In addition to the sea level plane we have to consider an {}``observation
level plane'' at some height $z$ above sea level. We define an observation
point $G$ in the observation level plane with coordinates $(x,y,z)$
with respect to the Earth frame. The point $C$ represents the center
of the air shower front at time $t'$, given as $\overrightarrow{SC\,}=\overrightarrow{SA\,}+\vec{V}\, t'$,
with respect to the basis $\vec{w}_{i}$. The vector $\vec{R}\equiv\overrightarrow{CG\,}$
is given as $\vec{R}=\overrightarrow{SG\,}-\overrightarrow{SC\,}$.
We may define the point $B$ of closest approach with respect to the
observer. The distance from $A$ is \begin{equation}
d_{B}=x\,\sin\theta+(a-z)\,\cos\theta.\end{equation}
The impact parameter is\begin{equation}
b=|\overrightarrow{BG\,}|=\sqrt{\big(x\,\cos\theta-(a-z)\,\sin\theta\big)^{2}+y^{2}}.\end{equation}
We also define for later use \begin{equation}
t_{B}=d_{B}/v\end{equation}
to be the time the shower passes at the position $B$.

Due to the magnetic field of the Earth, electrons and positrons move
on spiral trajectories with opposite orientations, representing a
transverse current \cite{olaf}. This means, even when considering
the shower as pointlike, the collective current $\vec{J}$ is not
simply a product of charge times velocity, there is in fact a component
transverse to the shower velocity. We compute this current by using
the {\small CONEX} shower simulation program, as\begin{equation}
\vec{J}=\sum_{\mathrm{positrons\, i}}e\vec{v}_{i}-\sum_{\mathrm{electrons\, j}}e\vec{v}_{j},\end{equation}
considering all electrons and positrons present at a given time. %
\begin{figure}[htb]
\begin{center}\includegraphics[%
  scale=0.4]{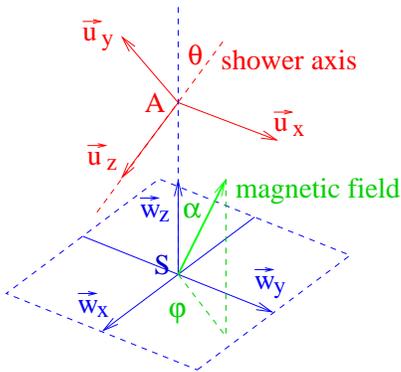}\end{center}

\caption{Reference frames. \label{cap:coord}}
\end{figure}
For calculating the currents, we employ the shower frame $\{ A,\vec{u}_{x},\vec{u}_{y},\vec{u}_{z}\}$,
see fig. \ref{cap:coord}, but the final results are expressed with
respect to the Earth frame $\{ S$,$\vec{w}_{x}$, $\vec{w}_{y}$,
$\vec{w}_{z}\}$.

We will discuss the currents corresponding to air showers in more
detail later, but it is useful already at this point to have some
idea about the times $t'$ corresponding to shower maxima. The shower
size and in particular the shower maximum depends primarily on the
atmospheric depth, given as \begin{equation}
X=X(t')=\int_{-vt'}^{\infty}\rho(h(x))\, dx,\end{equation}
integrating along the trajectory. Here, $h$ is the height above sea
level, and $\rho$ is the air density. %
\begin{figure}[htb]
\begin{center}\hspace*{-1cm}\includegraphics[%
  scale=0.35,
  angle=270]{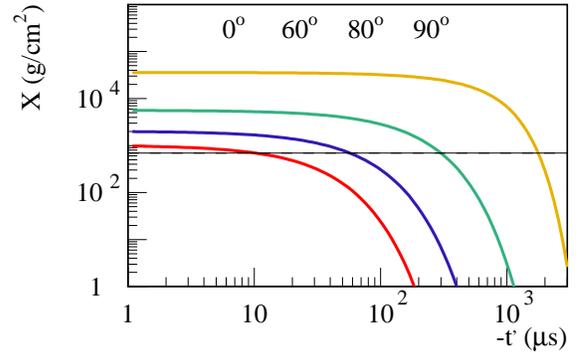}\end{center}
\vspace{-1cm}

\caption{The depth versus the negative time $-t'$, for different angles (using
always $a=140\,$m).\label{cap:depth} The horizontal line at $X=700\,\mathrm{g}/\mathrm{c}\mathrm{m}^{3}$
corresponds to the maximum of the charge distribution for showers
with an energy of $5\,10^{17}$eV.}
\end{figure}
The relation between $X$ and $t'$ (or $d$) depends of course strongly
on the angle $\theta$, as shown in fig. \ref{cap:depth}. Since we
are interested in negative time $t'$ (before the shower reaches $A$),
we plot $X$ versus $-t'$. The horizontal line corresponds to the
maximum of the charge distribution of showers with an energy of $5\,10^{17}$eV,
which can be used to obtain the times $t'$ corresponding to this
maximum. Varying the angle $\theta$ from $0^{0}$ to $90^{0}$, the
times range from some $10\,\mu$s up to almost $2000\,\mu$s. These
numbers define the interesting time scales for the later discussions
of retarded times.

\section{Electric field due to a general pointlike source}

Let us consider a point-like four-current,\begin{equation}
j(t',\vec{x})=J(t')\,\delta^{3}(\vec{x}-\vec{\xi}(t')),\end{equation}
for a given trajectory $\vec{\xi}(t')$, corresponding to the shower
position $C$ in the notation of the preceding chapters. A moving
charge corresponds to $J^{0}=q\, c$, $\vec{J}=q\,\vec{v},$ where
$c$ is the velocity of light. However, currents induced by the Earth's
magnetic field require a more general treatment, which we are going
to present in the following. In this chapter we investigate the case
an index of refraction $n$ equal to unity.

We first define the basic quantities of our approach. In order to
obtain four-vectors $x$ and $\xi$, we define $x^{0}=ct$ and $\xi^{0}=ct'$.
We will need the time derivative of the four-current $J$ and of the
four-vector $\xi$,and we define $K=dJ/dx'^{0}$ and $V=d\xi(t')/dx'^{0}$.
We consider here the case of a constant velocity $V$. Finally, we
define $R=(x-\xi(t'))$ to be the four-vector joining the position
of the current $\xi(t')$ and the position $x$ of the observer.

The four-potential at point $x$ is found to be\begin{equation}
A=\frac{\mu_{0}}{4\pi}\,\frac{J}{|RV|}\,,\end{equation}
(see appendix), where all quantities are meant to be evaluated at
the retarded time $t^{*}$. The three components of the electric field
can be expressed as\begin{equation}
E^{i}=c\left(\partial^{i}A^{0}-\partial^{0}A^{i}\right).\end{equation}
Taking also the $x$ dependence of $t^{*}$ into account, one gets
(see appendix)\begin{equation}
E^{i}=\frac{1}{4\pi\epsilon_{0}c}\,\frac{L^{i0}-L^{0i}}{|RV|^{3}}\,,\end{equation}
with\begin{equation}
L^{\alpha\beta}=RV\, R^{\alpha}K^{\beta}-RV\, J^{\beta}V^{\alpha}+VV\, R^{\alpha}J^{\beta}.\end{equation}
Concerning the denominator and possible singularities due to its roots,
it is useful to note that\begin{equation}
\frac{1}{|RV|}=\frac{\left|\partial t^{*}/\partial t\right|}{|\vec{R}|},\end{equation}
from which one can see that possible singularities of the field coincide
 with singularities of the derivative $\partial t^{*}/\partial t$.

\section{Realistic index of refraction}

\begin{figure}[htb]
\begin{center}\includegraphics[%
  scale=0.4]{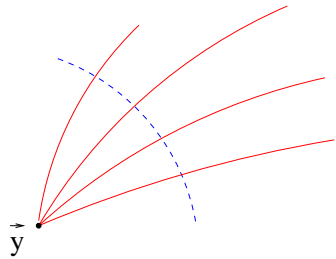}\end{center}
\vspace{-0.3cm}

\caption{Light curves (full lines) in a medium with varying index of refraction.
The dashed line represents a constant optical distance from the point
$\vec{y}$. \label{cap:curvilinear}}
\end{figure}

The index of refraction $n$ in air varies spatially, mainly via its
dependence on the altitude. We will study here the general case of
an index of refraction being a function $n(\vec{y})$ of the position
$\vec{y}$. In this case we define curvilinear coordinates via the
light curves passing through the position $\vec{y}$, see fig. \ref{cap:curvilinear}.%
The optical path length, $s$,
 is defined as the travel time of a light signal multiplied by $c$.
An elementary path length is expressed in terms of the components
$y^{i}$of $\vec{y}$ as $ds^{2}=n^{2}dy^{i}dy^{i}$. The equation
for the derivatives $v^{k}=dy^{k}/ds$ of the light curves is thus\begin{equation}
\frac{dv^{k}}{ds}=-\Gamma_{\, ij}^{k}v^{i}v^{j}=\frac{n_{k}}{n}v^{i}v^{i}-2\frac{n_{i}}{n}v^{k}v^{i},\end{equation}
where we used\begin{equation}
\Gamma_{\, ij}^{k}=\frac{1}{n}(\delta_{ik}n_{j}+\delta_{kj}n_{i}-\delta_{ij}n_{k}),\end{equation}
with $n_{i}=dn/dy^{i}$. The light curves can be obtained by integrating
this differential equation starting from some position $\vec{y}_{0}$.
One gets a family of curves, characterized by the initial velocity
vectors $\vec{v}_{0}$ (more precisely by the two angles characterizing
the direction of the initial velocity, since the module is given).

As the next step we discuss the effects of a varying index of refraction
$n(\vec{x})$ on the calculation of the electric field. The main difference
compared to the case of $n=1$ is the fact that the condition for
the retarded time, expressed via \begin{equation}
\delta\big(\left.R^{0}\right.^{2}-\vec{R}^{2}\big),\label{eq:lc1}\end{equation}
with $R^{0}=ct-ct'$ and $\vec{R}=\vec{x}-\vec{\xi}$, has to be modified
by taking into account the curved light trajectories. We define a
{}``distance'' $L=L(\vec{\xi},\vec{x})$ to be the optical path
length from the source position $\vec{\xi}$ to the observer position
$\vec{x}$, in other words $L$ is the integral $\int ds$ along the
light curve from $\vec{\xi}$ to $\vec{x}$. Instead of eq. (\ref{eq:lc1}),
we now have\begin{equation}
\delta\big(\left.R^{0}\right.^{2}-L(\vec{\xi}(t'),\vec{x})^{2}\big),\label{eq:lcn}\end{equation}
which leads to \begin{equation}
A=\frac{\mu_{0}}{4\pi}\,\frac{J}{|\widetilde{R}V|}\,,\end{equation}
with \begin{equation}
\widetilde{R}{}^{0}=c(t-t^{*}),\quad\widetilde{R}{}^{i}=-L\frac{\partial}{\partial\xi^{i}}L.\end{equation}
All quantities are meant to be evaluated at $\vec{\xi}(t^{*}$), with
the retarded time $t^{*}$ corresponding to the roots of the argument
of the $\delta$ function in eq. (\ref{eq:lcn}). We thus get formally
the same results as in the case $n=1$, but with $R$ replaced by
$\widetilde{R}$, and a different definition of $t^{*}$. The computation
of the retarded times will be discussed in the next chapter. 

Taking the derivative of $\left.R^{0}\right.^{2}-L^{2}$, we get the
useful relation\begin{equation}
\partial^{\alpha}ct^{*}=\frac{\bar{R}{}^{\alpha}}{\widetilde{R}V},\label{eq:dtt}\end{equation}
with \begin{equation}
\bar{R}{}^{0}=c(t-t^{*}),\quad\bar{R}{}^{i}=L\frac{\partial}{\partial x^{i}}L,\end{equation}
which again relates possible singularities of the potential $A$ (roots
of the denominator) to singularities of $dt^{*}/dt$. 

To compute the derivative $\partial^{\alpha}A^{\beta}$ of the potential,
we use\begin{equation}
\partial^{\alpha}\widetilde{R}V=\bar{V}^{\alpha}-\widetilde{V}V\,\partial^{\alpha}ct^{*}\label{eq:drv}\end{equation}
with the four-vectors $\bar{V}$ and $\widetilde{V}$ given as\begin{equation}
\bar{V}^{\nu}=\bar{g}^{\nu\mu}V_{\mu},\quad\widetilde{V}^{\nu}=\widetilde{g}^{\mu\nu}V_{\mu},\end{equation}
where the tensors $\bar{g}$ and $\widetilde{g}$ are defined as\begin{equation}
\bar{g}^{0\mu}=\bar{g}^{\mu0}=\widetilde{g}^{0\mu}=\widetilde{g}^{\mu0}=g^{\mu0}\end{equation}
\begin{eqnarray}
\bar{g}^{ij} & = & \frac{\partial}{\partial x^{i}}L\frac{\partial}{\partial\xi^{j}}L+L\frac{\partial}{\partial x^{i}}\frac{\partial}{\partial\xi^{j}}L,\\
\widetilde{g}^{ij} & = & -\frac{\partial}{\partial\xi^{i}}L\frac{\partial}{\partial\xi^{j}}L-L\frac{\partial}{\partial\xi^{i}}\frac{\partial}{\partial\xi^{j}}L.\end{eqnarray}
 Making use of eq. (\ref{eq:drv}) together with eq. (\ref{eq:dtt}),
we finally can write the electric field as\begin{equation}
E^{i}=\frac{1}{4\pi\epsilon_{0}c}\,\frac{L^{i0}-L^{0i}}{|\widetilde{R}V|^{3}}\,,\end{equation}
with\begin{equation}
L^{\alpha\beta}=\widetilde{R}V\,\bar{R}^{\alpha}K^{\beta}-\widetilde{R}V\,\bar{V}^{\alpha}J^{\beta}+\widetilde{V}V\,\bar{R}^{\alpha}J^{\beta}.\end{equation}
In case of $n=1$, we have $\widetilde{R}=\bar{R}=R$, as well as
$\widetilde{g}=\bar{g}=g$, and therefore $\widetilde{V}=\bar{V}=V$,
and we recover the formula derived earlier for the case $n=1$. 

It is of course not surprising to rediscover the $n=1$ case, but
it is important to notice that the equation for the the field $E^{i}$
in case of $n>1$ has the same structure as the one for $n=1$, just
with {}``~$\widetilde{\:}$~'' and {}``~$\bar{}$~'' decorated
quantities instead of the {}``bare'' ones. The former ones depend
continuously and smoothly on a parameter $\delta$ which characterizes
the deviation of $n$ from unity. We have, for example, $\widetilde{R}=\widetilde{R}(\delta)$,
with $\lim_{\delta\to0}\widetilde{R}=R$, and thereafter $\widetilde{R}=R+R^{(1)}\delta+...$.
Since $\delta<0.0003$ we can actually already ignore to a good precision
the linear term, in other words we can forget about all the {}``~$\widetilde{\:}$~''
and {}``~$\bar{}$~'' symbols, with two exceptions: first the
term $|\widetilde{R}V|$ in the denominator can become very small,
and here even a very small value of $\delta$ could matter. Second
the term $\widetilde{V}V$ is certainly very small ($VV=0$), but
can be comparable to $\widetilde{R}V$, close to a singularity. We
finally obtain\begin{equation}
E^{i}=\frac{1}{4\pi\epsilon_{0}c}\,\frac{L^{i0}-L^{0i}}{|\widetilde{R}V|^{3}}\,,\end{equation}
with\begin{equation}
L^{\alpha\beta}=\widetilde{R}V\, R^{\alpha}K^{\beta}-\widetilde{R}V\, V^{\alpha}J^{\beta}+\widetilde{V}V\, R^{\alpha}J^{\beta}.\end{equation}
In vector notation, we may write\begin{equation}
\vec{E}=\vec{E}_{RK}+\vec{E}_{JV}+\vec{E}_{RJ},\end{equation}
with\begin{equation}
\vec{E}_{RK}=\frac{\vec{W}_{RK}}{D},\quad\vec{E}_{JV}=\frac{\vec{W}_{JV}}{D},\quad\vec{E}_{RJ}=\frac{\widetilde{V}V}{\widetilde{R}V}\,\frac{\vec{W}_{RJ}}{D},\label{eq:efield}\end{equation}
using $D=4\pi\epsilon_{0}c(\!\widetilde{R}V\!)^{2}$, and with\begin{equation}
\vec{W}_{AB}=\vec{A}B^{0}-A^{0}\vec{B}.\end{equation}
Here, we assume $\widetilde{R}V>0$, to be verified later. The first
two terms in eq. ( \ref{eq:efield}) are the contribution $\vec{E}_{JV}$
due to the current $J$, and the contribution $\vec{E}_{RK}$ due
to the variation $K=c^{-1}\, dJ/dt'$ of the current. The third contribution
$\vec{E}_{RJ}$ is small, unless one is close to a singularity, due
to the smallness of the ratio $\widetilde{V}V/\widetilde{R}V$, for
high energy showers with $|\vec{V}|\approx1$. 

As already mentioned (eq. (\ref{eq:dtt})), $|\widetilde{R}V|$ may
be expressed in terms of the derivatives of the retarded time, as
$|\widetilde{R}V|=c(t-t^{*})\,/\,(dt^{*}/dt)$.  We therefor should
be very careful about the evaluation of the retarded  time $t^{*}(t)$
for a given observer time $t$, where the precise values of $n$ may
be important even though $n-1$ is very small.

\section{Retarded time}

The retarded time is obtained from solving $L=R^{0}$, which may be
written as \begin{equation}
L(\vec{\xi}(t'),\vec{x})-c(t_{B}-t')=c(t-t_{B}),\label{eq:tstar}\end{equation}
where a term $-ct_{B}$ has been added on both sides, for later convenience.
The time $t_{B}$ corresponds to the time when the shower is closest
to the observer position $\vec{x}$. %
\begin{figure}[htb]
\begin{center}\includegraphics[%
  scale=0.4]{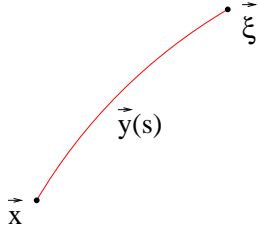}\end{center}
\vspace{-0.3cm}

\caption{Light curve joining $\vec{x}$ and $\vec{\xi}$. \label{cap:light}}
\end{figure}

We first evaluate the l.h.s. of eq. (\ref{eq:tstar}), for a given
arbitrary value of $t'$, corresponding to some shower position $\vec{\xi}(t')$.
To obtain $L$, we solve the differential equation \begin{equation}
\frac{dy^{k}}{ds}=v^{k},\quad\frac{dv^{k}}{ds}=\frac{n_{k}}{n}v^{i}v^{i}-2\frac{n_{i}}{n}v^{k}v^{i},\label{eq:de}\end{equation}
with the boundary conditions \begin{equation}
\vec{y}(0)=\vec{x},\quad\vec{y}(s^{*})=\vec{\xi},\end{equation}
with $s^{*}$ to be determined (see fig. \ref{cap:light}). The index
of refraction $n$ is\begin{equation}
n(\vec{y})=1+0.226\frac{\mathrm{cm^{3}}}{g}\rho(h(\vec{y})),\end{equation}
with $\rho$ being the density of air and $h(\vec{y})$ the atmospheric
height $h$ corresponding to a position $\vec{y}$. We employ an adaptive
step-size Bulirsch-Stoer algorithm to integrate very efficiently the
differential equation, and the three-dimensional Newton-Raphson formula
to accommodate the boundary conditions. Having solved this problem
means having found an $s^{*}$ which corresponds to the optical path
length joining $\vec{\xi}$ and $\vec{x}$, so \begin{equation}
L(\vec{\xi}(t'),\vec{x})=s^{*}.\end{equation}

\begin{figure*}[htb]
\begin{center}~\vspace*{-2cm}\end{center}

\begin{center}\hspace*{-1cm}\includegraphics[%
  scale=0.35,
  angle=270]{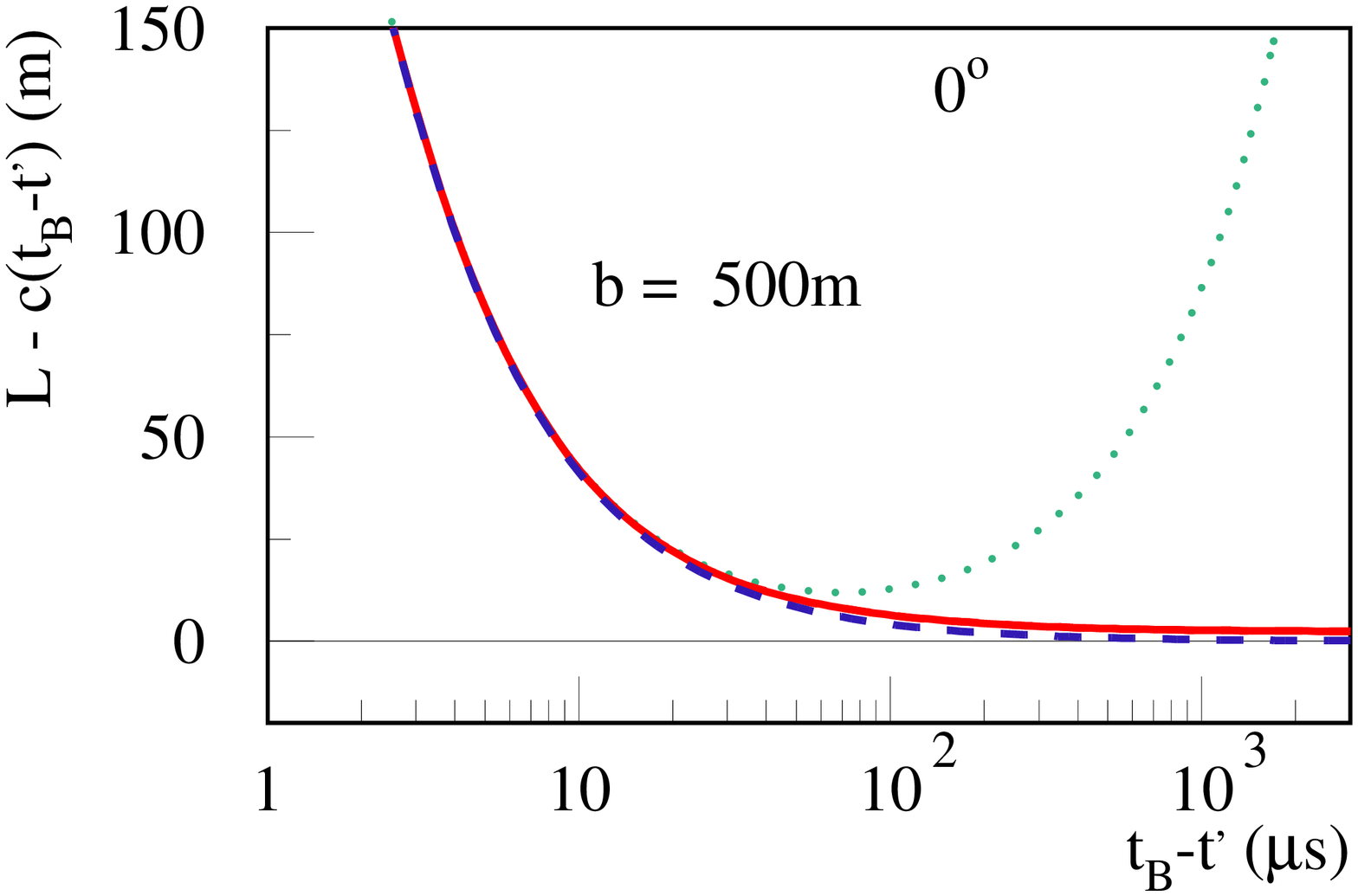}\hspace*{-1cm}\includegraphics[%
  scale=0.35,
  angle=270]{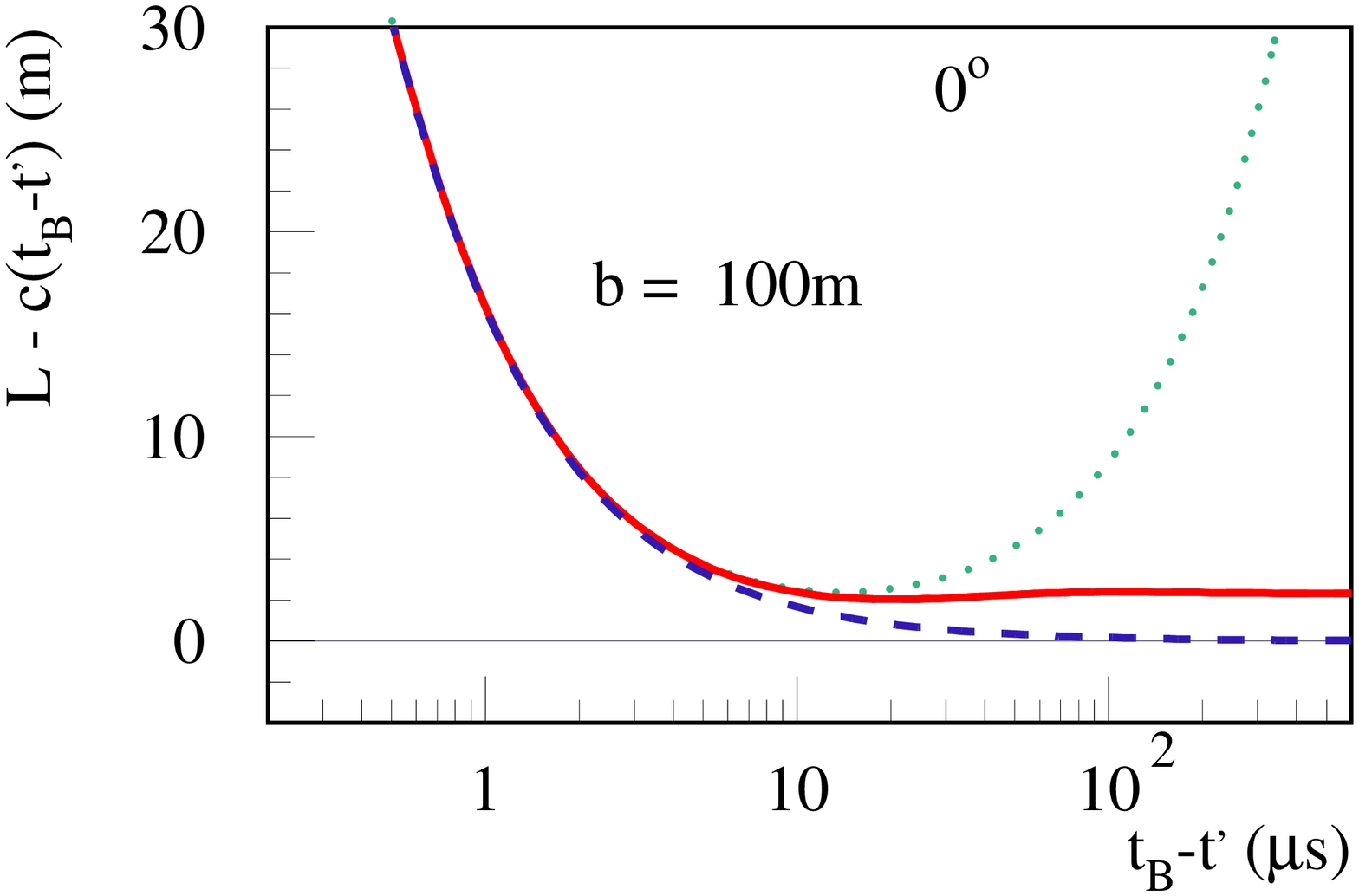}\end{center}

\vspace{-3.5cm}
\begin{center}\hspace*{-1cm}\includegraphics[%
  scale=0.35,
  angle=270]{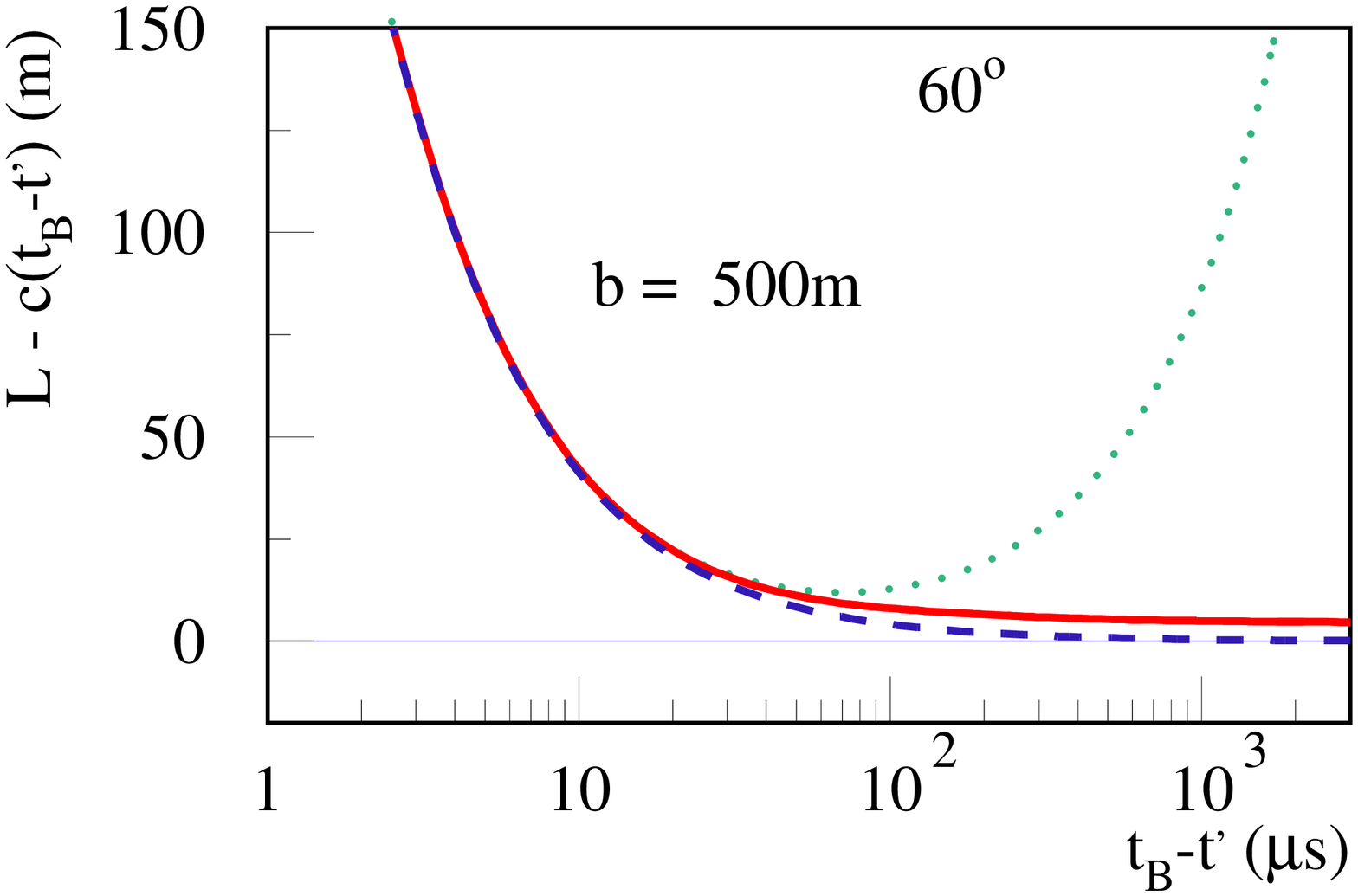}\hspace*{-1cm}\includegraphics[%
  scale=0.35,
  angle=270]{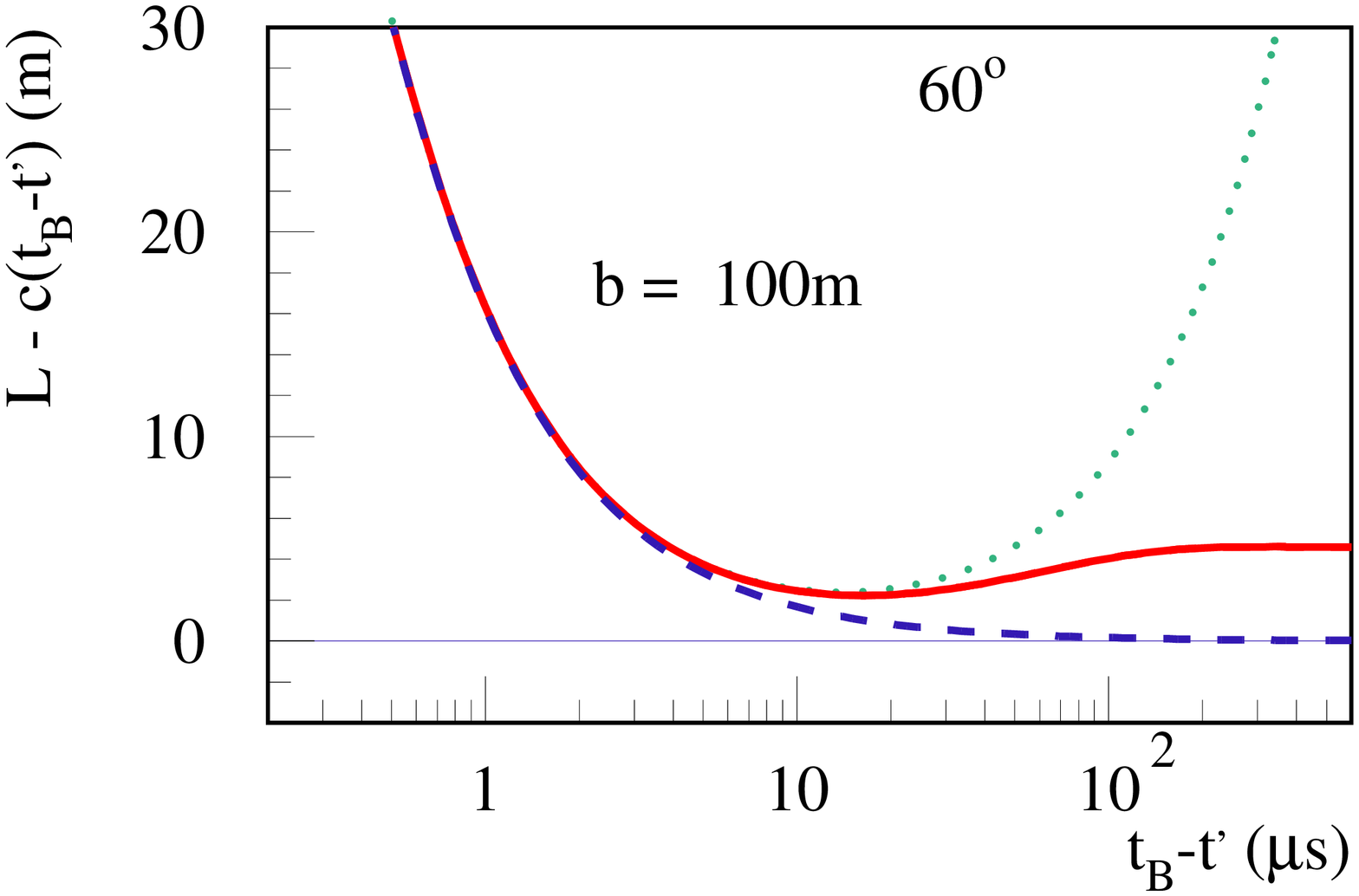}\end{center}
\vspace{-3.5cm}

\begin{center}\hspace*{-1cm}\includegraphics[%
  scale=0.35,
  angle=270]{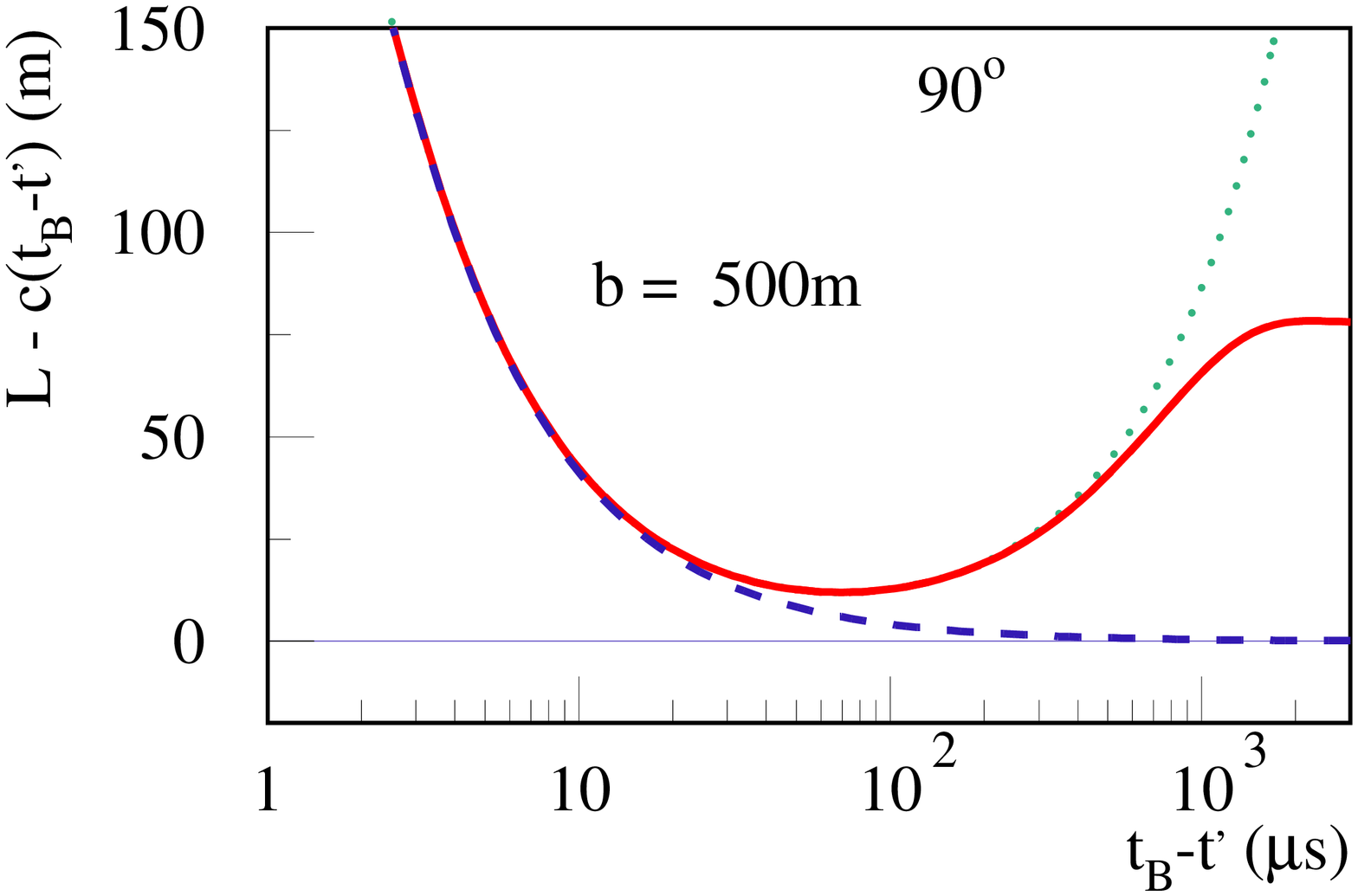}\hspace*{-1cm}\includegraphics[%
  scale=0.35,
  angle=270]{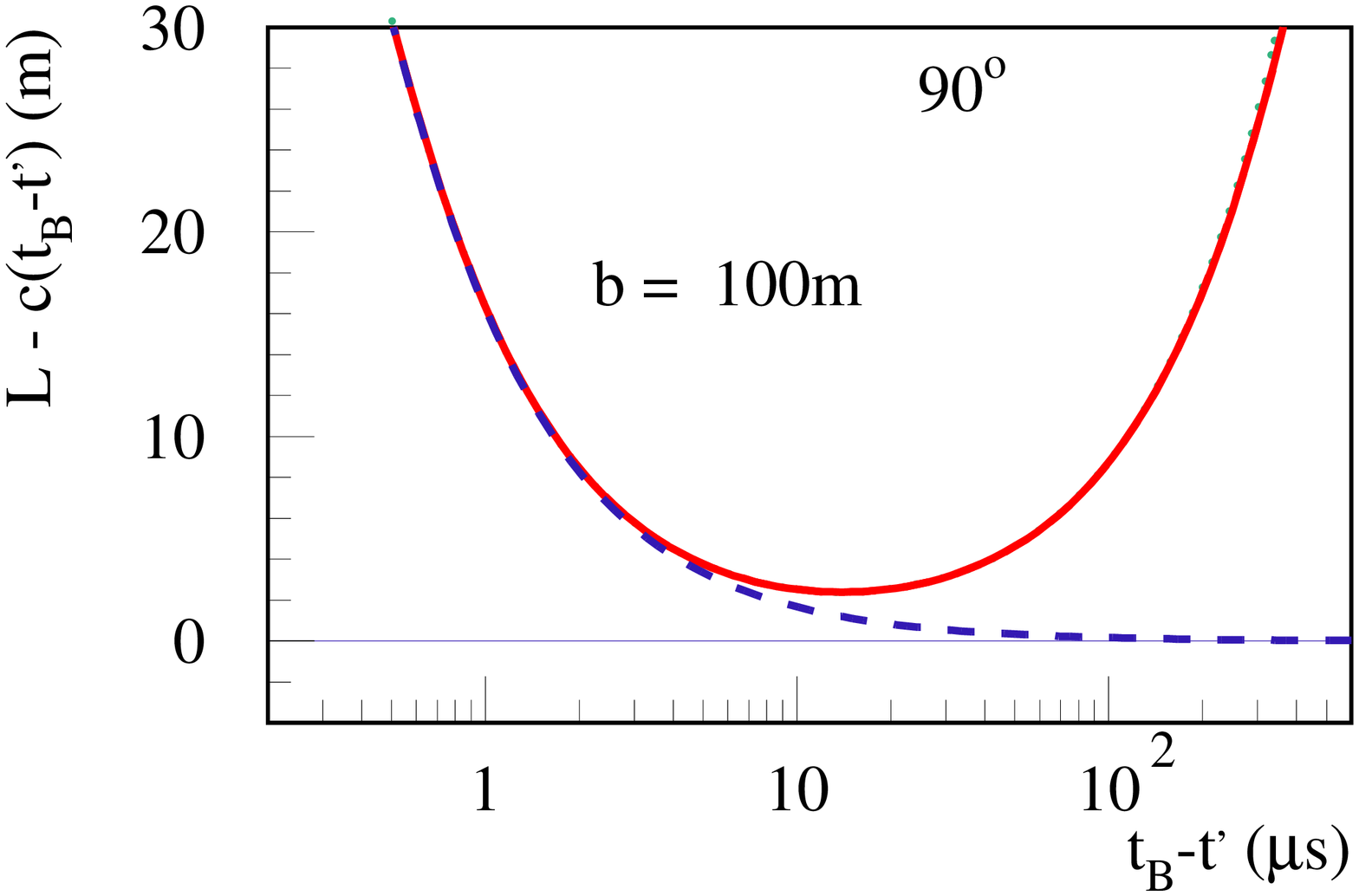}\end{center}
\vspace{-1.4cm}

\caption{The optical path length $L(\vec{\xi}(t'),\vec{x})$ minus $c(t_{B}-t')$
versus $t_{B}-t'$, for impact parameters of $500\,\mathrm{m}$ and
$100\,\mathrm{m}$, for different angles . The solid lines represent
the results for a realistic variable index of refraction.The other
curves refer to calculations for a constant $n$: namely $n=1$ (dashed)
and $n=n_{\mathrm{ground}}\approx1.0003$ (dotted). \label{cap:f1mif2}}
\end{figure*}

\begin{figure*}[htb]
\begin{center}~\vspace*{-2cm}\end{center}

\begin{center}\hspace*{-1cm}\includegraphics[%
  scale=0.35,
  angle=270]{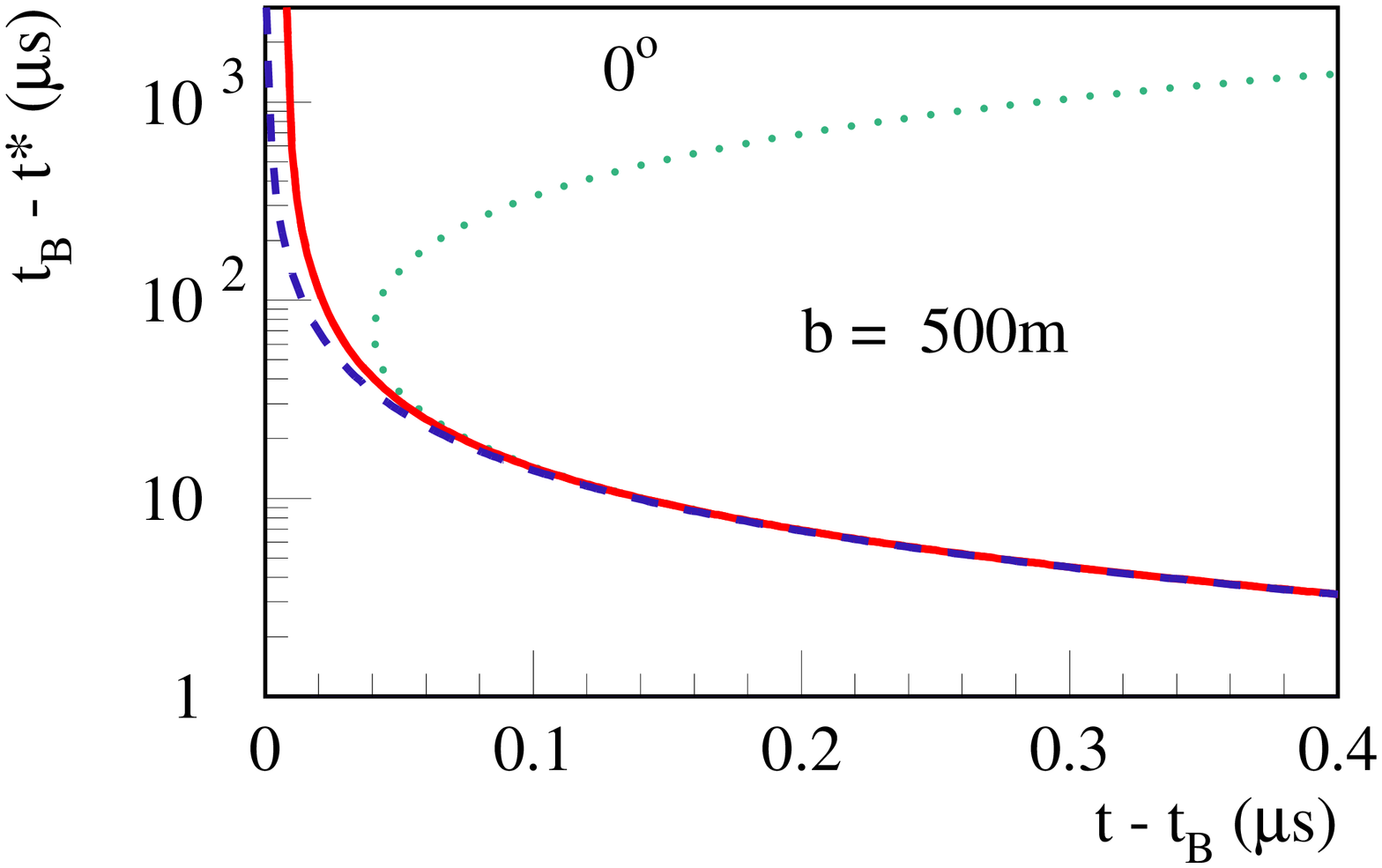}\hspace*{-1cm}\includegraphics[%
  scale=0.35,
  angle=270]{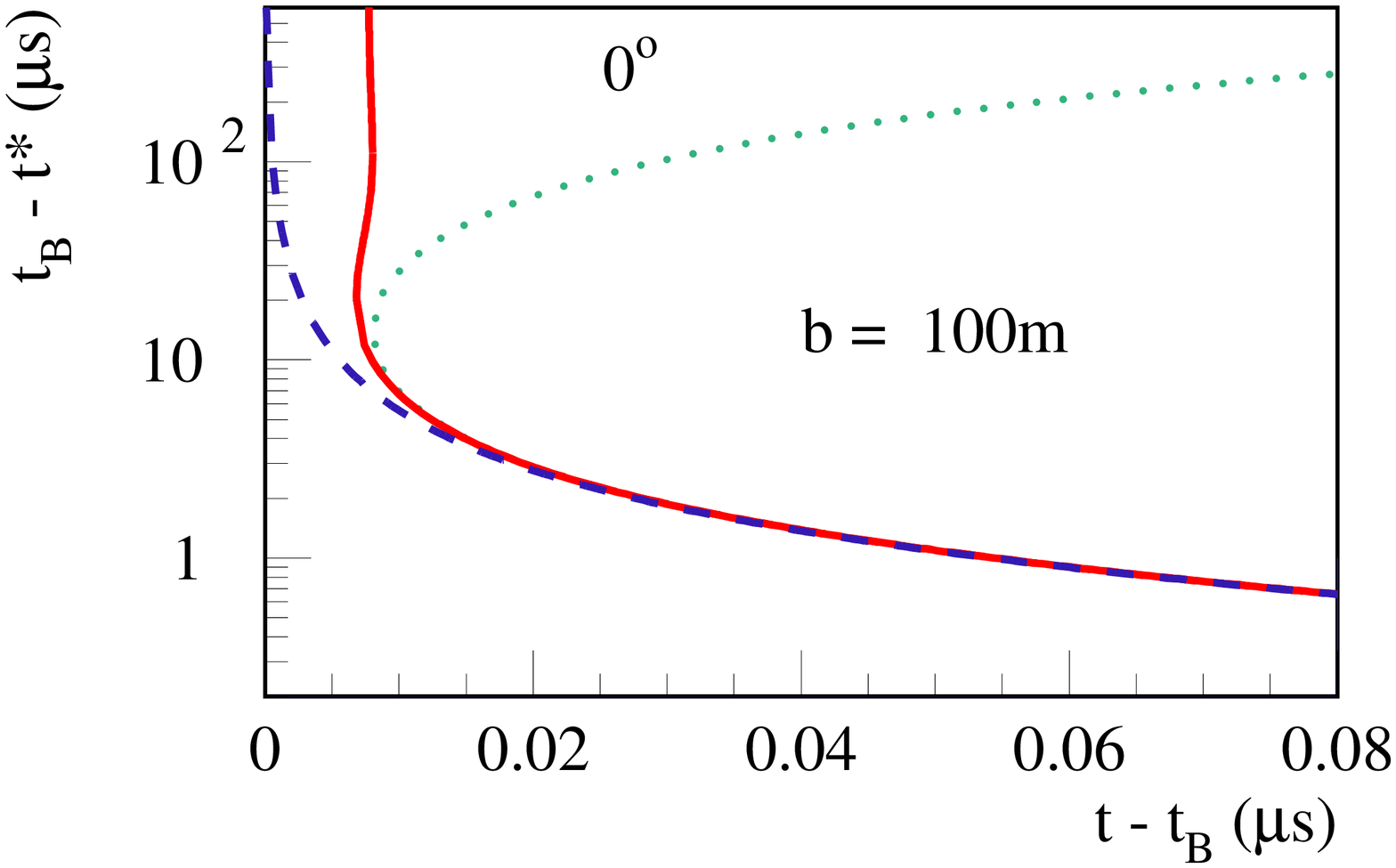}\end{center}

\vspace{-3.5cm}
\begin{center}\hspace*{-1cm}\includegraphics[%
  scale=0.35,
  angle=270]{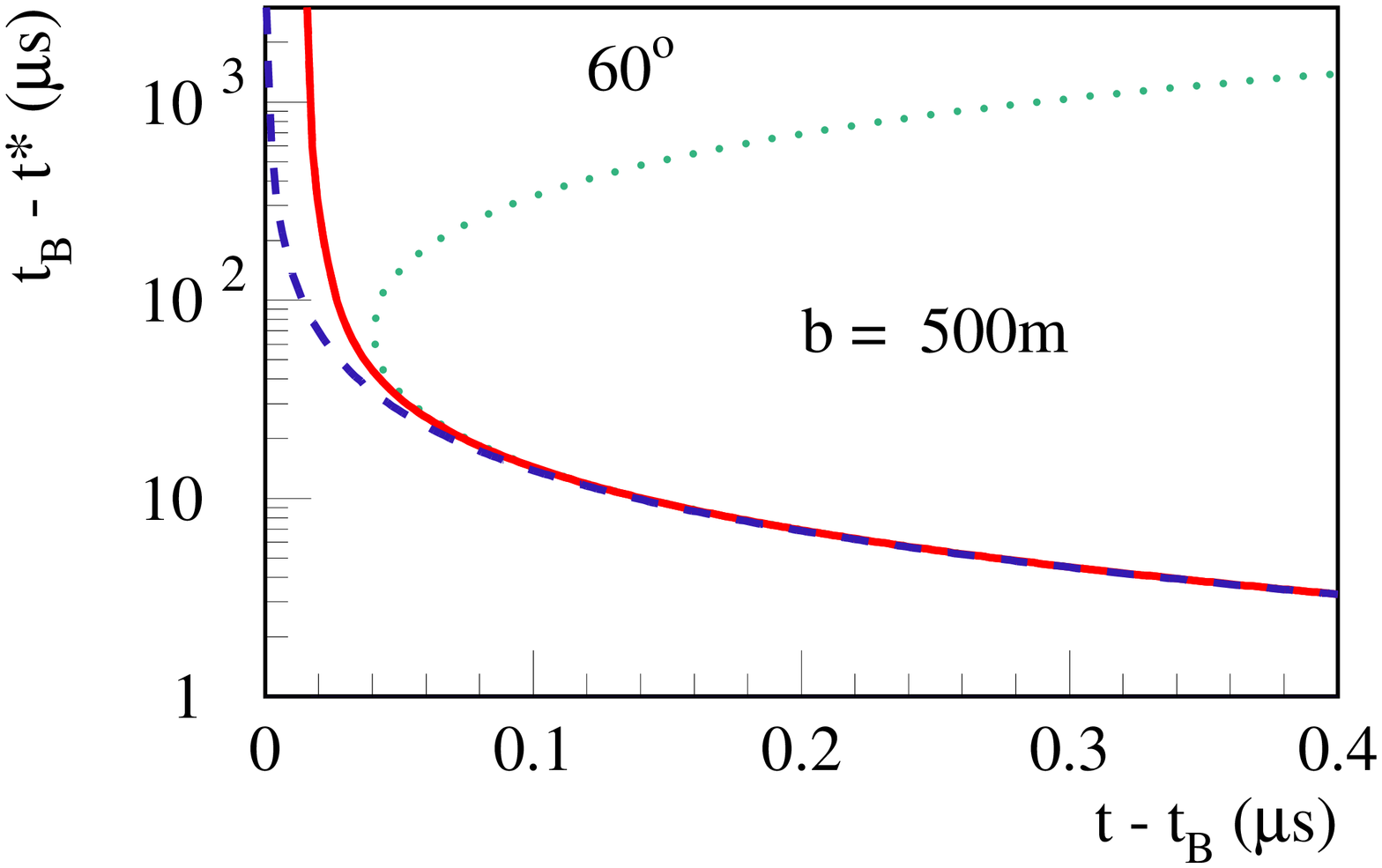}\hspace*{-1cm}\includegraphics[%
  scale=0.35,
  angle=270]{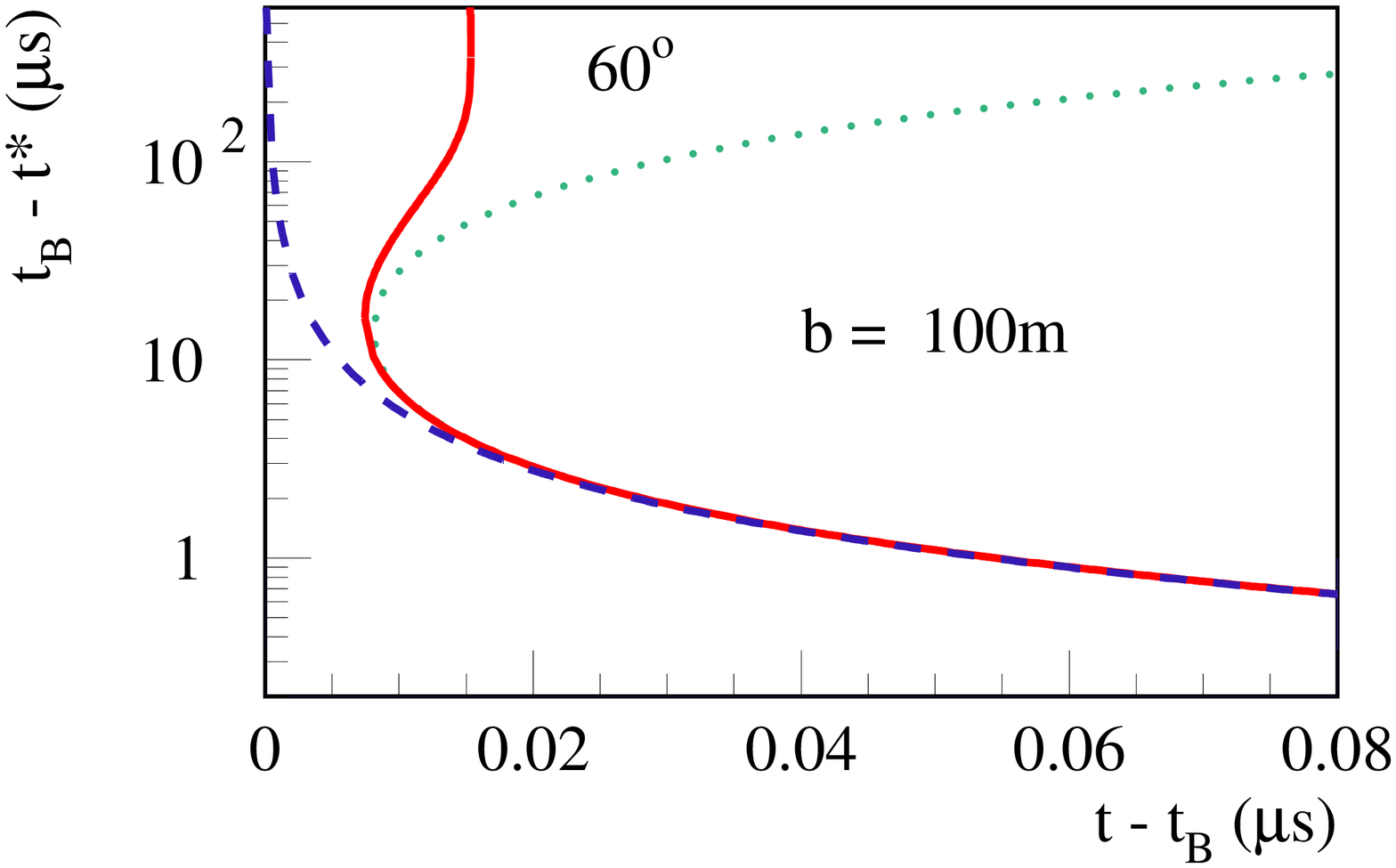}\end{center}
\vspace{-3.5cm}

\begin{center}\hspace*{-1cm}\includegraphics[%
  scale=0.35,
  angle=270]{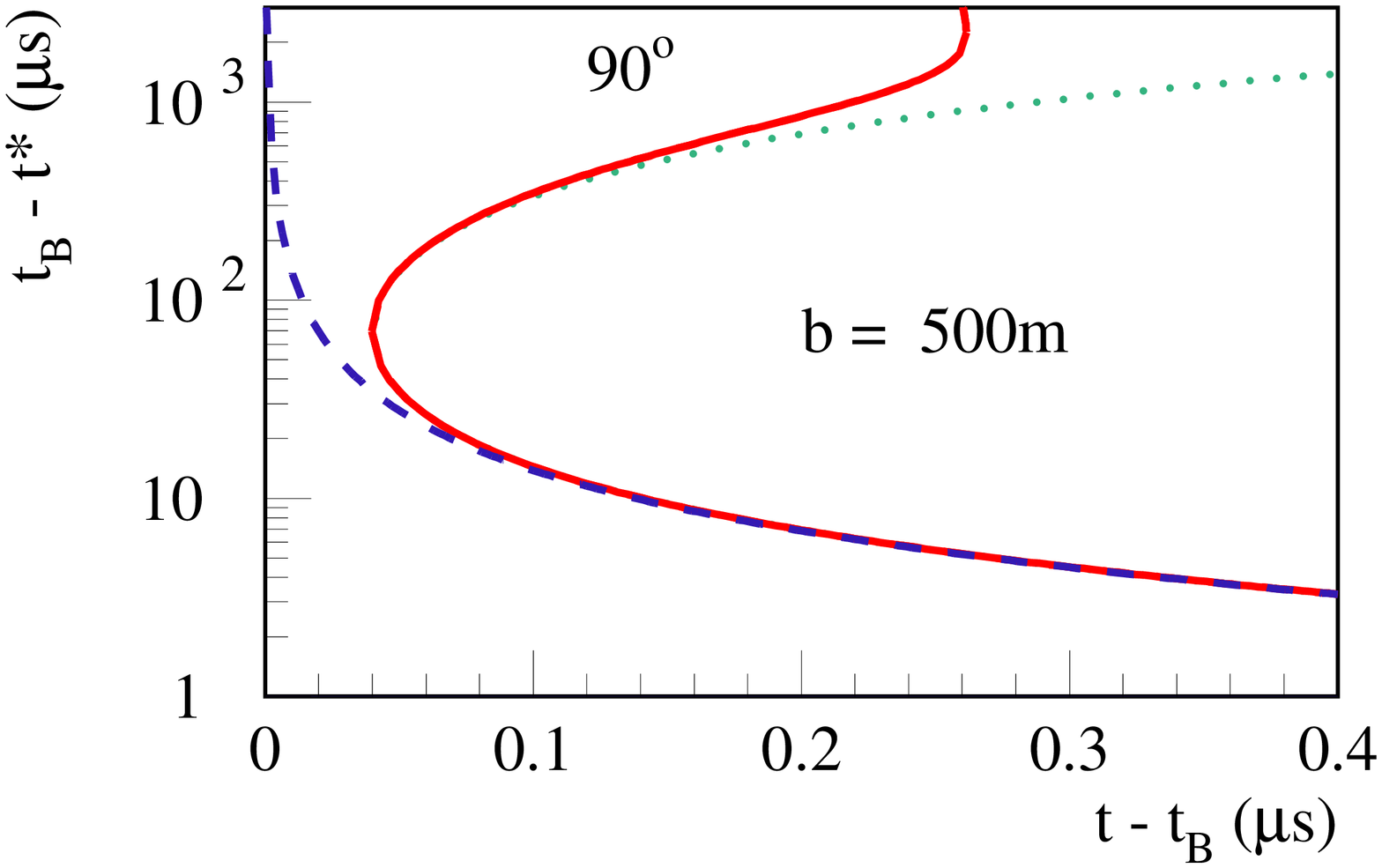}\hspace*{-1cm}\includegraphics[%
  scale=0.35,
  angle=270]{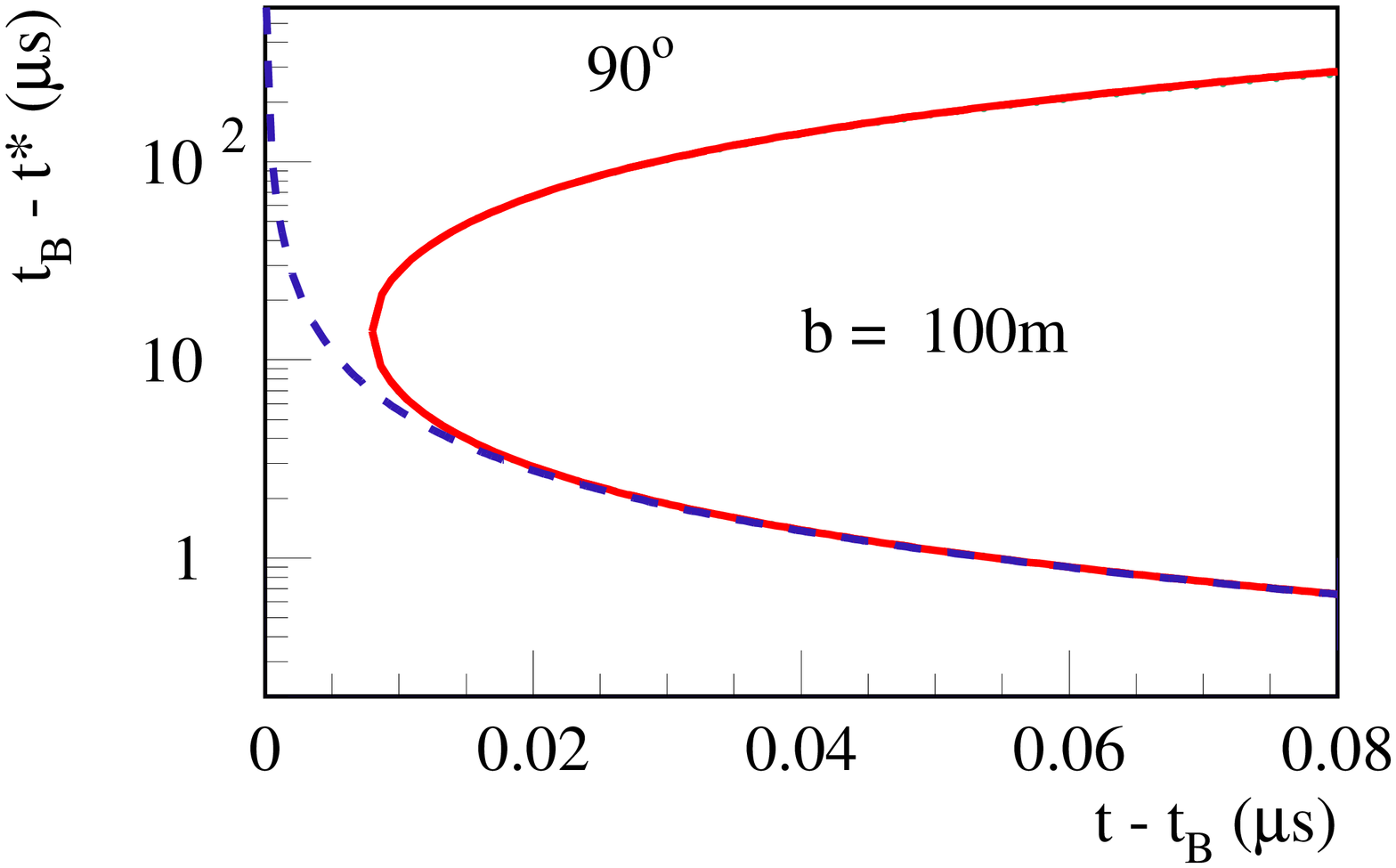}\end{center}
\vspace{-1.4cm}

\caption{The retarded time expressed via $t_{B}-t^{*}$ versus observer time
$t-t_{B}$, for impact parameters of $500\,\mathrm{m}$ and $100\,\mathrm{m}$,
for different angles . The solid lines represent the results for a
realistic variable index of refraction.The other curves refer to calculations
for a constant $n$: namely $n=1$ (dashed) and $n=n_{\mathrm{ground}}\approx1.0003$
(dotted). \label{cap:f1mif2b}}
\end{figure*}
\begin{figure*}[htb]
\begin{center}~\vspace*{-2cm}\end{center}

\begin{center}\hspace*{-1cm}\includegraphics[%
  scale=0.35,
  angle=270]{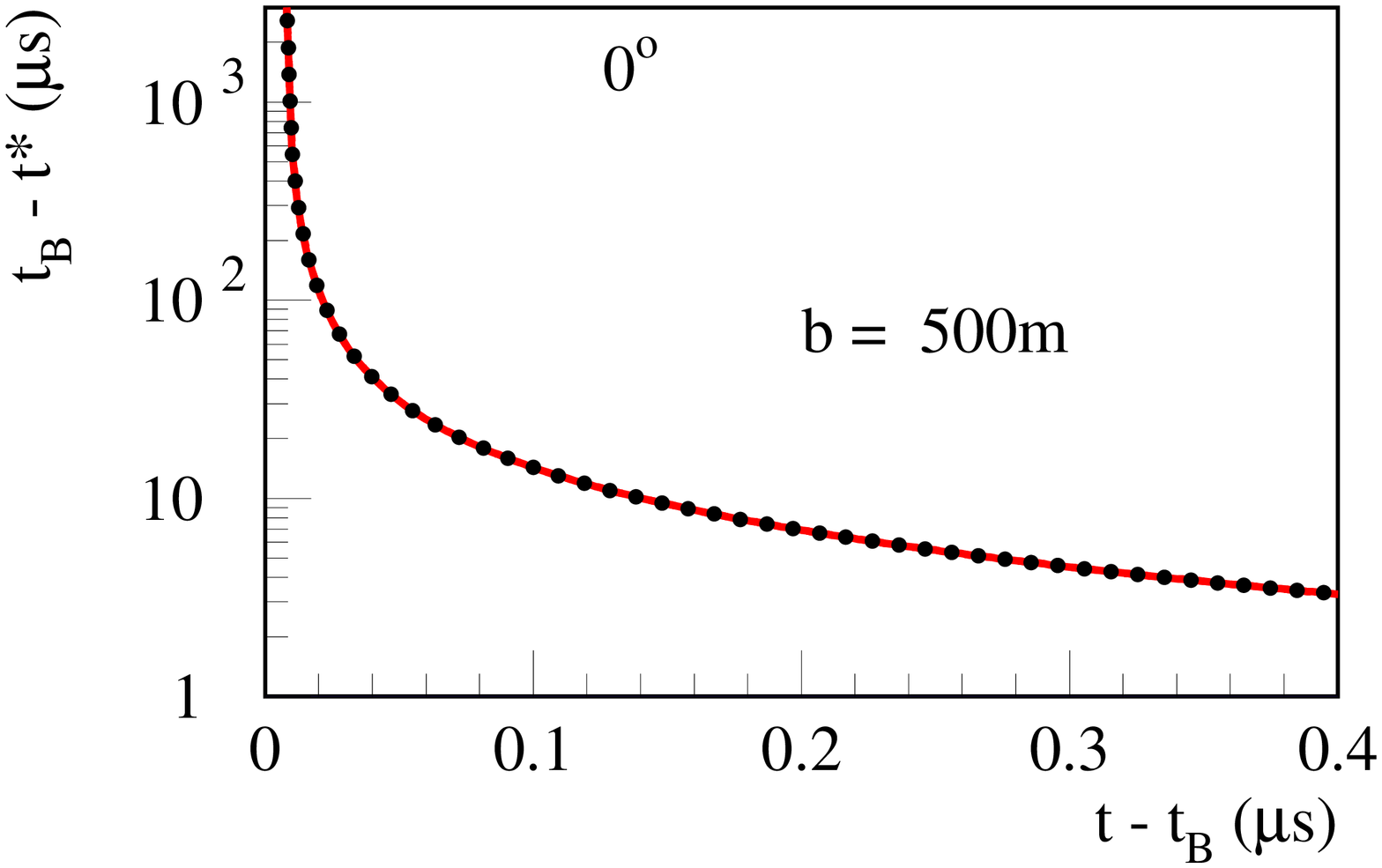}\hspace*{-1cm}\includegraphics[%
  scale=0.35,
  angle=270]{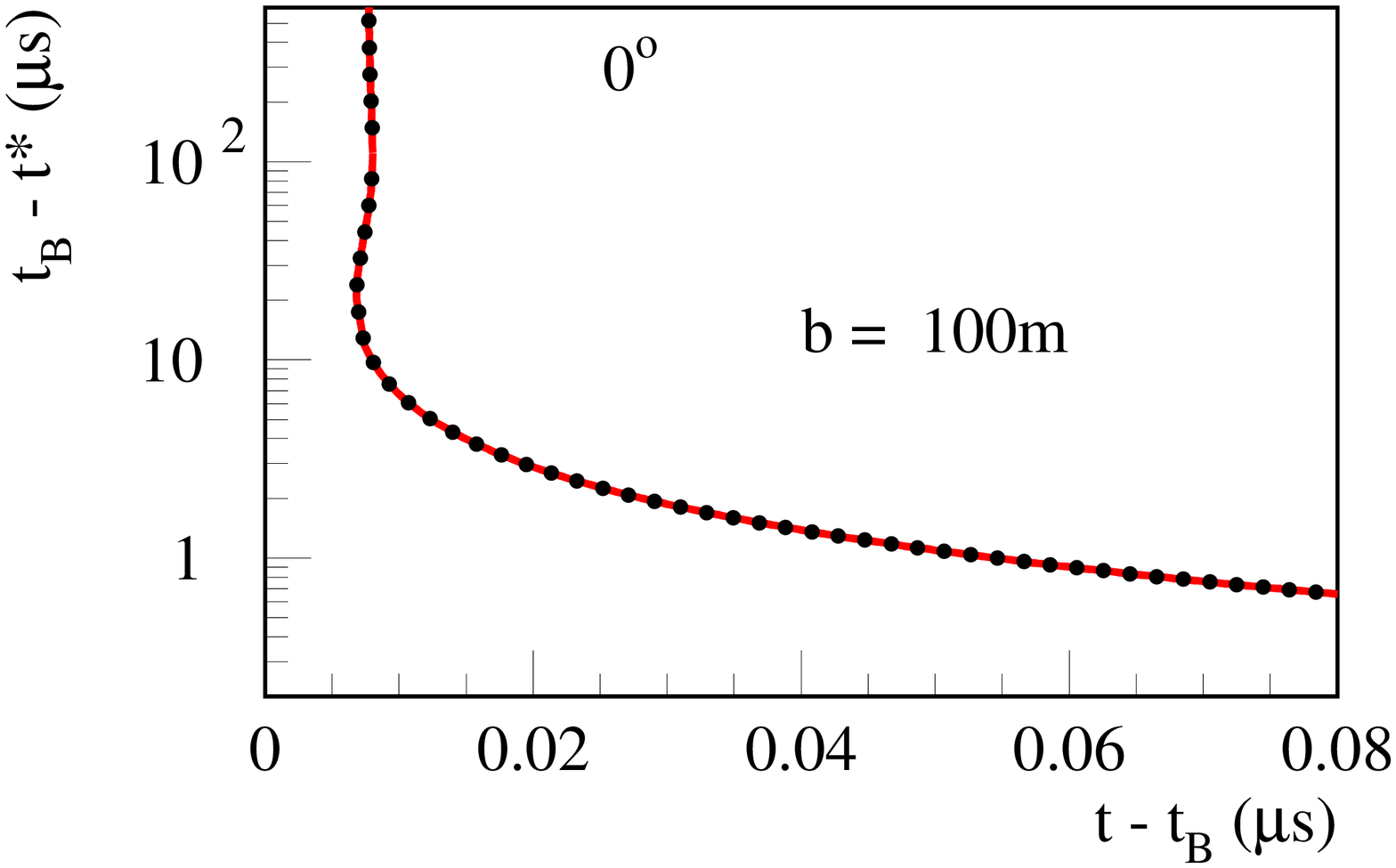}\end{center}

\vspace{-3.5cm}
\begin{center}\hspace*{-1cm}\includegraphics[%
  scale=0.35,
  angle=270]{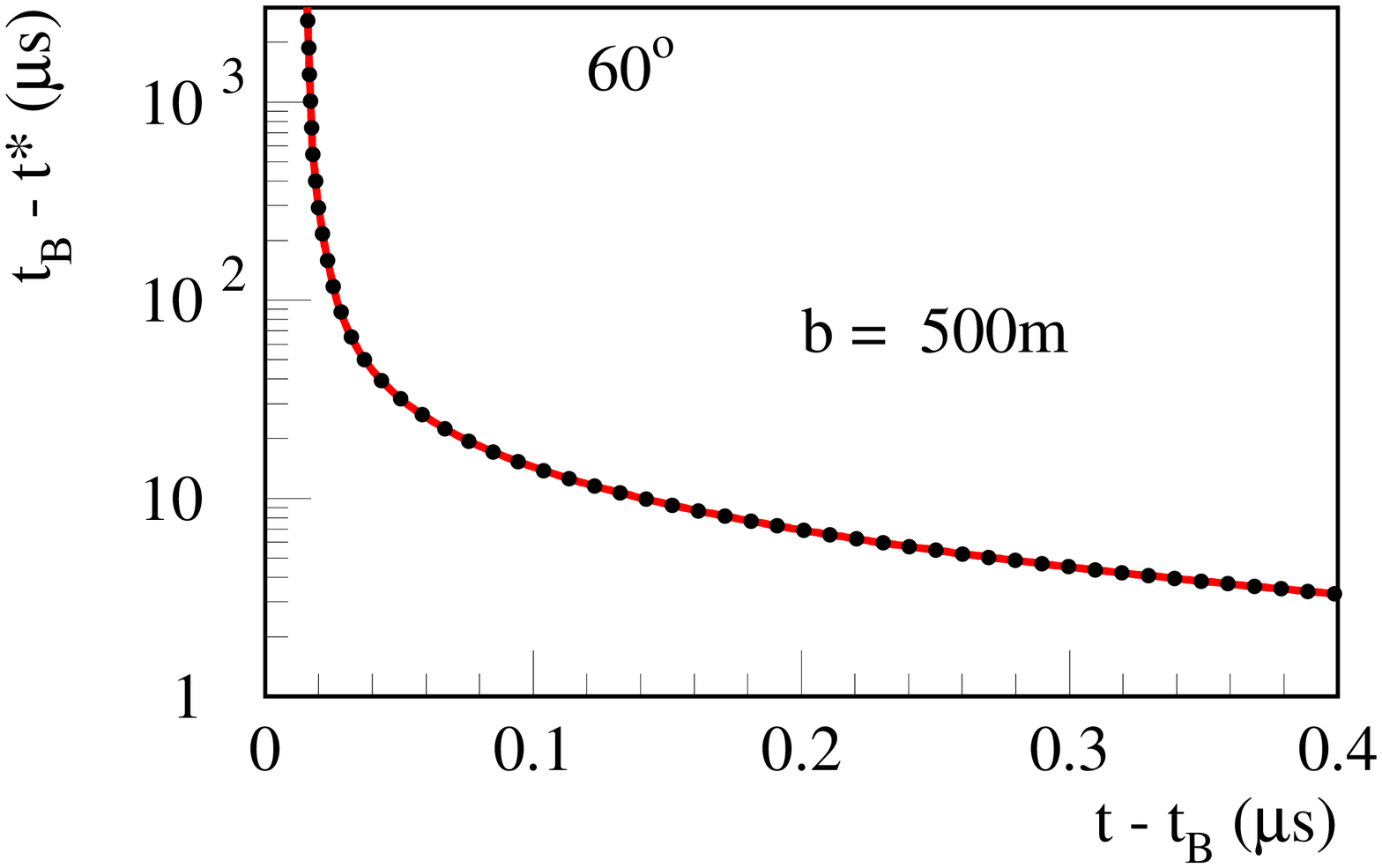}\hspace*{-1cm}\includegraphics[%
  scale=0.35,
  angle=270]{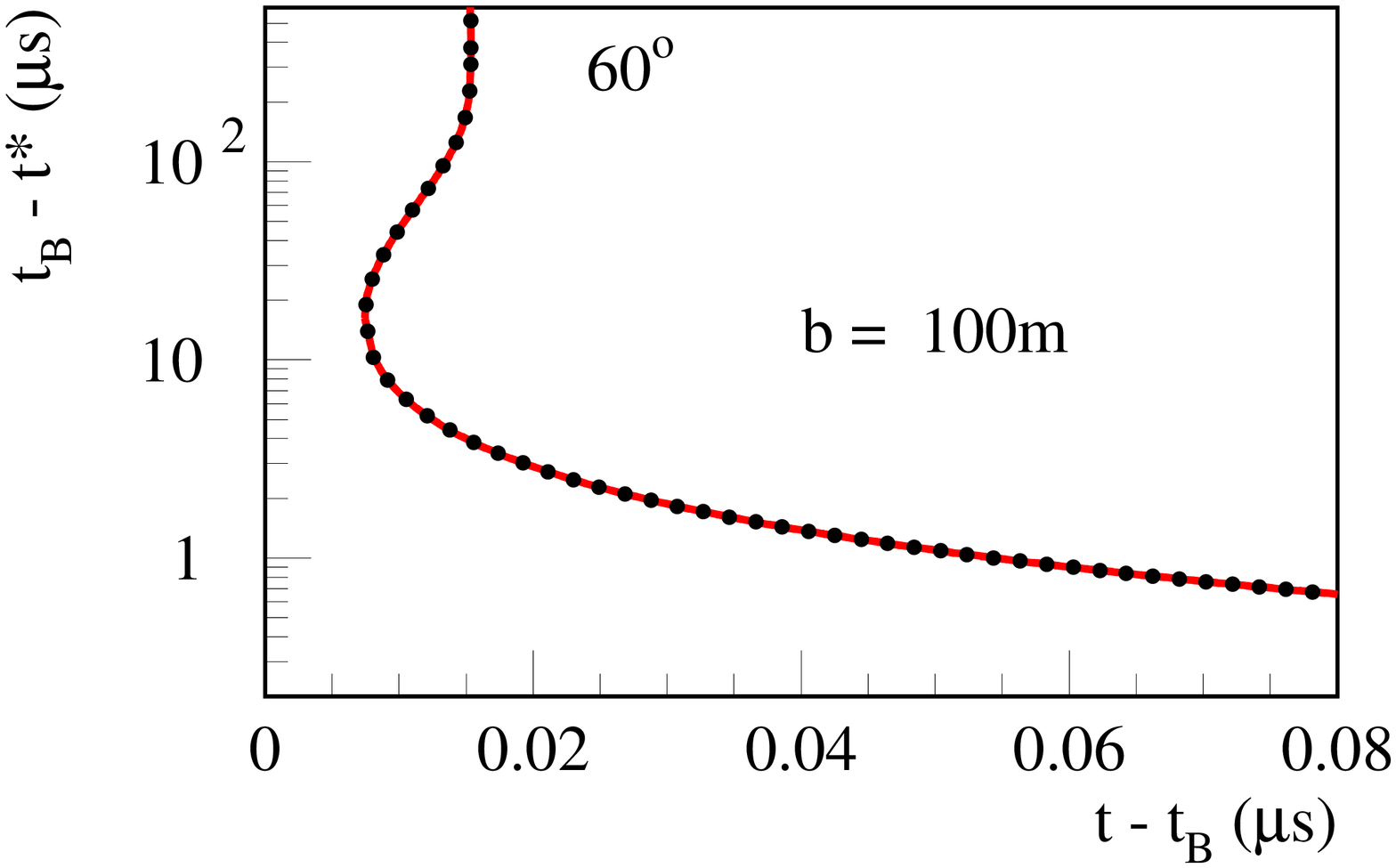}\end{center}
\vspace{-3.5cm}

\begin{center}\hspace*{-1cm}\includegraphics[%
  scale=0.35,
  angle=270]{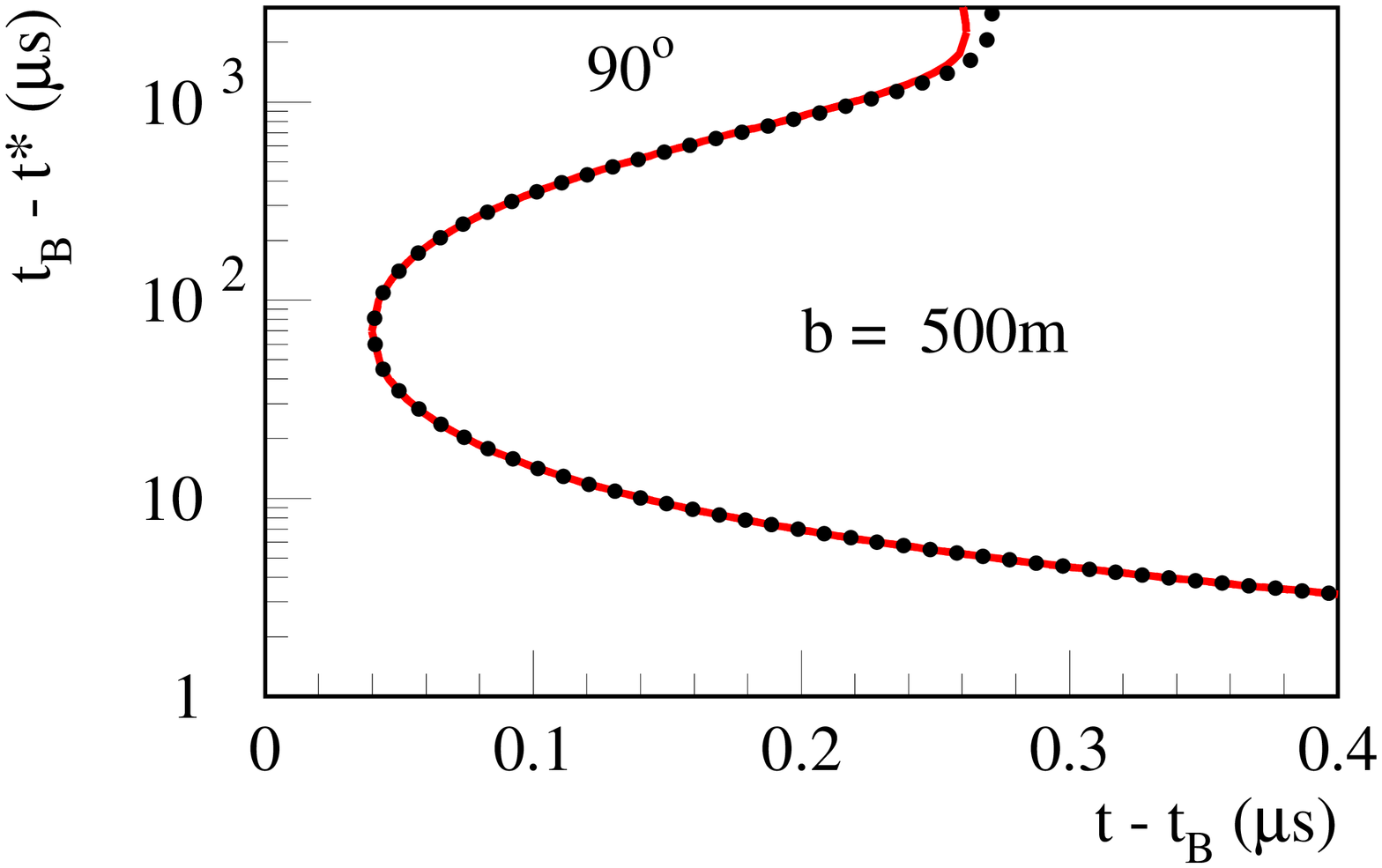}\hspace*{-1cm}\includegraphics[%
  scale=0.35,
  angle=270]{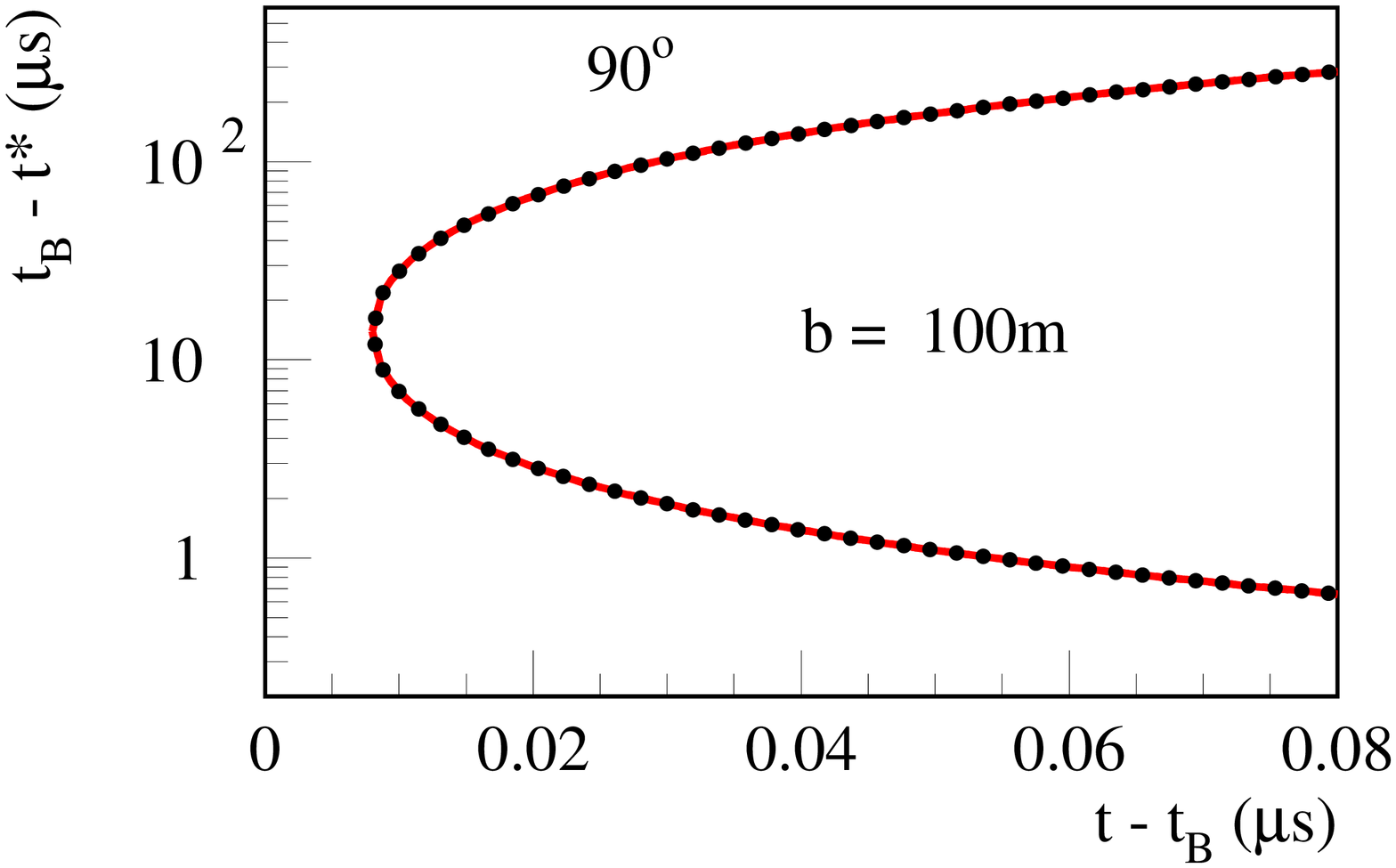}\end{center}
\vspace{-1.4cm}

\caption{The retarded time expressed via $t_{B}-t^{*}$ versus observer time
$t-t_{B}$, for impact parameters of $500\,\mathrm{m}$ and $100\,\mathrm{m}$,
for different angles, considering a realistic variable index of refraction.
The solid lines represent the exact results, the dotted lines refer
to the {}``straight line approximation''. \label{cap:f1mif2c}}
\end{figure*}

To understand the results, it is useful to consider two limiting cases,
namely $n=1$ and $n=n_{\mathrm{ground}}\approx1.0003$ (the reality
lying in between). For constant $n$, the l.h.s. of eq. (\ref{eq:tstar})
reads\begin{equation}
L-c(t_{B}-t')=n|\vec{R}|-c(t_{B}-t'),\end{equation}
where $\vec{R}$ has been defined earlier to be the vector joining
the emission point on  the shower and the position of the observer.
 One gets\begin{equation}
L-c(t_{B}-t')=n\sqrt{v^{2}(t_{B}-t')^{2}+b^{2}}-c(t_{B}-t').\label{eq:f1f2}\end{equation}
The expression $L-c(t_{B}-t')$ divided by $b$ is an universal function
of $\tau=c(t_{B}-t')/b$, namely\begin{equation}
n\sqrt{\beta^{2}\tau^{2}+1}-\tau,\end{equation}
with $\beta=v/c$. For $n=1$ and $\beta<1$, we have a monotonically
decreasing function with no lower limit. In all the following considerations,
we always consider a fast shower, with $v=c$, so $\beta=1$. In this
case, we still have a monotonically decreasing function, but with
a lower limit equal to zero (for $\tau\to\infty$). Keeping $\beta=1$,
but taking $n>1$, the curve reaches a minimum at some finite value
of $\tau$, and then increases. We show the corresponding results
in fig. \ref{cap:f1mif2}, where we plot $L-c(t_{B}-t')$ versus $t_{B}-t'$,
for two impact parameters $b$ and several inclinations. The dashed
curve refers to $n=1$, the dotted one to $n>1$. To underline the
scaling behavior, we scaled abscissas and ordinates as $1/b$, and
indeed we observe, for all angles and impact parameters, the same
 curves, for the two limiting cases with constant $n$. 

Also shown in fig. \ref{cap:f1mif2} are the curves $L-c(t_{B}-t')$
versus $t_{B}-t'$, for the case of a realistic index of refraction
(full lines), obtained  from the numerical procedures discussed above.
 As expected, these curves are well in between the two limiting cases
$n=1$ and $n=n_{\mathrm{ground}}$. The scaling is violated: depending
on impact parameter and inclination, the exact curve follows more
closely either the $n=1$ or the $n=n_{\mathrm{ground}}$ case. Whereas
for vertical showers the exact curves are close to the $n=1$ case,
we see an increasing deviation with increasing inclination, up to
 the extreme at $90^{0}$, where the curves are close to the results
for $n=n_{\mathrm{ground}}$. All this is understandable, since, with
increasing angle, even light emitted at early times (large $t_{B}-t'$)
travels more and more at $n\approx n_{\mathrm{ground}}$. It should
be noted that at very large $t_{B}-t'$, the curves $L-c(t_{B}-t')$
approach a constant, for obvious reasons, and there will be a (at
least one) turning point.

The retarded time $t'=t^{*}$ is the solution of \begin{equation}
L-c(t_{B}-t')=c(t-t_{B}),\end{equation}
for a given observer time $t$. In other words, $t^{*}$ is the intersection
of the curves $L-c(t_{B}-t')$, shown in fig. \ref{cap:f1mif2}, with
a horizontal line with ordinate $c(t-t_{B})$. In the range of $t_{B}-t'$
up to $3000\,\mu\mathrm{s}$ (useless to go to earlier times), we
have no solution for $t<t_{\mathrm{min}}$, while for larger values
there are one, two, or even more solutions. For the minimum time we
have $t_{\mathrm{min}}>t_{B}$, where the precise value depends on
angle and impact parameter. We compute the retarded times $t^{*}$
numerically. A first estimate is obtained via interpolation using
tables of $L-c(t_{B}-t')$, a refinement is obtained by again solving
the differential equation eq. (\ref{eq:de}), but now with the boundary
condition \begin{equation}
\vec{y}(0)=\vec{x},\quad\vec{y}(t-t^{*})=\vec{\xi}.\end{equation}
Only few Newton-Raphson iterations are needed to get a very high precision.
The results are shown in fig. \ref{cap:f1mif2b}, where we plot $t_{B}-t^{*}$
versus $t-t_{B}$, with $t^{*}$ being the retarded time. We get of
course what we expect after the above discussion. In case of $n=1$,
we have no solution for negative $t-t_{B}$, for positive values there
is exactly one solution, representing a monotonically decreasing curve.
The other limiting case $n=n_{\mathrm{ground}}$ provides a two-valued
curve and we have two solutions for $t-t_{B}$ being bigger than some
minimum value, no solutions otherwise. The results for a realistic
index of refraction (full lines) lies in between the two limits. For
large impact parameters (like $500\,\mathrm{m}$), the exact curves
are close to the $n=1$ case, with increasing angle and decreasing
impact parameter, the curves get more and more close to the other
limiting case, leading to two or even more retarded times, for a given
observer time $t$. This means that for the electric fields one has
to sum up the contributions due to several retarded times, if the
currents $J(t^{*})$ at these times are non-zero. 

From the above discussion -- and in particular from the comparison
of the exact calculation with the limiting cases of constant index
of refraction -- one gets the impression that the main effect is not
really due to curved light trajectories but to the fact that the index
changes along the trajectory. In order to verify the conjecture, we
compute the optical path length in {}``straight line approximation''
as\begin{equation}
L_{\mathrm{straight}}(\vec{\xi}(t'),\vec{x})=\int_{GC}n(\vec{y})ds,\end{equation}
where $GC$ is the straight line joining abserver position $\vec{x}$
and charge $\vec{\xi}(t')$. Subsequently we compute the retarded
times by solving $L_{\mathrm{straight}}=R^{0}$. The results are shown
in fig. \ref{cap:f1mif2c} as dotted lines, together with the exact
curves (full lines). The two results are almost indistinguishable,
with the only exception of a small deviation for $b=500\,\mathrm{m}$
and $\theta=90^{o}$, at very early retarded times. The straight line
approximation is therefore for most applications largely sufficient.

\begin{figure}[htb]
\begin{center}\hspace*{-1cm}\includegraphics[%
  scale=0.35,
  angle=270]{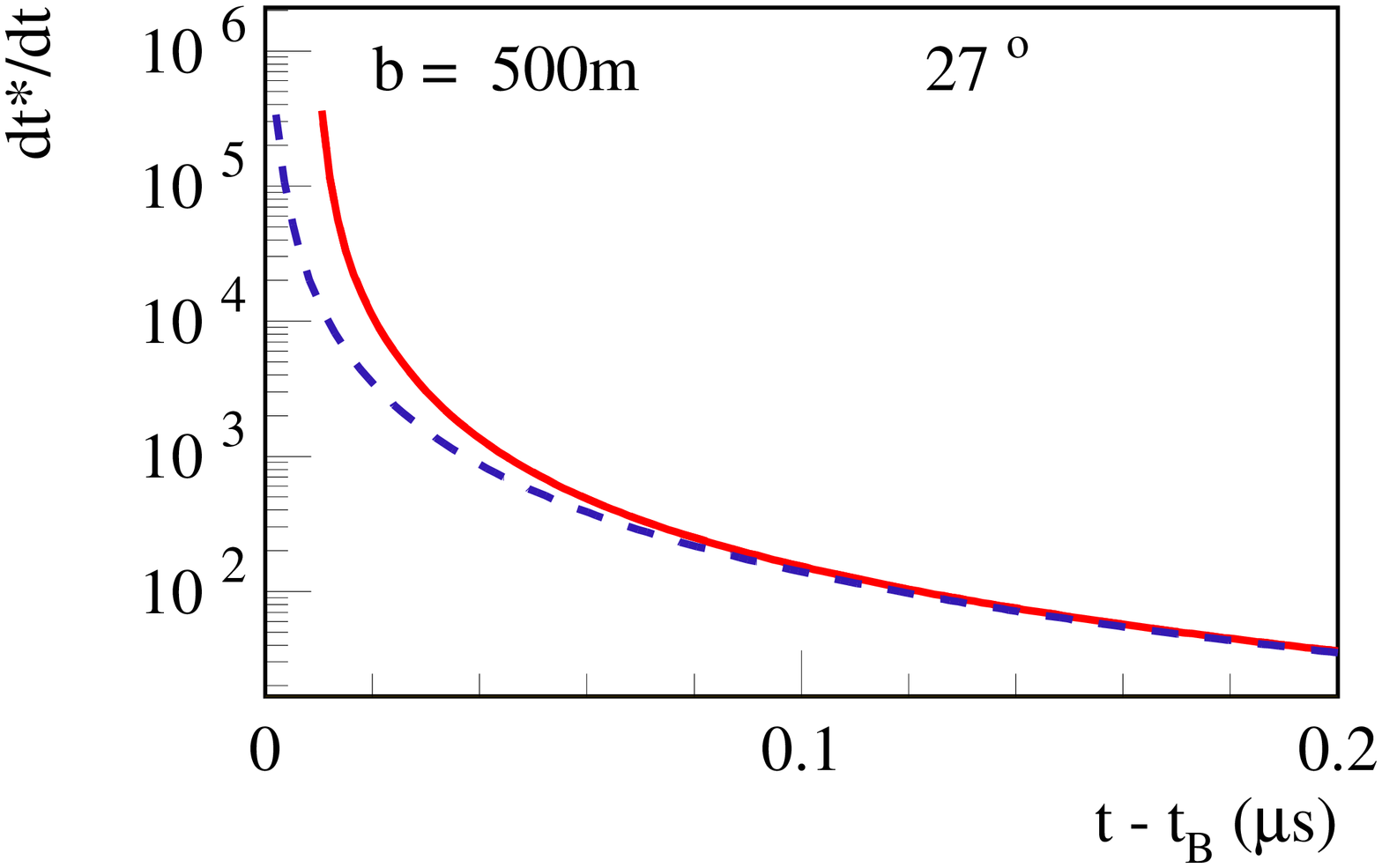}\end{center}
\vspace{-1cm}

\caption{The derivative $dt^{*}/dt$ versus $t-t_{B}$\label{cap:dtt-dt},
for \textbf{$b=500\,$}m and $\theta=27^{0}$, in case of a realistic
index of refraction (full line) and the case $n=1$ (dashed).}
\end{figure}

A delicate point is related to the derivative $\partial t^{*}/\partial t$.
We have \begin{equation}
\frac{1}{\widetilde{R}V}=\frac{1}{c(t-t^{*})}\,\frac{\partial t^{*}}{\partial t},\end{equation}
showing that possible singularities of the expression for the electric
field,  corresponding to the roots of $\widetilde{R}V$, are related
to singularities of $\partial t^{*}/\partial t$. In the simplest
case of a large impact parameter and small angles, $t_{B}-t^{*}$
is a monotonic function of $t-t_{B}$, and thus $\partial t^{*}/\partial t$
is as well, as shown in fig. \ref{cap:dtt-dt}. Compared to the case
$n=1$ (dashed) the exact curve (full line) grows more rapidly for
$t-t_{B}\to0$. For larger angles, we obtain singular points, where
$\partial t^{*}/\partial t$ diverges. This means that whenever we
have a situation where $dt^{*}/dt$ is singular, and in addition the
current is non-vanishing at the corresponding retarded time, we have
to consider the finite extension of the source, which amounts to an
integration over the (square root) singularity, which gives a finite
result. 

It should be noted that the singularities of $\partial t^{*}/\partial t$
correspond to the singularities that gives rise to Cherenkov radiation.
 It reflects the fact that electromagnetic waves emitted at different
times  arrive simultaneously at the observer position. 

The retarded time where the first singularity may occur depends on
the impact parameter: it is proportional to $b$. It is somewhat larger
than $10\,\mu\mathrm{s}$ for $b=100\,\mathrm{m}$, corresponding
to a source more that $3\,\mathrm{km}$ away.

To avoid this discussion related to singularities for the moment (to
be treated in a forthcoming publication), we restrict ourselves to
situations which are {}``well behaved'' in the sense that $\partial t^{*}/\partial t$
is nonsingular there where the currents are non-vanishing.

\section{The role of the positive ions}

For the individual charged paticles the magnitude of $J^{0}$ turns
out to be always bigger than that of $J^{z}$, even if one turns off
the magnetic field. This is due to the fact that the electrons move
slower than the shower front. They are constantly loosing energy and
eventually drop out. Close to the shower maximum we have a steady
state since here the number  of net charge created is equal to the
net charge dropped.  On the other hand, the net charge created must
correspond to a corresponding positive charge in terms of ionized
atoms, and therefore the net charge behind the shower (and in front
of it anyway) is zero. So we have effectively a moving charge, moving
with the speed of the shower front, which means that the $z$ and
the $0$ component of the total, collective, effective current must
be equal.  As mentioned above, we do not get this identity if we would
only consider  electrons and positrons, and not the ions.  The ions
will contribute to $J^{0}$, not to the other components of the current
(they are not moving) to yield  finally $J_{\mathrm{tot}}^{0}=J_{\mathrm{tot}}^{z}\equiv J^{z}$,
where {}``tot'' refers to the fact that we consider all, electrons
and ions. In the following calculations, we will always use $J_{\mathrm{tot}}^{0}=J^{z}$,
although strictly speaking this is only valid close to the maximum,
but on the other hand the most important contribution to the fields
is due to the charges close to the maximum. For the rest of the paper
$J$ will refer to the total current, including the ion part. 

\begin{figure}[htb]
\begin{center}\includegraphics[%
  scale=0.27]{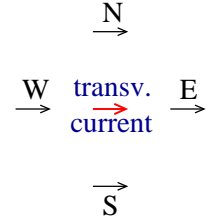}\end{center}
\vspace{-0.3cm}

\caption{The electric field vectors from the contribution $\vec{E}_{JV}$,
being proportional to the transverse current $\vec{J}^{\bot}$. \label{cap:polar1}}
\end{figure}

The property $J^{0}=J^{z}$ has important consequences concerning
the polarity of the signal. The electric field is given as a sum of
three terms, \begin{equation}
\vec{E}_{JV}=\frac{\vec{W}_{JV}}{D},\quad\vec{E}_{RK}=\frac{\vec{W}_{RK}}{D},\quad\vec{E}_{RJ}=\frac{\widetilde{V}V}{\widetilde{R}V}\,\frac{\vec{W}_{RJ}}{D},\end{equation}
with $\vec{W}_{AB}=\vec{A}B^{0}-A^{0}\vec{B}.$ We may split the vectors
$\vec{W}$ into components parallel and orthogonal to the shower axis.
The components of the shower velocity are $V^{0}=1,$ $\vec{V}^{\bot}=0$,
$V^{\Vert}=\beta$, with $\beta$ being very close to unity. We thus
find\begin{equation}
W_{JV}^{\Vert}=J^{0}(1-\beta)\approx0\end{equation}
and\begin{equation}
\vec{W}_{JV}^{\bot}=\vec{J}^{\bot}.\end{equation}
So the field $\vec{E}_{JV}$ is almost purely transverse, parallel
to the transverse current, as sketched in figs. \ref{cap:polar1}
, where we show the electric field vectors for four different observers
situated in a plane transverse to the shower axis. %

\begin{figure}[htb]
\begin{center}~\end{center}

\begin{center}\includegraphics[%
  scale=0.27]{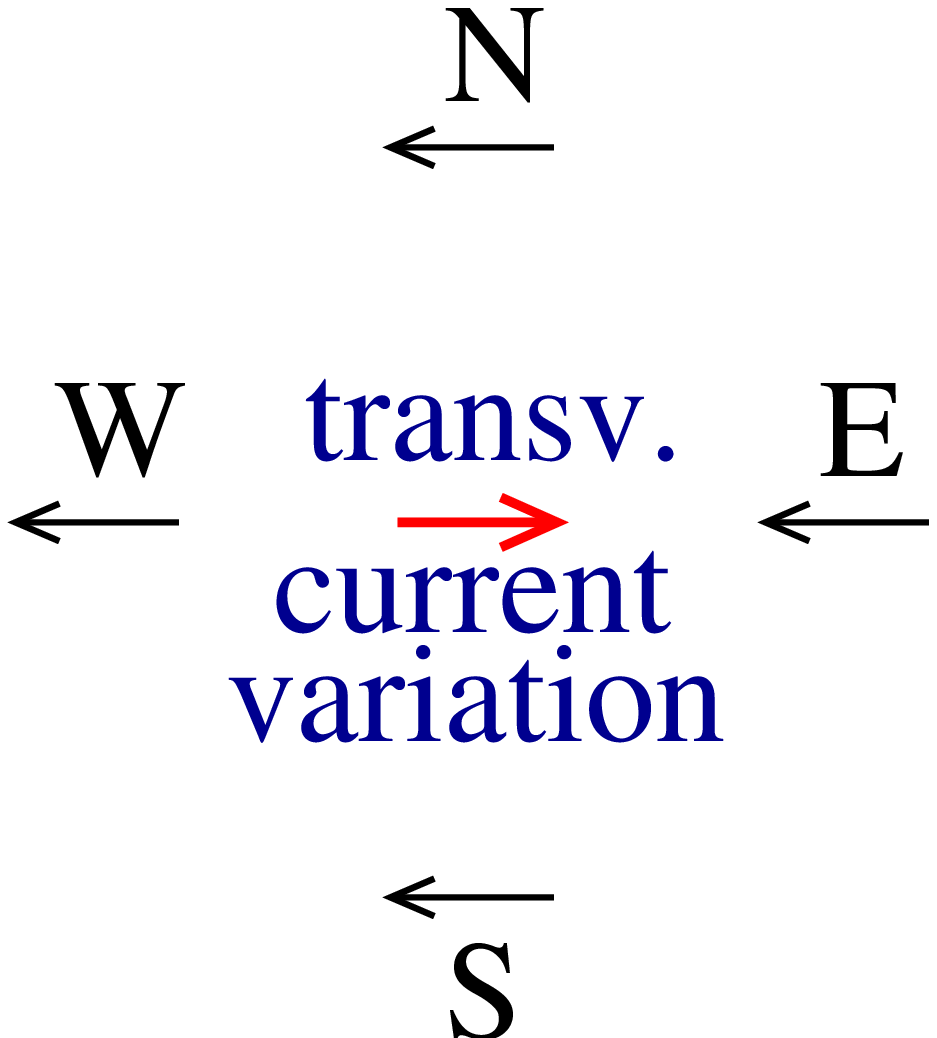}$\qquad$ \includegraphics[%
  scale=0.27]{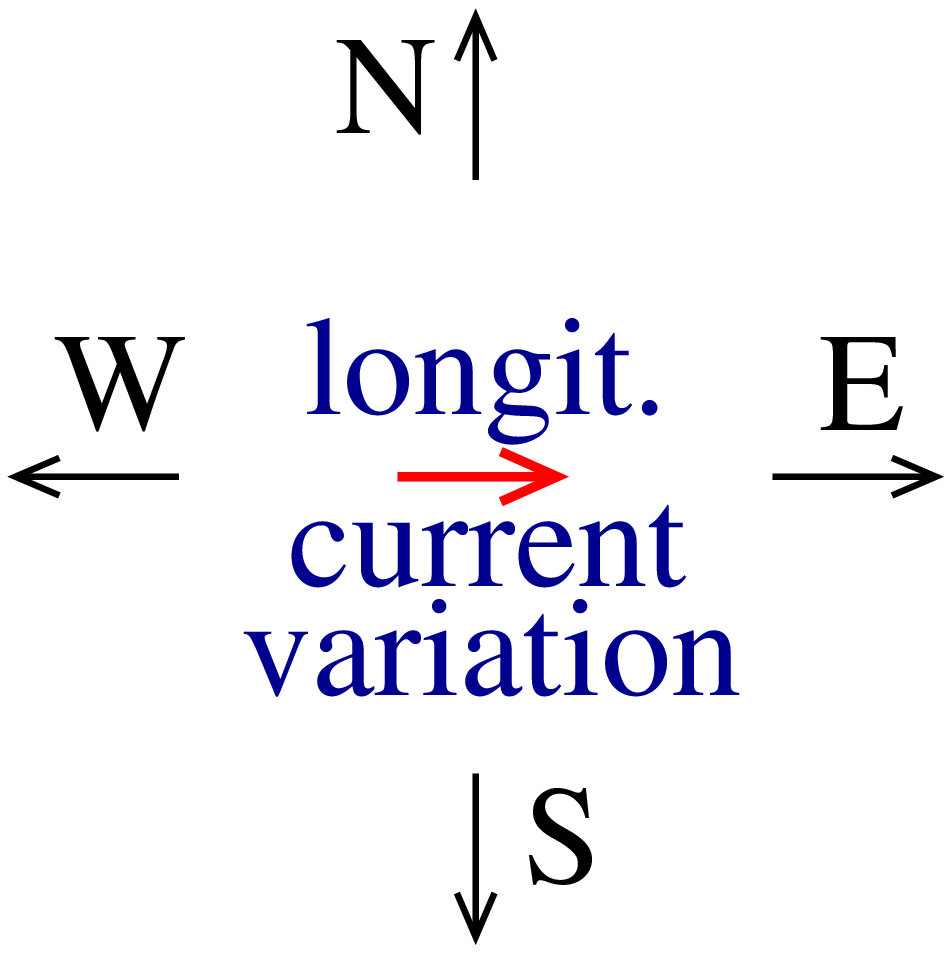}\end{center}
\vspace{-0.3cm}

\caption{The electric field vectors from the contribution $\vec{E}_{RK}$;
one part is proportional to the time variation $\vec{K}^{\bot}$ of
the transverse current $\vec{J}^{\bot}$(left plot), a second part
is proportional to the variation $K^{0}$ of the longitudinal current,
and proportional to $\vec{R}^{\bot}$ and therefore radial (right
plot). \label{cap:polar2}}
\end{figure}

Furthermore, we find $W_{RK}^{\Vert}\approx\widetilde{R}V\, K^{0}$,
(where {}``$\approx$'' here means correct to a precision of $n-1$)
which can be proven by using $1=(\partial^{0}-\partial^{\Vert})ct^{*}$$\approx(R^{0}-R^{\Vert})/\widetilde{R}V$,
due to time translation invariance. Since the field in general is
large when $\widetilde{R}V$ is small, we may conclude that $\vec{W}_{RK}^{\Vert}$
is small compared to the transverse component, which we express as
\begin{equation}
W_{RK}^{\Vert}=\widetilde{R}V\, K^{0}\approx0.\end{equation}
For the transverse part we find\begin{equation}
\vec{W}_{RK}^{\bot}=\vec{R}^{\bot}K^{0}-R^{0}\vec{K}^{\bot},\end{equation}
which means that here not only the transverse current (or better:
its time derivative) contributes, but also the longitudinal one via
$K^{0}$. Whereas the latter one is radial with respect to the axis,
the former one is anti-parallel to the direction of the transverse
current, see fig. \ref{cap:polar2}. %

We said earlier that the third term $\vec{E}_{RJ}$ is small, away
from singularities; let us investigate this term more closely. Using
the same arguments as for $W_{RK}$, we find\begin{equation}
E_{RJ}^{\Vert}=\frac{\widetilde{V}V}{\widetilde{R}V}\,\frac{\widetilde{R}V\, J^{0}}{D}=\widetilde{V}V\,\frac{\, J^{0}}{D}\approx0,\end{equation}
where {}``$\approx0$'' means small compared to the transverse component,
which is\begin{equation}
\vec{E}_{RJ}^{\bot}=\frac{\widetilde{V}V}{\widetilde{R}V}\,\frac{\vec{R}^{\bot}J^{0}-R^{0}\vec{J}^{\bot}}{D}.\end{equation}
Lets us consider (just here, for an estimate) a constant index of
refraction $n>1$ and let us assume that $\beta>n^{-1}$. The prefactor
$\widetilde{V}V/\widetilde{R}V$ would then be proportional to $\gamma^{-2}(1-n\beta\,\mathrm{cos}\alpha)^{-1}$,
with $\alpha$ being the angle between the shower axis and $\vec{R}$.
Even though $\gamma$ is large, the term contributes when $n\beta\,\mathrm{cos}\alpha$
approaches unity, which corresponds to the above-mentioned singularity.
This is actually the well known {}``Cherenkov radiation''. It should
be noted that these singularities have to be treated very carefully,
using the realistic position dependent index of fraction $n(\vec{y})$,
and naively replacing $\widetilde{V}$ $\widetilde{R}$ by $V$ and
$R$ is certainly not permitted. In this paper we restrict ourselves
to not too small impact parameters, where the contribution $\vec{E}_{RJ}$
is negligible.

\section{Dipole field}

So far we discussed electric fields created by the current due to
the charge excess -- taking into account the positive ions -- and
the fields due to transverse current induced by the geomagnetic field.
However, this magnetic field does not only create a transverse current,
but also a dipole moment $\vec{P}$ originating from the separation
of the electric charges \cite{Kah66,olaf}. We define $q$ to be the
product $Ne$ of the number of electrons and positrons multiplied
by the elementary charge $e$, and we define the derivatives of $q$
as $q_{n}=c^{-n}\partial^{n}q/dt'^{n}$, for $n=1,2$. The ratios
$M^{k}=P^{k}/q$ are slowly varying (compared to the variation of
$q$) along the trajectory. The potential due to this dipole moment
may be written as\begin{equation}
A_{\mathrm{dip}}^{\beta}=-\vec{M}\vec{\nabla}A_{q}^{\beta}=-M^{k}\partial_{k}A_{q}^{\beta}\;,\end{equation}
where $A_{q}^{\beta}$ is the potential due to the current $J_{q}=c\, q\, V$,
representing a moving charge $q$. The electric dipole field is then
given as \begin{equation}
E_{\mathrm{dip}}^{i}=-M^{k}\partial_{k}E_{\mathrm{q}}^{i},\end{equation}
with $E_{q}^{i}$ being the electric field due to a moving charge
$q$. We may use the formulas derived earlier, where the electric
field is given as a sum of three contributions. Only one term ($\vec{E}_{RK}$)
is dominant, if we are not close to a singularity, with $K=c\, q_{1}V$.
The dipole field can then be written as \begin{equation}
E_{\mathrm{dip}}^{i}=-\frac{1}{4\pi\epsilon_{0}}\, M^{k}\partial_{k}\, q_{1}\,\frac{R^{i}V^{0}-R^{0}V^{i}}{(\!\widetilde{R}V\!)^{2}}\qquad\qquad\qquad\end{equation}
\begin{eqnarray}
 & = & -\frac{1}{4\pi\epsilon_{0}}\, M^{k}\left\{ q_{2}(\partial_{k}\, ct^{*})\,\frac{R^{i}V^{0}-R^{0}V^{i}}{(\!\widetilde{R}V\!)^{2}}\right.\\
 &  & \qquad\qquad\qquad\left.+q_{1}\,\partial_{k}\,\frac{R^{i}V^{0}-R^{0}V^{i}}{(\!\widetilde{R}V\!)^{2}}\right\} .\end{eqnarray}
 The identities\begin{eqnarray}
\partial_{k}R^{\alpha} & = & =g_{k}^{\:\alpha}-V^{\alpha}\partial_{k}ct^{*},\\
\partial_{k}\widetilde{R}V & = & \bar{V}_{k}-\widetilde{V}V\,\partial_{k}ct^{*},\\
\partial_{k}ct^{*} & = & \bar{R}{}_{k}/\widetilde{R}V,\end{eqnarray}
allow us to obtain \begin{equation}
\vec{E}_{\mathrm{dip}}=-q_{1}\frac{\vec{M}}{D_{2}}+q_{2}\frac{(\vec{M}\!\vec{R})\:\vec{W}_{RV}}{D_{3}}+q_{1}\frac{2\,\widetilde{V}V\,(\vec{M}\!\vec{R})\:\vec{W}_{RV}}{D_{4}}\,,\label{eq:ee}\end{equation}
with the denominators given as\begin{equation}
D_{n}=4\pi\epsilon_{0}\,(\widetilde{R}V)^{n},\end{equation}
and with$\vec{W}_{AB}=\vec{A}B^{0}-A^{0}\vec{B}$. The last term in
eq. (\ref{eq:ee}) only contributes close to a singularity, and for
all the applications in the following chapters we only consider the
first two terms.

\begin{figure}[htb]
\begin{center}\includegraphics[%
  scale=0.27]{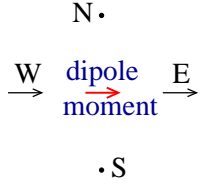}\end{center}
\vspace{-0.3cm}

\caption{The electric field vectors from the dominant dipole contribution,
being proportional to the projection of $\vec{M}$ to the radial vector
pointing to the observer. \label{cap:polar4}}
\end{figure}
We may again split the vectors into longitudinal and transverse parts,
with respect to the shower axis. Let us consider the dominant second
term. We have $\vec{W}_{RV}\approx0$, following the same arguments
which lead to the conclusion that $\vec{W}_{RK}$ is small. Therefore
also the dipole field is essentially transverse. Using $\vec{W}_{RV}^{\bot}=\vec{R}^{\bot}$and
the fact that $\vec{M}$ is purely transverse, we get \begin{equation}
\vec{E}_{\mathrm{dip}}\approx q_{2}\frac{(\vec{M}\vec{R}^{\bot})\:\vec{R}^{\bot}}{D_{3}}=q_{2}\frac{(\vec{M}\vec{R}^{\bot})\:\vec{R}^{\bot}}{D_{3}},\end{equation}
which means that the field is proportional to the projection of $\vec{M}$
to the radial vector pointing to the observer, as shown in fig. \ref{cap:polar4}.

\section{Shower parallel to the magnetic field}

\begin{figure}[htb]
\begin{center}\includegraphics[%
  scale=0.3]{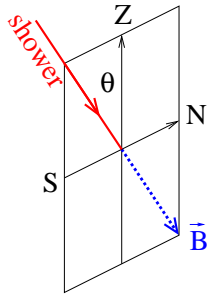}\end{center}

\caption{South-north shower ($\theta=27^{0}$, $\psi=180^{0}$), being parallel
to the geo-magnetic field. \label{cap:South-north-shower}}
\end{figure}

In the following, we consider the case of a shower with an initial
energy of $5\cdot10^{17}$eV, an inclination $\theta=27^{0}$, and
an azimuth angle $\psi=180^{0}$, defined with respect to the magnetic
north pole, see fig. \ref{cap:South-north-shower}. The angle $\psi$
refers to the origin of the shower, it is related to the angle $\varphi$
from the preceding chapter as $\psi=180^{0}-\varphi$. So $\psi=180^{0}$
thus implies that the shower moves from south to north.  We consider
the magnetic field at the {\small CODALEMA} site, i.e. $|\vec{B}|=47.3\mu T$
and $\alpha=153^{0}$, so the shower moves parallel to the magnetic
field. We consider an observer at $x=0$, \textbf{$y=500\,$}m, $z=140\,$m.
We suppose $a=z$ (so the shower hits the ground at $x=y=0$). The
coordinates $x$, $y$, $z$ are defined in the Earth system.

\begin{figure}[htb]
\begin{center}~\end{center}

\vspace{-2cm}
\begin{center}\hspace*{-1cm}\includegraphics[%
  scale=0.35,
  angle=270]{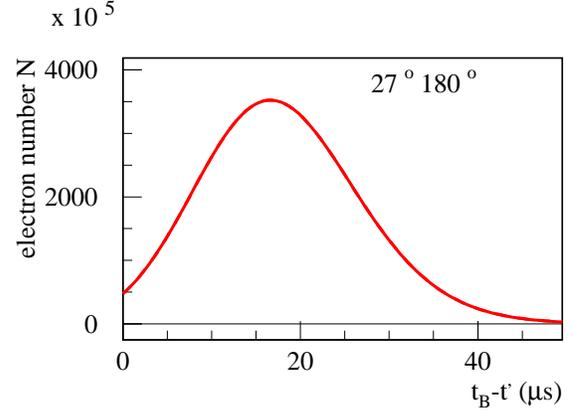}\end{center}
\vspace{-1.4cm}

\caption{The number $N$ of electrons and positrons, as a function of the
time $t_{B}-t'$.\label{cap:nel00}}
\end{figure}

The electron number $N$ reaches its maximum at about $-15\,\mu\mathrm{s}$,
reaching almost $4\cdot10^{8}$ particles, as shown in fig. \ref{cap:nel00}.
\begin{figure}[htb]
\begin{center}~\end{center}

\vspace{-2cm}
\begin{center}\hspace*{-1cm}\includegraphics[%
  scale=0.35,
  angle=270]{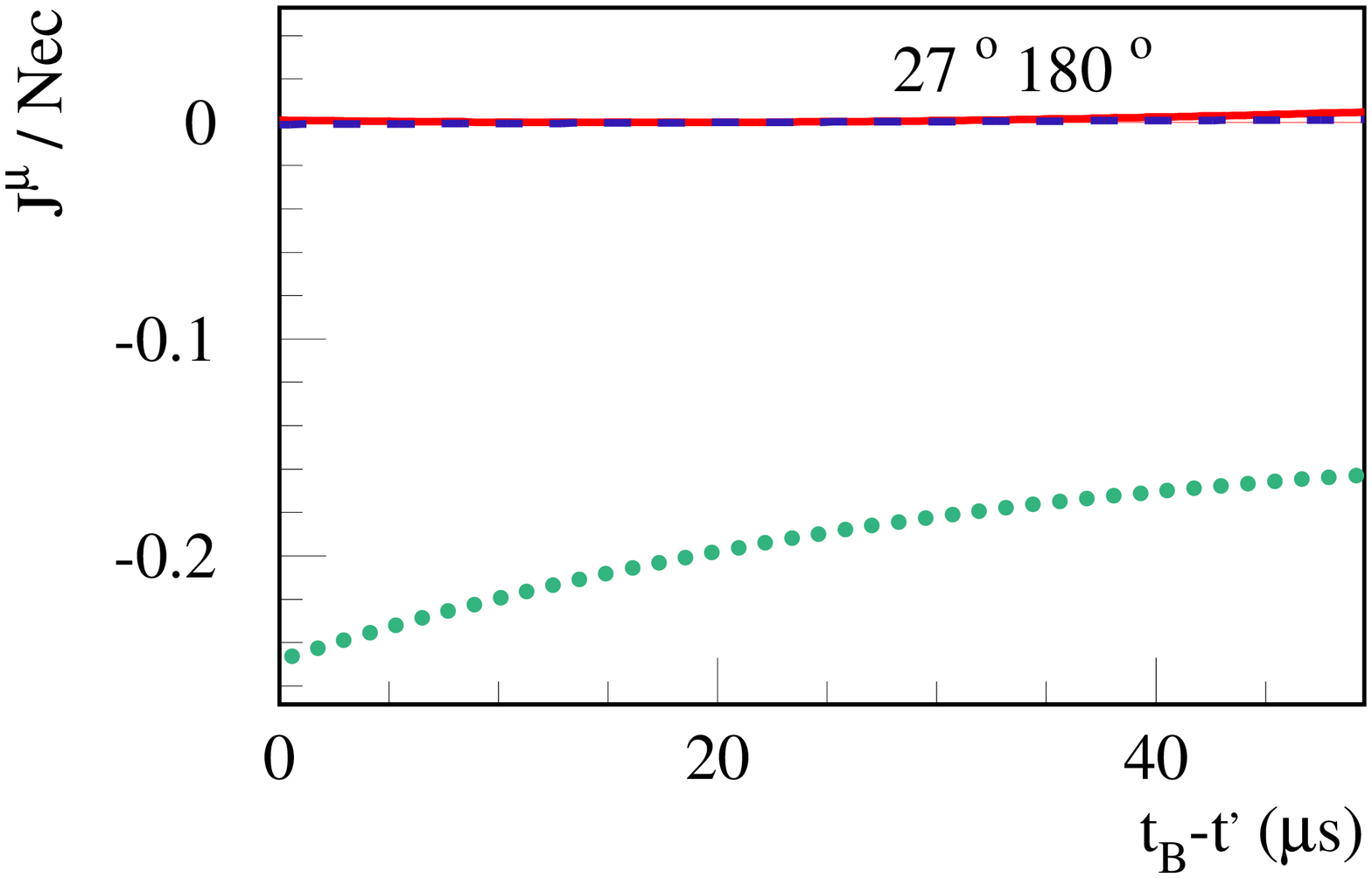}\end{center}

\vspace{-3cm}
\begin{center}\hspace*{-1cm}\includegraphics[%
  scale=0.35,
  angle=270]{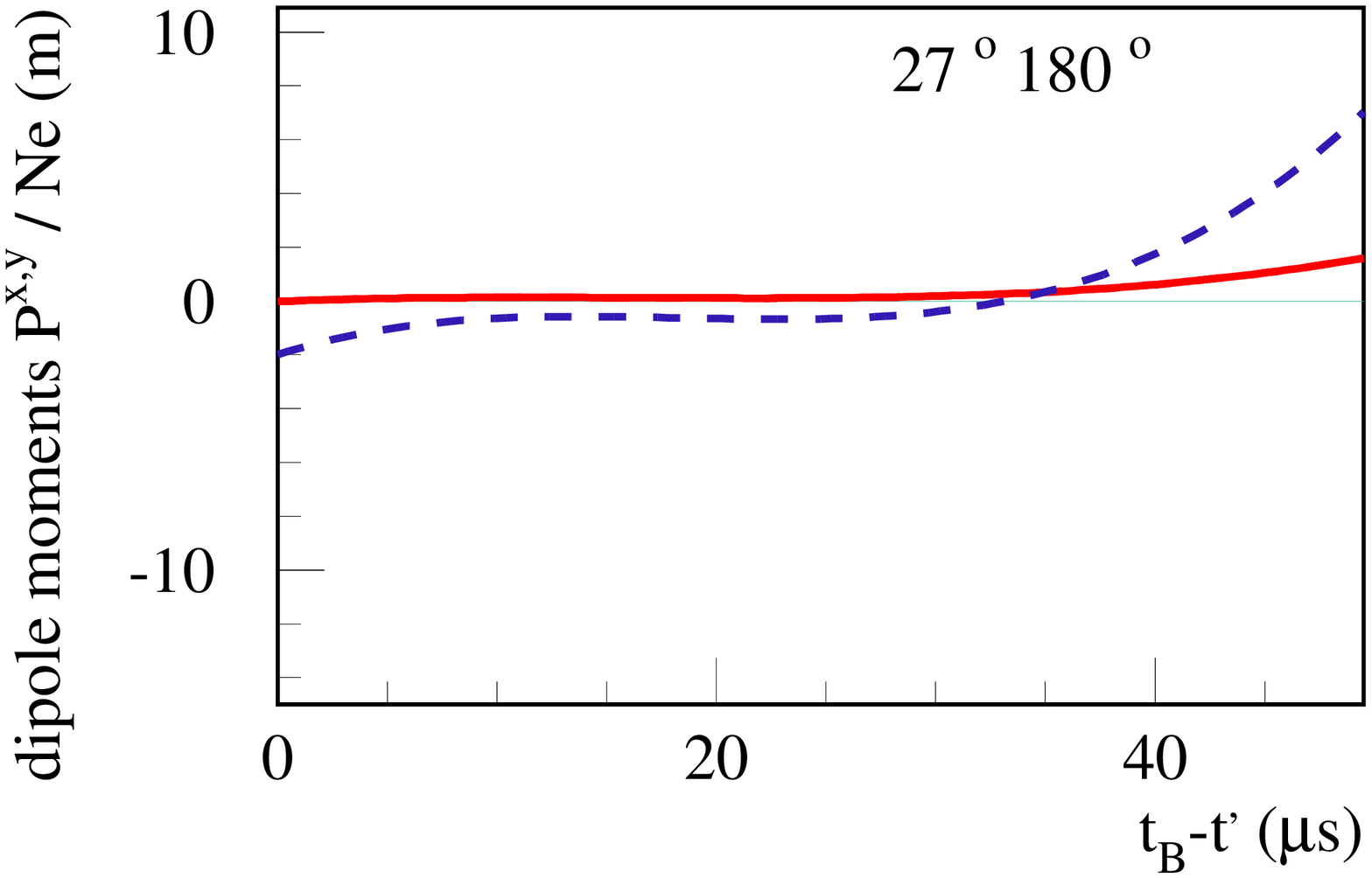}\end{center}
\vspace{-1.4cm}

\caption{The $x$- (full line), $y$- (dashed) and $z$-component (dotted)
of of the current $J$ divided by $Nec$ (upper plot), and the $x$-
(full line), and $y$-component (dashed) of of the dipole moment $P$
divided by $Ne$, (lower plot), as a function of the retarded time.
The vector components refer to the shower frame.\label{cap:current180} }
\end{figure}
\begin{figure}[htb]
\begin{center}~\end{center}

\vspace{-2cm}
\begin{center}\hspace*{-1cm}\includegraphics[%
  scale=0.35,
  angle=270]{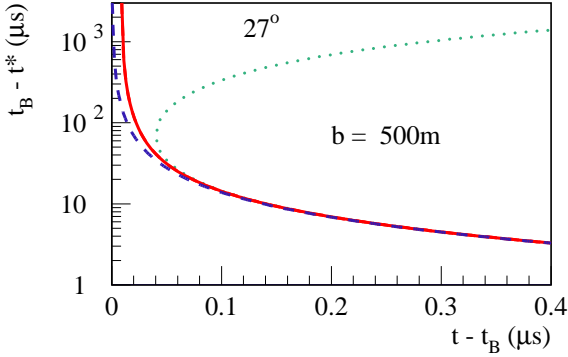}\end{center}
\vspace{-1.4cm}

\caption{The retarded time expressed via $t_{B}-t^{*}$ versus observer time
$t-t_{B}$, for $b=500\,\mathrm{m}$ and $\theta=27^{0}$ . The solid
lines represent the results for a realistic variable index of refraction,
the two other curves to $n=1$ and $n=n_{\mathrm{ground}}$.\label{cap:tstar27}}
\end{figure}
\begin{figure}[htb]
\begin{center}\hspace*{-1cm}\includegraphics[%
  scale=0.35,
  angle=270]{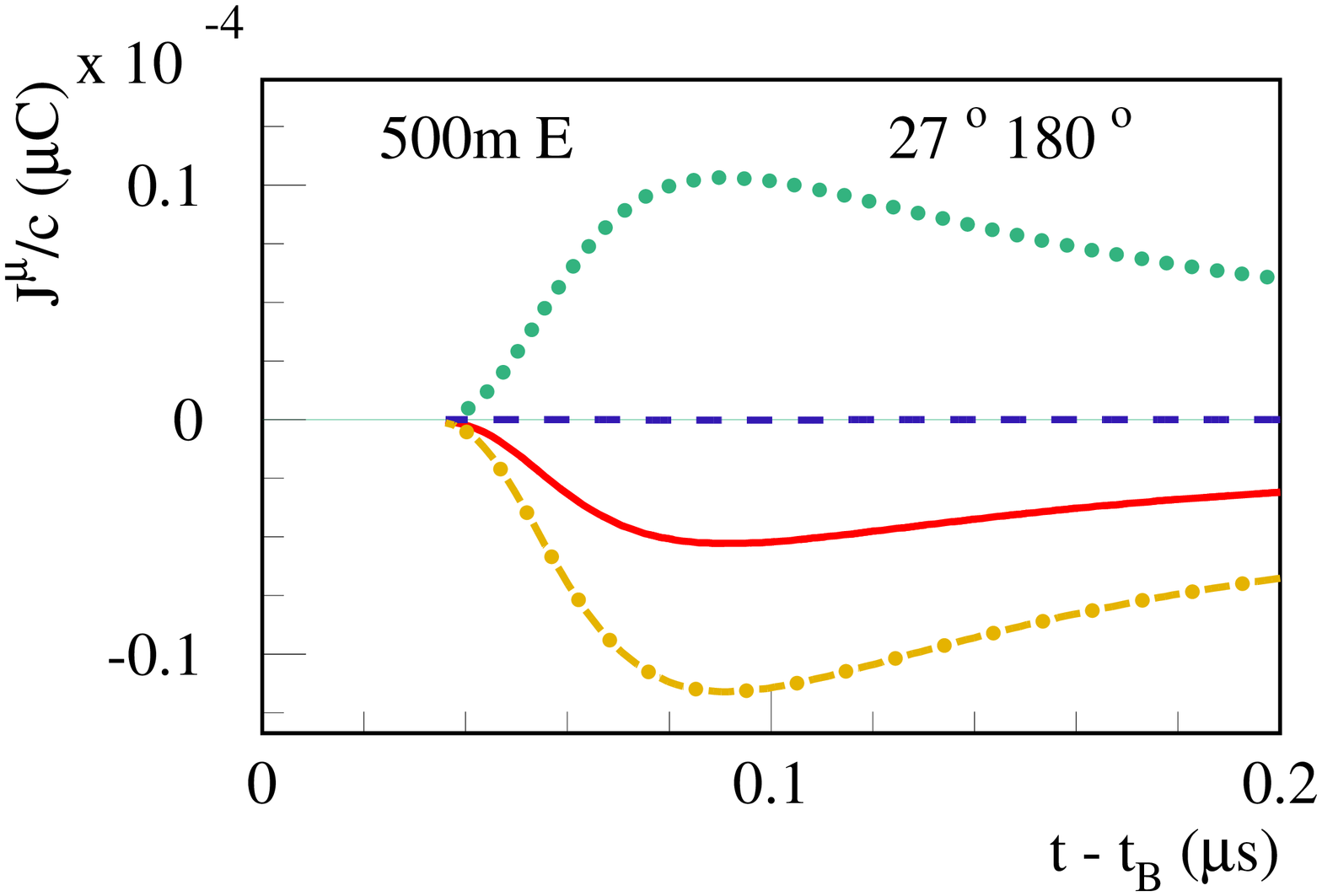}\end{center}

\vspace{-3cm}
\begin{center}\hspace*{-1cm}\includegraphics[%
  scale=0.35,
  angle=270]{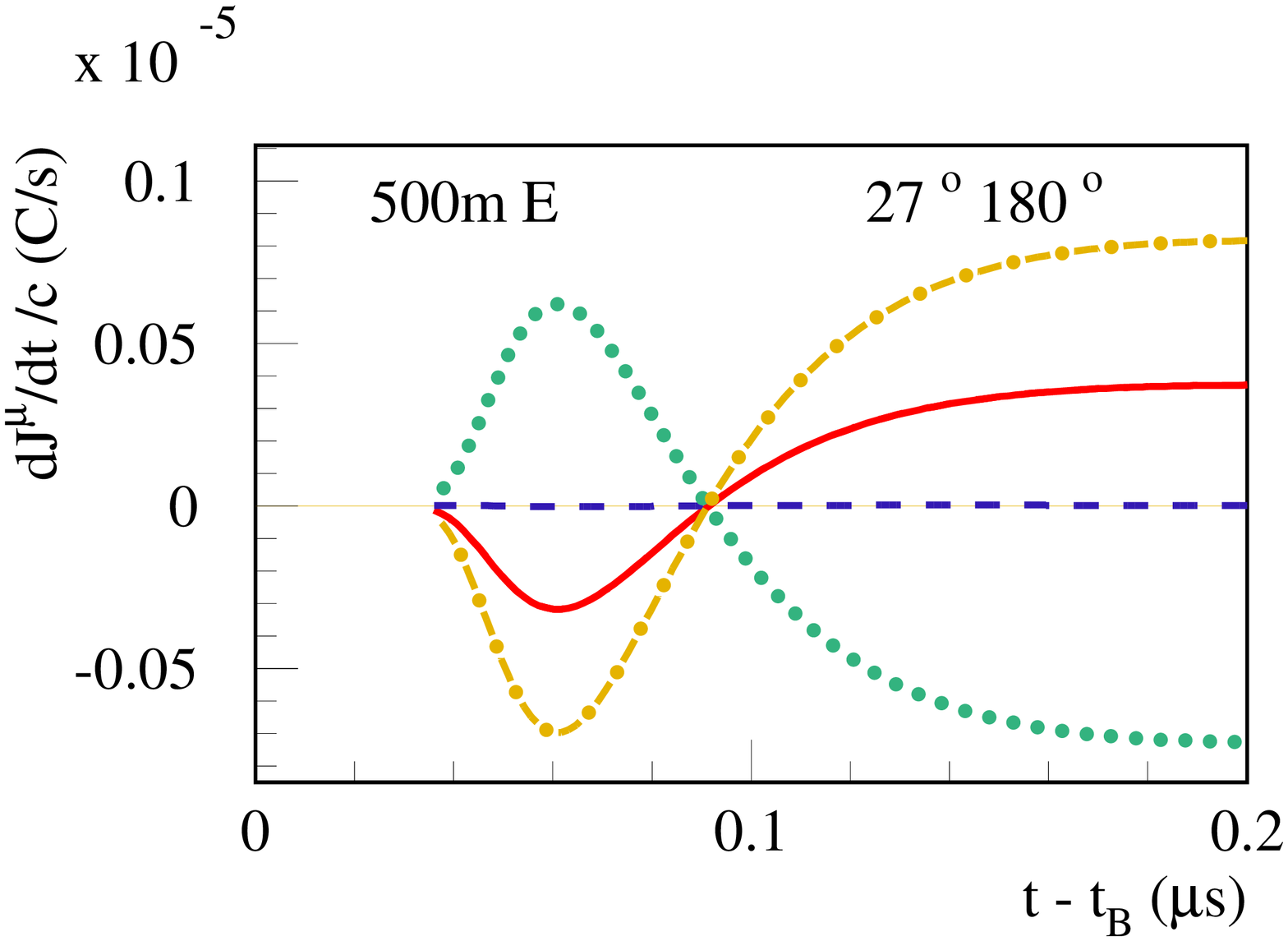}\end{center}
\vspace{-1.4cm}

\caption{The $x$- (full line), $y$- (dashed), $z$- (dotted), and $0$-component
(dashed-dotted) of the current $J$ and their derivatives (bottom)
versus the observer time $t$. Here we consider the total current,
including the ion contribution. \label{cap:curr-earth-180} The vector
components refer to the Earth frame. }
\end{figure}
 It is useful to plot the components of the current, divided by $Nec$,
with $e$ being the electron charge, and $c$ the velocity of light,
see fig. \ref{cap:current180}. The components $J^{x}/Nec$ and $J^{y}/Nec$
represent the average transverse drift velocity, in units of $c$,
caused by the magnetic field. For the present example, the components
$J^{x}/Nec$ and $J^{y}/Nec$ are zero. This is understandable, since
the average transverse drift velocity is zero for showers moving parallel
to the magnetic field. The component $J^{z}/Nec$ represents the charge
excess which stays almost constant at a value around  -0.2.  The time
variation of the currents is mainly due to the time variation of $N$.
For the numerical calculations, also the small time variations of
$J^{z}/Nec$ are taken into account. In fig. \ref{cap:current180},
we also plot the components $P^{x}$ and $P^{y}$ of the dipole moment
divided by $Ne$. As expected, the dipole moment is close to zero.
It is not exactly zero due to statistical fluctuations in regions
where $Ne$ is small. The actual Monte Carlo results are strongly
fluctuating around zero, the curves shown in fig. \ref{cap:current180}
is obtained after smoothening the {}``raw data''.

In the expressions for the electric fields at some observer time $t$,
the currents $J$ and their derivatives $K$ appear, evaluated at
the retarded time $t^{*}$. Following the procedure discussed earlier,
we compute $t^{*}(t)$, as shown in fig.  \ref{cap:tstar27}. We see
that a shower maximum at $t_{B}-t^{*}\approx15\,\mu\mathrm{s}$ corresponds
to $t-t_{B}\approx0.1\,\mu\mathrm{s}$. Given the time dependence
$J(t')$ of the current and the relation $t^{*}(t)$ between the retarded
time and the observation time $t$, we may present the current as
$J\big(t^{*}(t)\big)$ versus the observer time.In fig. \ref{cap:curr-earth-180},
we show the components of the current and their derivatives in the
Earth frame. The $y-$component vanishes, as expected, since this
is the component perpendicular to the shower plane (containing the
shower axis and the vertical axis). One should note that the width
of $J(t')$ is of the order of $10\,\mu\mathrm{s}$, whereas the width
of $J$ versus observer time is just a fraction of a micro second.

\begin{figure*}[htb]
\begin{flushleft}{\large (a)\hspace*{7.5cm}(b)}\end{flushleft}{\large \par}

\vspace{-2.3cm}
\begin{center}{\large \hspace*{-1cm}}\includegraphics[%
  scale=0.32,
  angle=270]{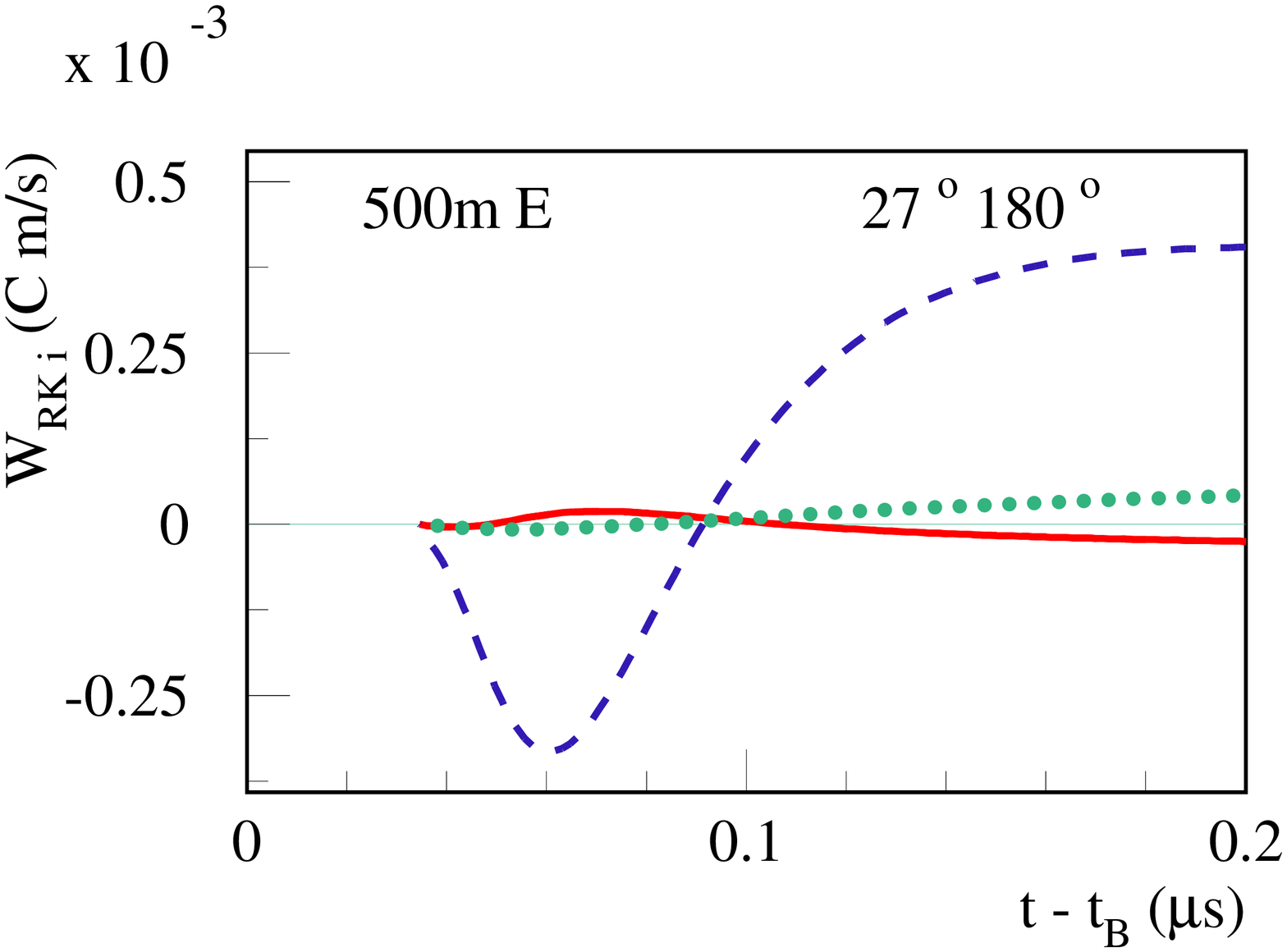}{\large \hspace*{-1.0cm}}\includegraphics[%
  scale=0.32,
  angle=270]{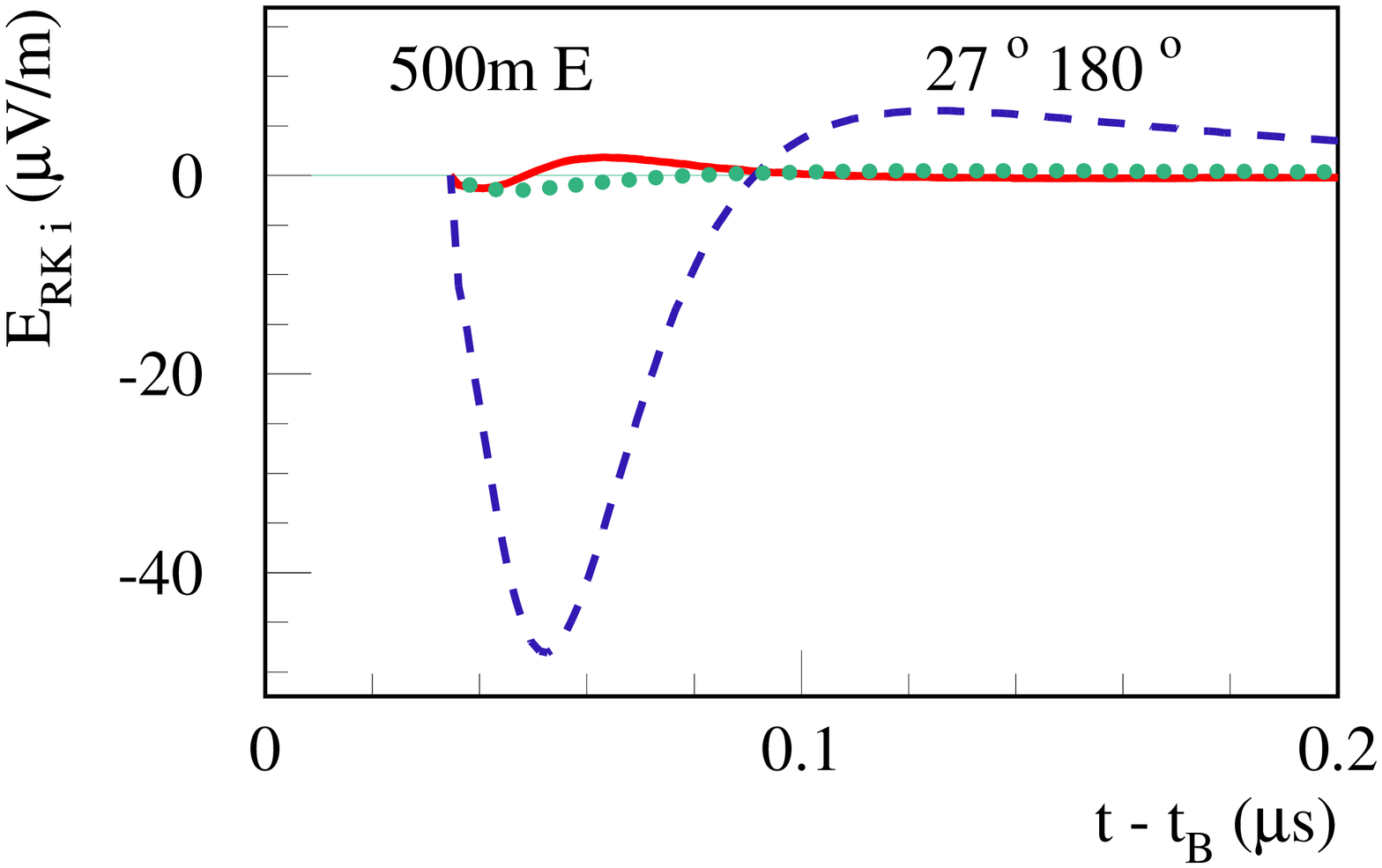}\end{center}
\vspace{-1.4cm}

\noindent \begin{flushleft}{\large (c)\hspace*{7.5cm}(d)}\end{flushleft}{\large \par}

\vspace{-2.3cm}
\begin{center}{\large \hspace*{-1cm}}\includegraphics[%
  scale=0.32,
  angle=270]{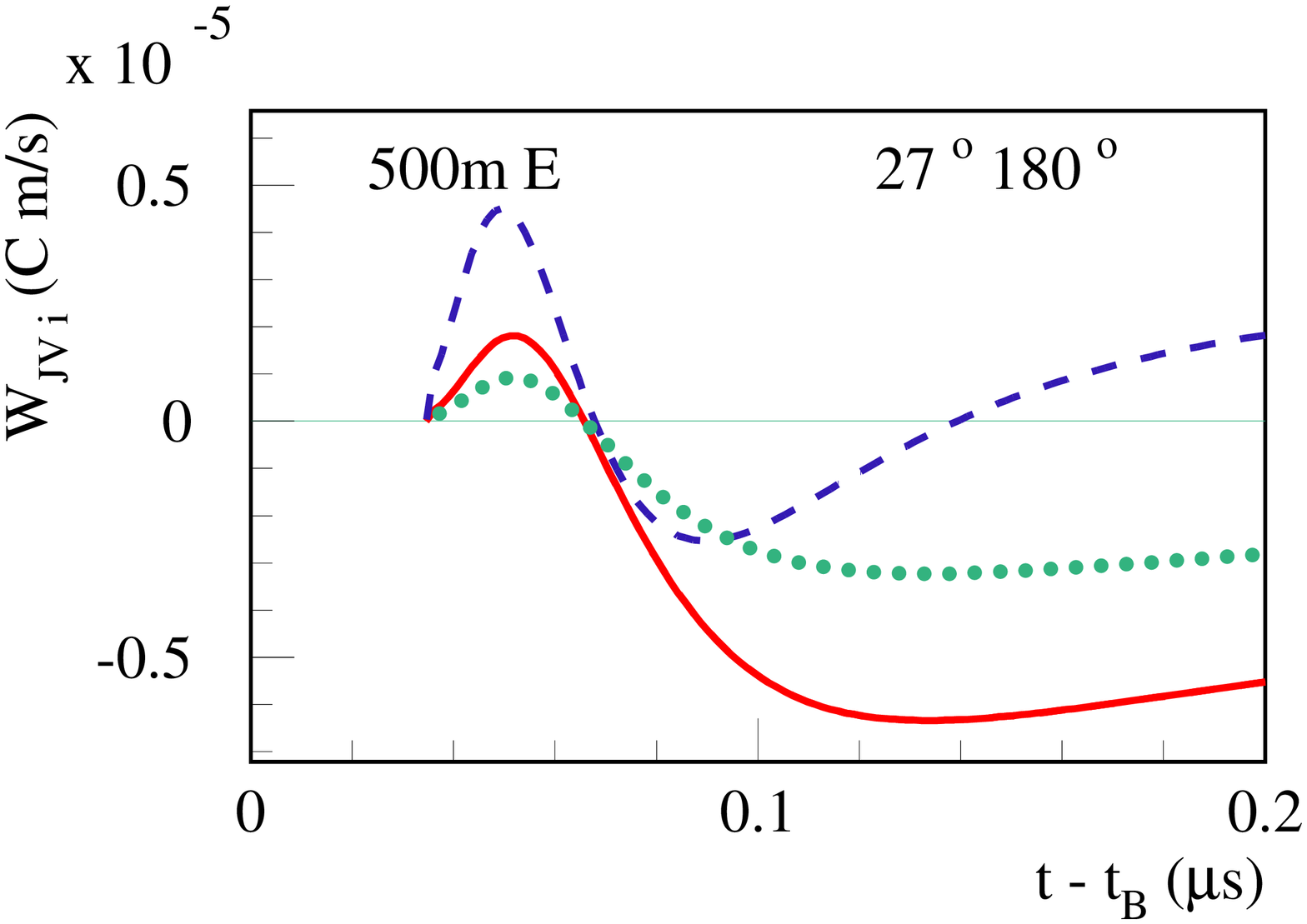}{\large \hspace*{-1.0cm}}\includegraphics[%
  scale=0.32,
  angle=270]{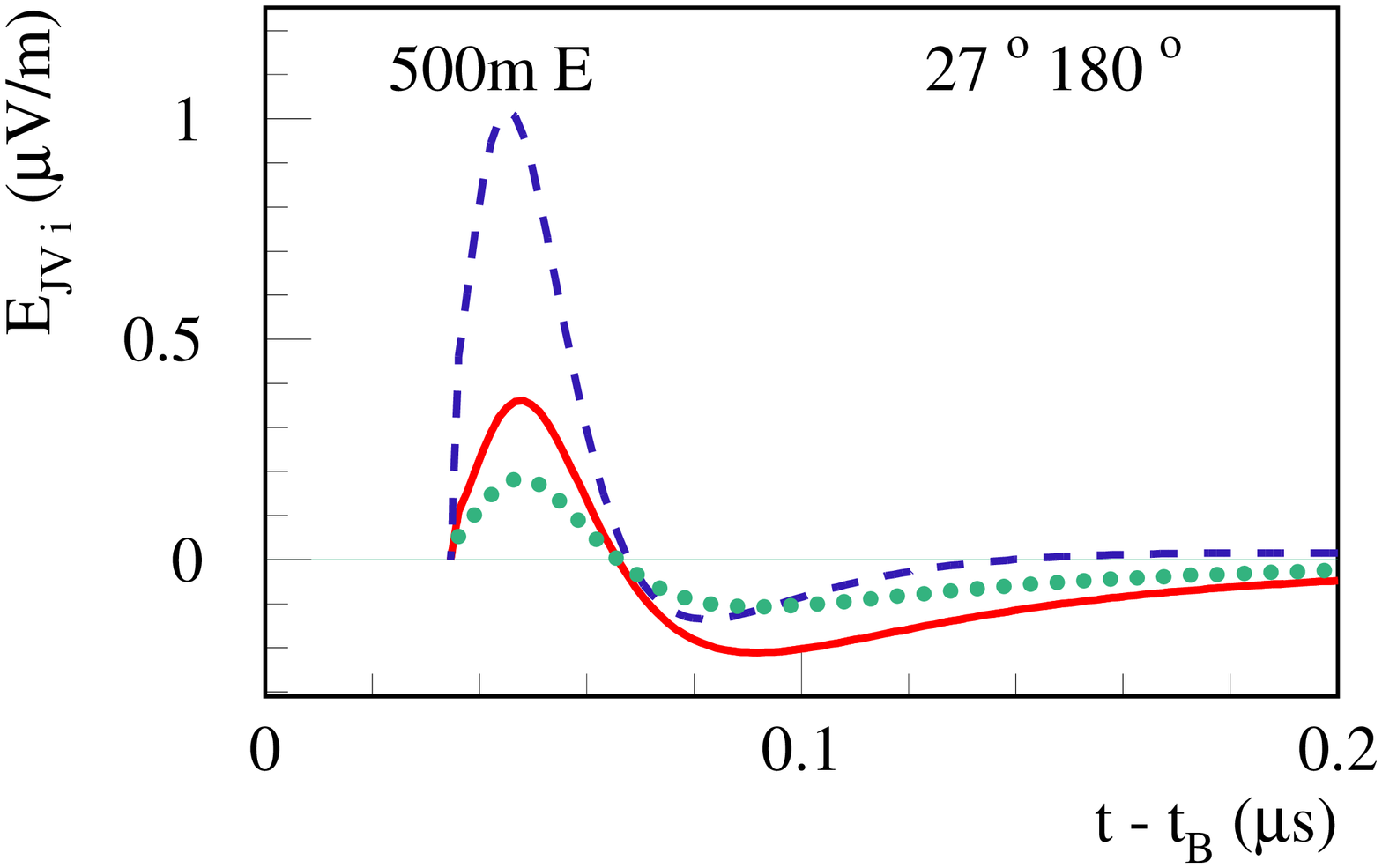}\end{center}
\vspace{-1.4cm}

\noindent \begin{flushleft}{\large (e)\hspace*{7.5cm}(f)}\end{flushleft}{\large \par}

\vspace{-2.5cm}
\begin{center}{\large \hspace*{-1cm}}\includegraphics[%
  scale=0.32,
  angle=270]{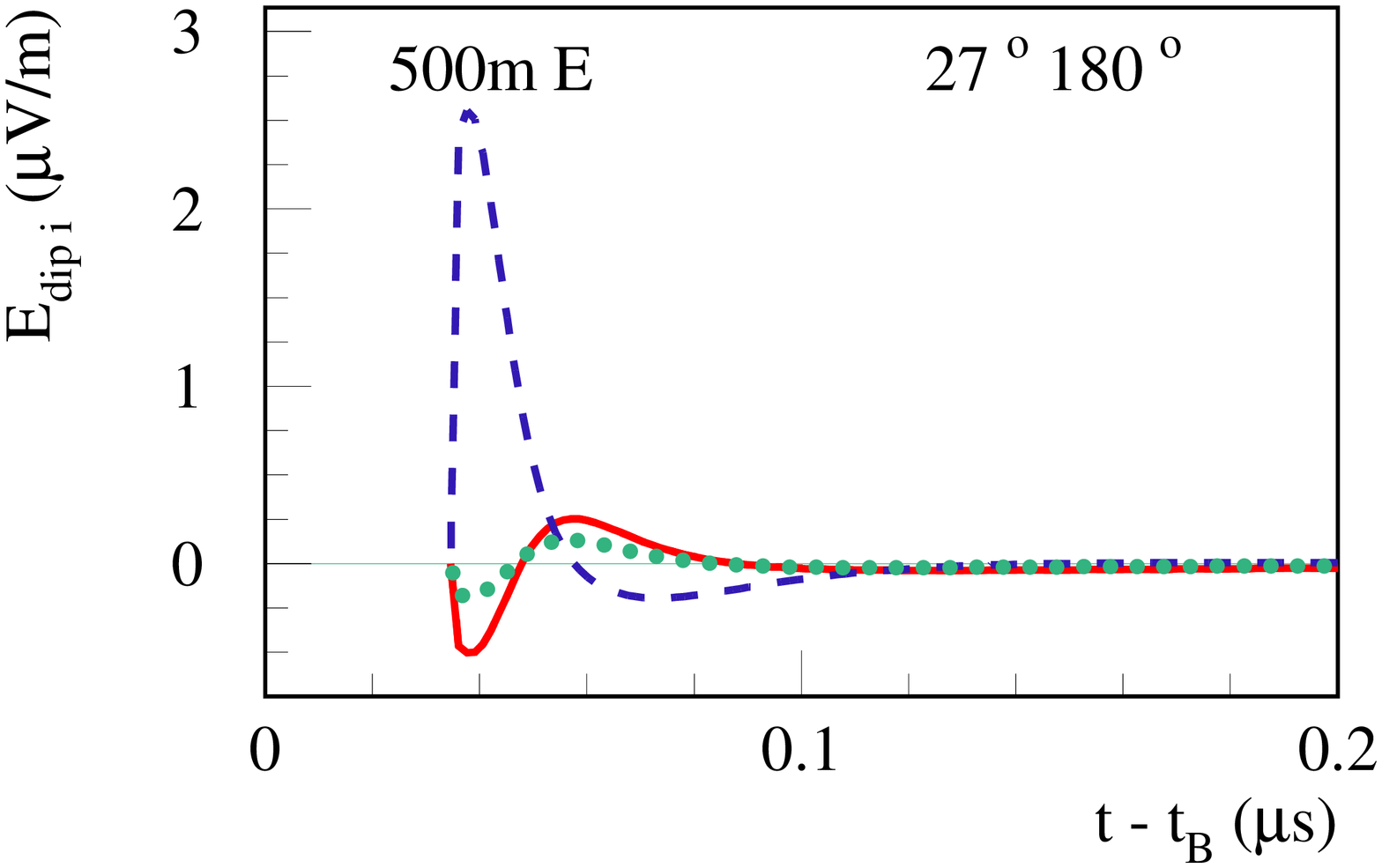}{\large \hspace*{-1.0cm}}\includegraphics[%
  scale=0.32,
  angle=270]{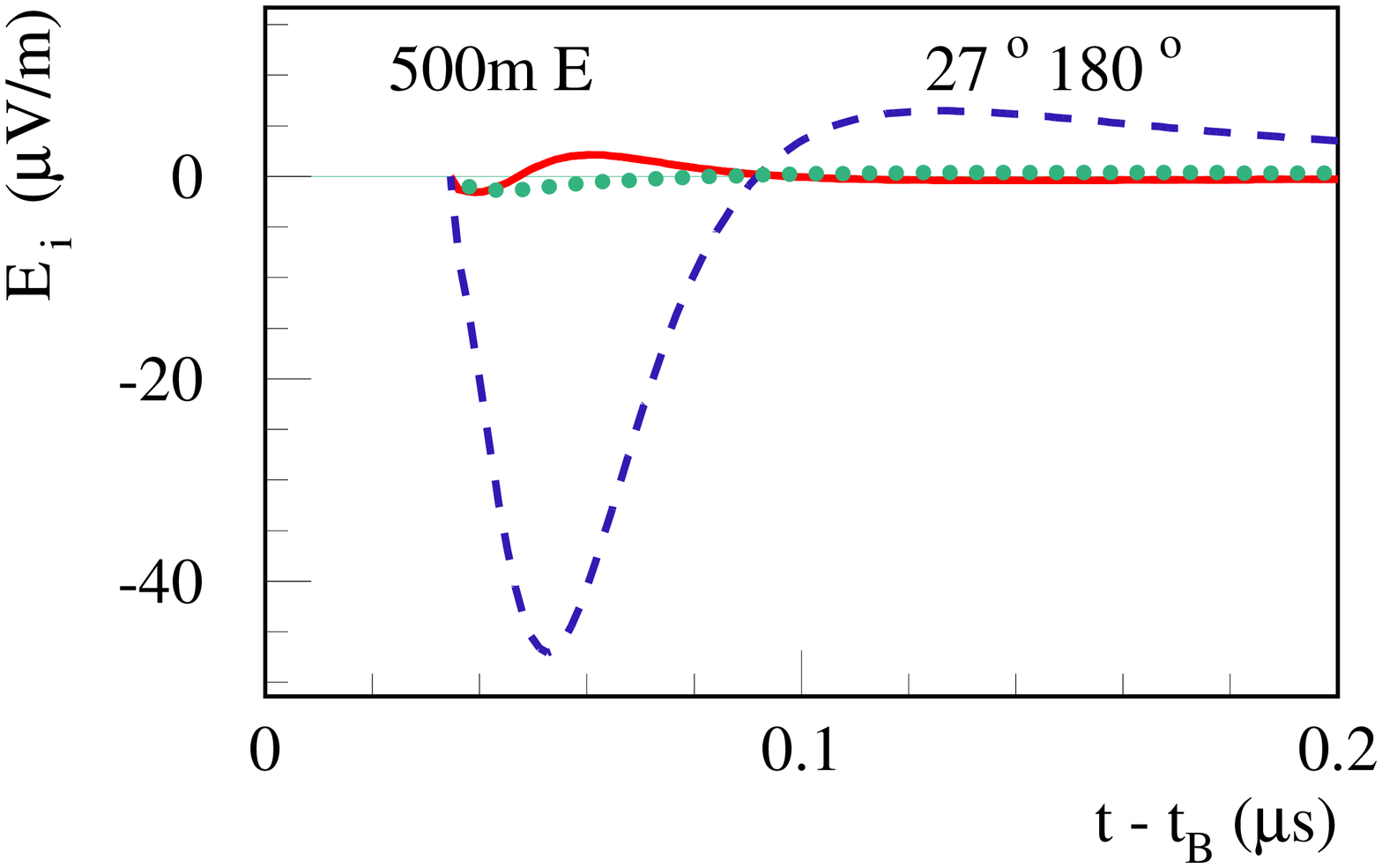}\end{center}
\vspace{-1.4cm}

\caption{The $x$- (full line), $y$- (dashed) and $z$-component (dotted)
of $\vec{W}_{RK}$ (a), $\vec{W}_{JV}$ (c), \label{cap:econtr180}and
of the corresponding fields $\vec{E}_{RK}$ (b), $\vec{E}_{JV}$(d),
the dipole contribution $\vec{E}_{dip}$ (e), and the components of
the total electric field, with all contributions (f). The vector components
refer to the Earth frame.}
\end{figure*}

We are now ready to discuss the electric fields. As shown earlier,
we have \begin{equation}
\vec{E}=\vec{E}_{\mathrm{dip}}+\vec{E}_{RK}+\vec{E}_{JV},\end{equation}
with the dipole field $\vec{E}_{\mathrm{dip}}$and with\begin{equation}
\vec{E}_{RK}=\frac{\vec{W}_{RK}}{D},\quad\vec{E}_{JV}=\frac{\vec{W}_{JV}}{D},\end{equation}
using\[
D=4\pi\epsilon_{0}n^{2}(\!\widetilde{R}V\!)^{2}\]
and\begin{equation}
\vec{W}_{AB}=\vec{A}B^{0}-A^{0}\vec{B},\end{equation}
for two four-vectors $A$ and $B$. The three components of the vectors
$\vec{W}$ and the corresponding field contributions, as well as the
three components of the complete field $\vec{E}$ are shown in fig.
\ref{cap:econtr180}. The dipole field is of course negligible. The
contribution $\vec{E}_{RK}$ -- having its origin in the time variation
$K$ of the currents -- is by far dominant. Only the $y-$component
is sizable, meaning that the field is oriented radially compared  to
the shower axis (the observer sits on the $y-$axis). The contribution
$\vec{E}_{JV}$ is very small, in comparison. This can easily be understood
since here, without transverse drift, we  have essentially a longitudinally
moving charge, and in such a case $\vec{J}V^{0}-J^{0}\vec{V}$ vanishes.

\section{Shower with a large angle with respect to the magnetic field}

\begin{figure}[htb]
\begin{center}~\end{center}

\begin{center}\includegraphics[%
  scale=0.3]{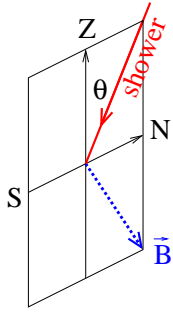}\end{center}
\vspace{-0.4cm}

\caption{North-south shower ($\theta=27^{0}$, $\psi=0^{0}$).\label{cap:North-south-shower}}
\end{figure}

In the following, we consider a shower with an initial energy of $5\cdot10^{17}$eV,
an inclination $\theta=27^{0}$, and an azimuth angle $\psi=0^{0}$,
defined with respect to the magnetic north pole, see fig. \ref{cap:North-south-shower}.

\begin{figure}[htb]
\begin{center}{\large ~}\end{center}{\large \par}
\vspace{-2cm}

\begin{center}{\large \hspace*{-1cm}}\includegraphics[%
  scale=0.35,
  angle=270]{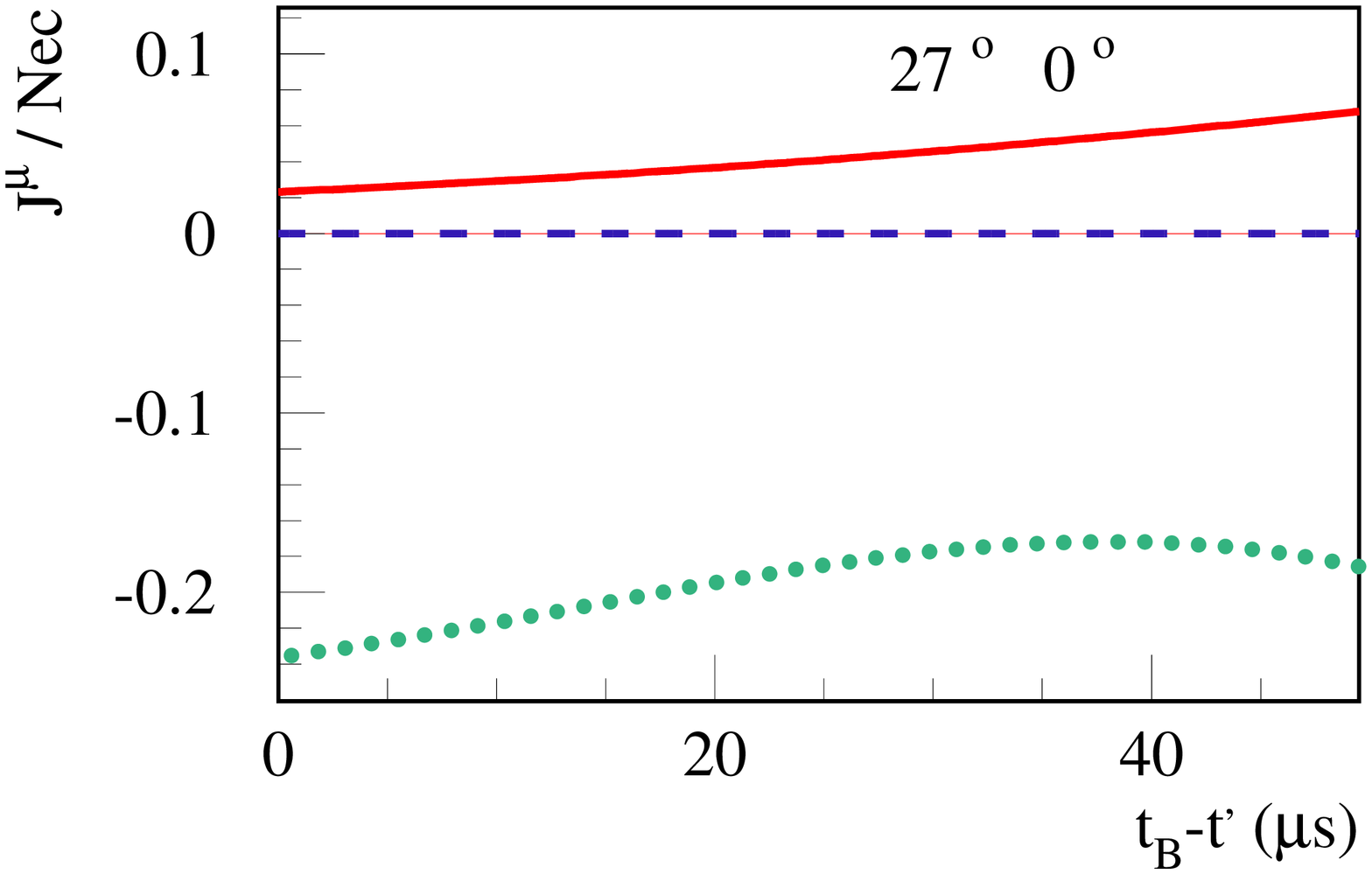}\end{center}

\vspace{-3cm}
\begin{center}\hspace*{-1cm}\includegraphics[%
  scale=0.35,
  angle=270]{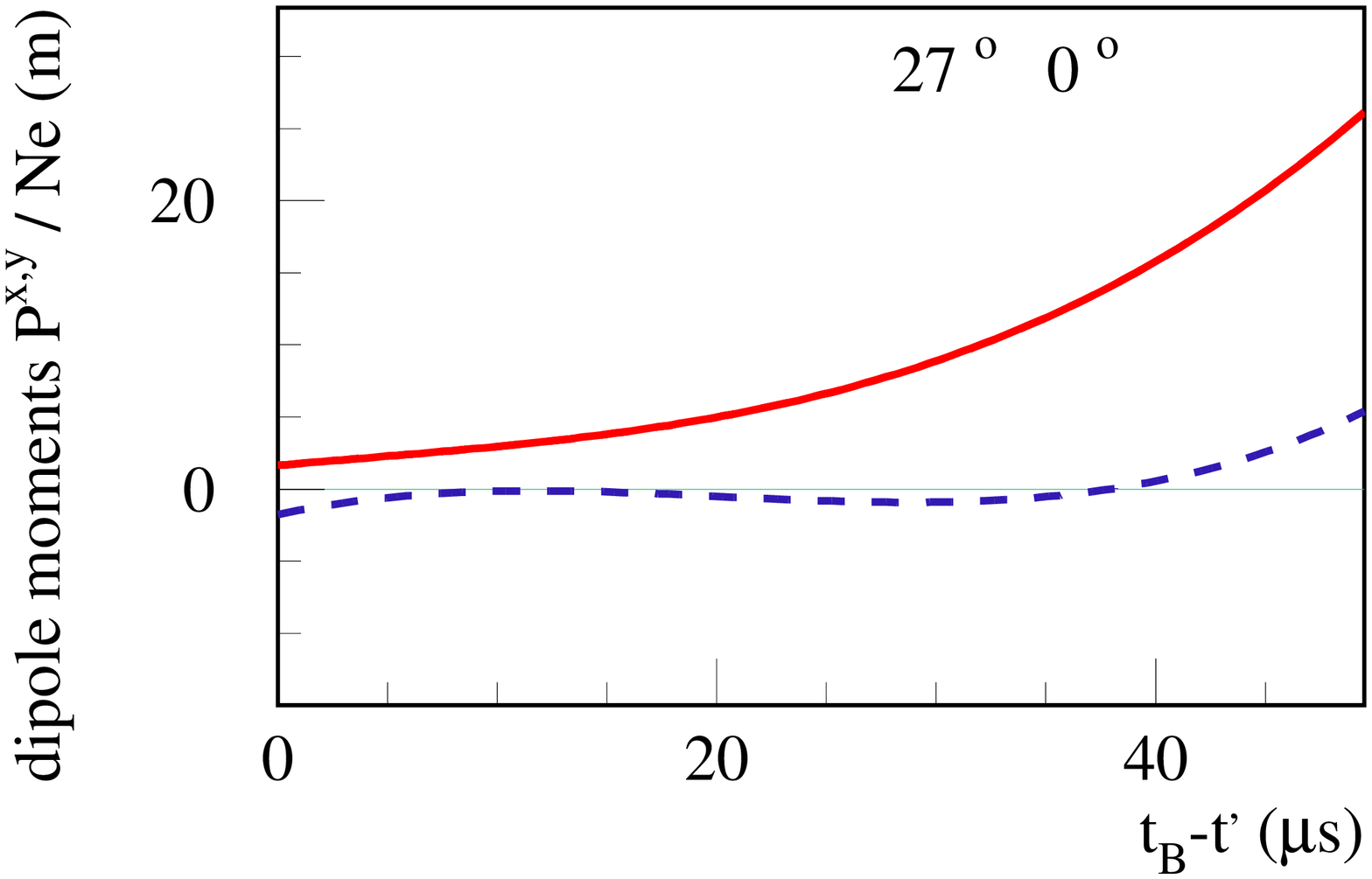}\end{center}
\vspace{-1.4cm}

\caption{The $x$- (full line), $y$- (dashed) and $z$-component (dotted)
of of the current $J$ divided by $Nec$ (upper plot), and the $x$-
(full line), and $y$-component (dashed) of of the dipole moment $P$
divided by $Ne$ (lower plot), as a function of of the time $t_{B}-t'$.
The vector components refer to the shower frame.\label{cap:current0x} }
\end{figure}
\begin{figure}[htb]
~
\vspace{-2cm}

\begin{center}{\large \hspace*{-1cm}}\includegraphics[%
  scale=0.35,
  angle=270]{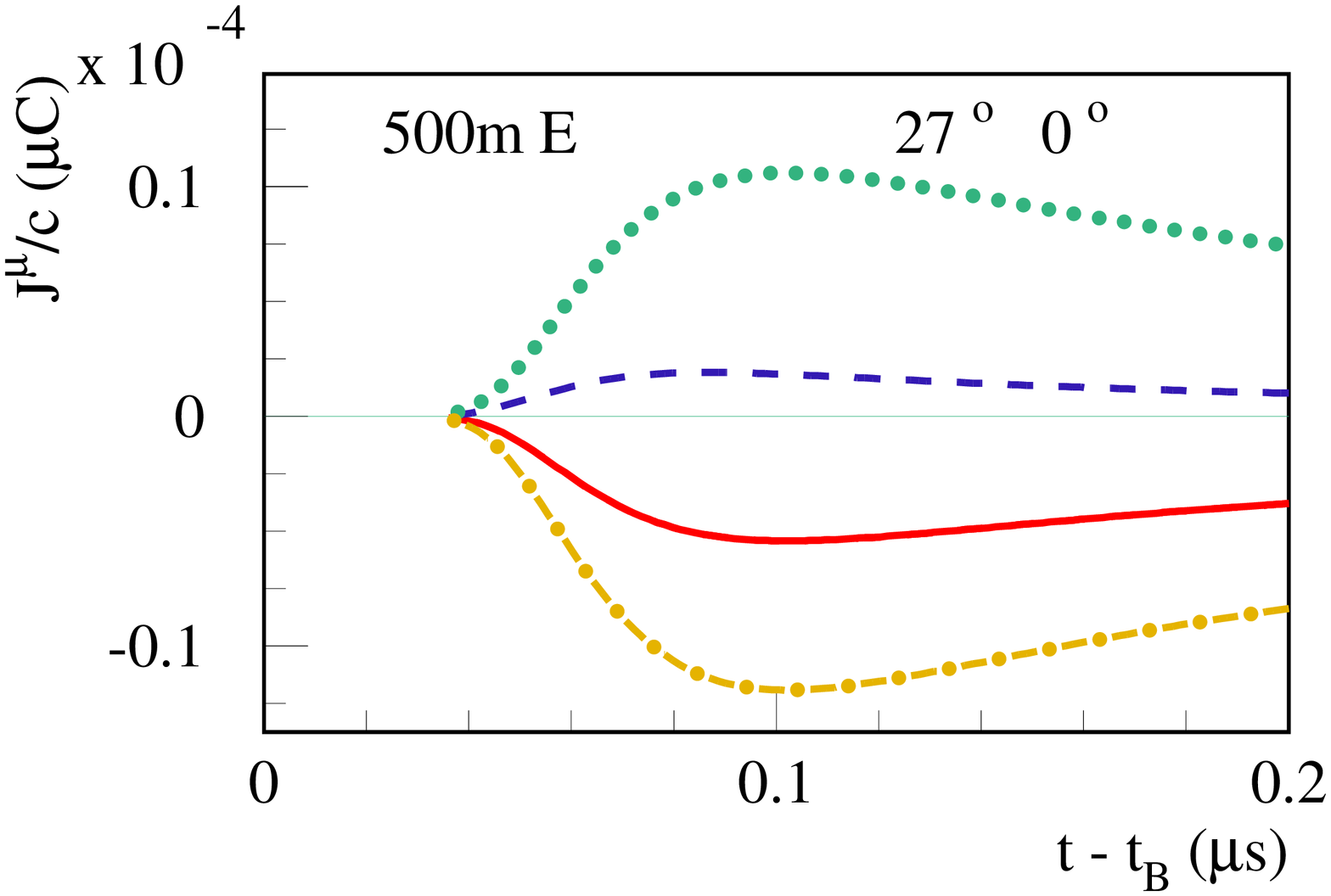}\end{center}

\vspace{-3cm}
\begin{center}{\large \hspace*{-1cm}}\includegraphics[%
  scale=0.35,
  angle=270]{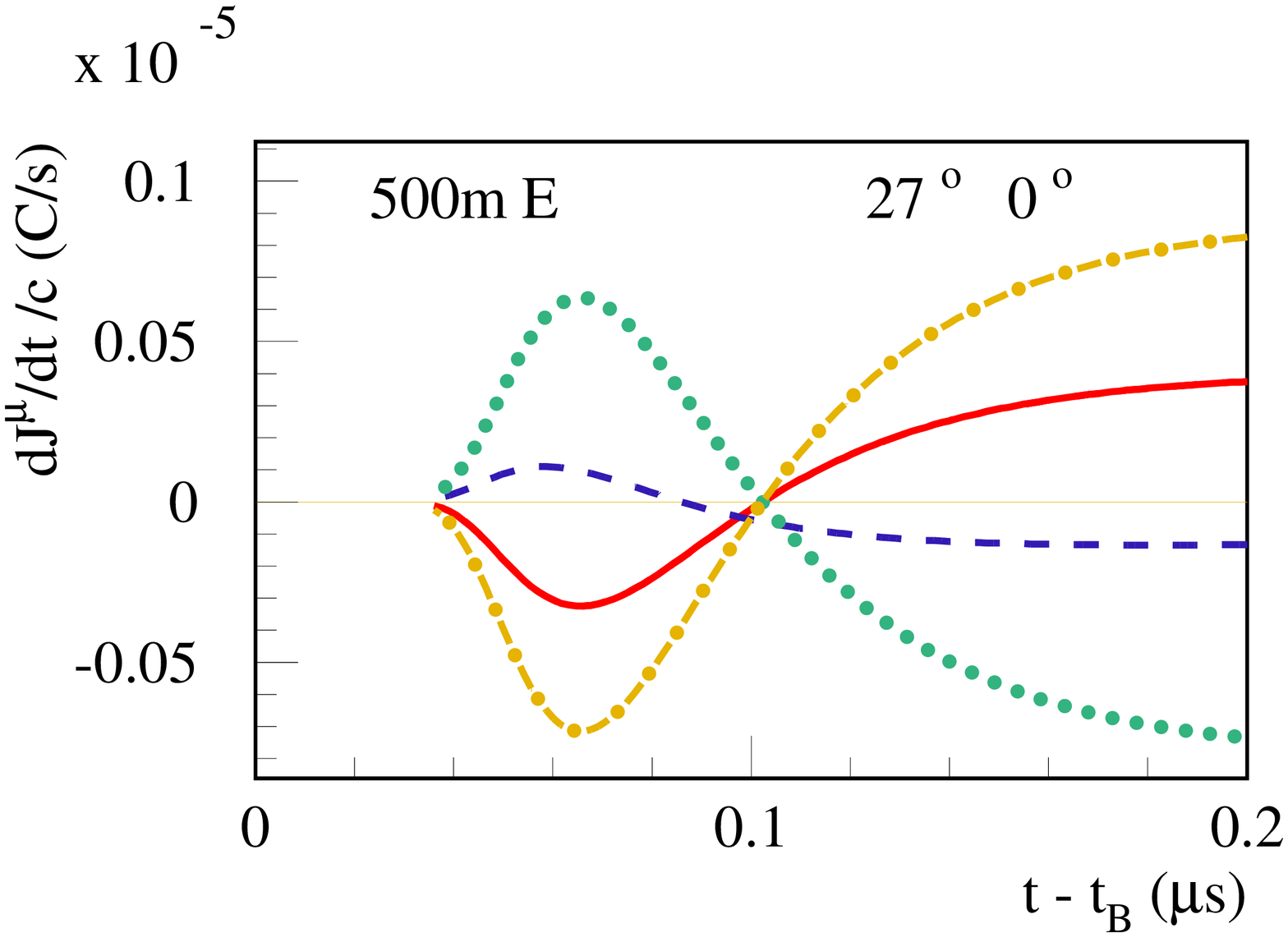}\end{center}
\vspace{-1.4cm}

\caption{The $x$- (full line), $y$- (dashed), $z$- (dotted), and $0$-component
(dashed-dotted) of the current $J$ (top) and their derivatives (bottom)
versus the observer time $t-t_{B}$. Here we consider the total current,
including the ion contribution. \label{cap:curr-earth} The vector
components refer to the Earth frame. }
\end{figure}
\begin{figure*}[htb]
\begin{flushleft}{\large (a)\hspace*{7.5cm}(b)}\end{flushleft}{\large \par}

\vspace{-2.3cm}
\begin{center}{\large \hspace*{-1cm}}\includegraphics[%
  scale=0.32,
  angle=270]{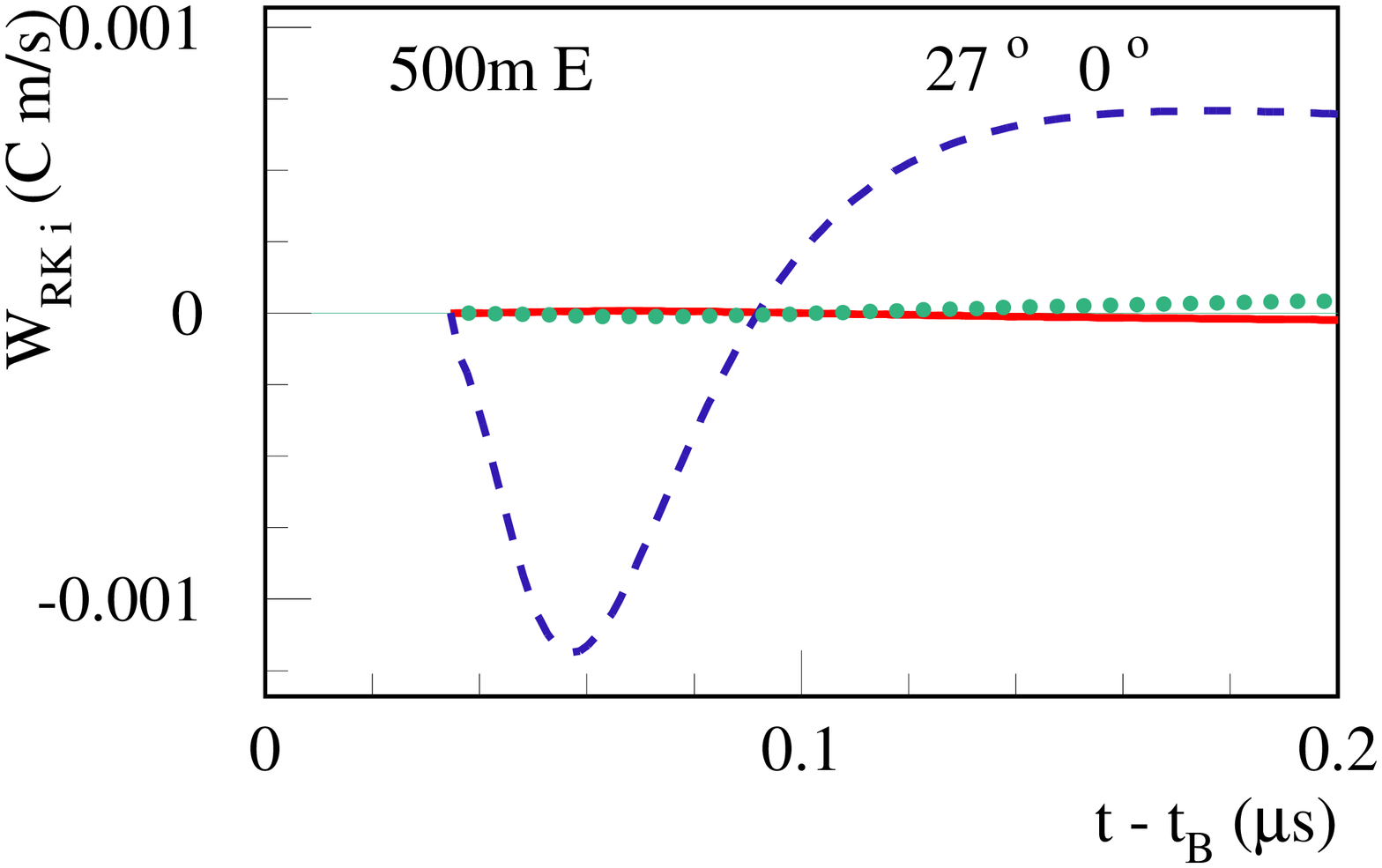}{\large \hspace*{-1.0cm}}\includegraphics[%
  scale=0.32,
  angle=270]{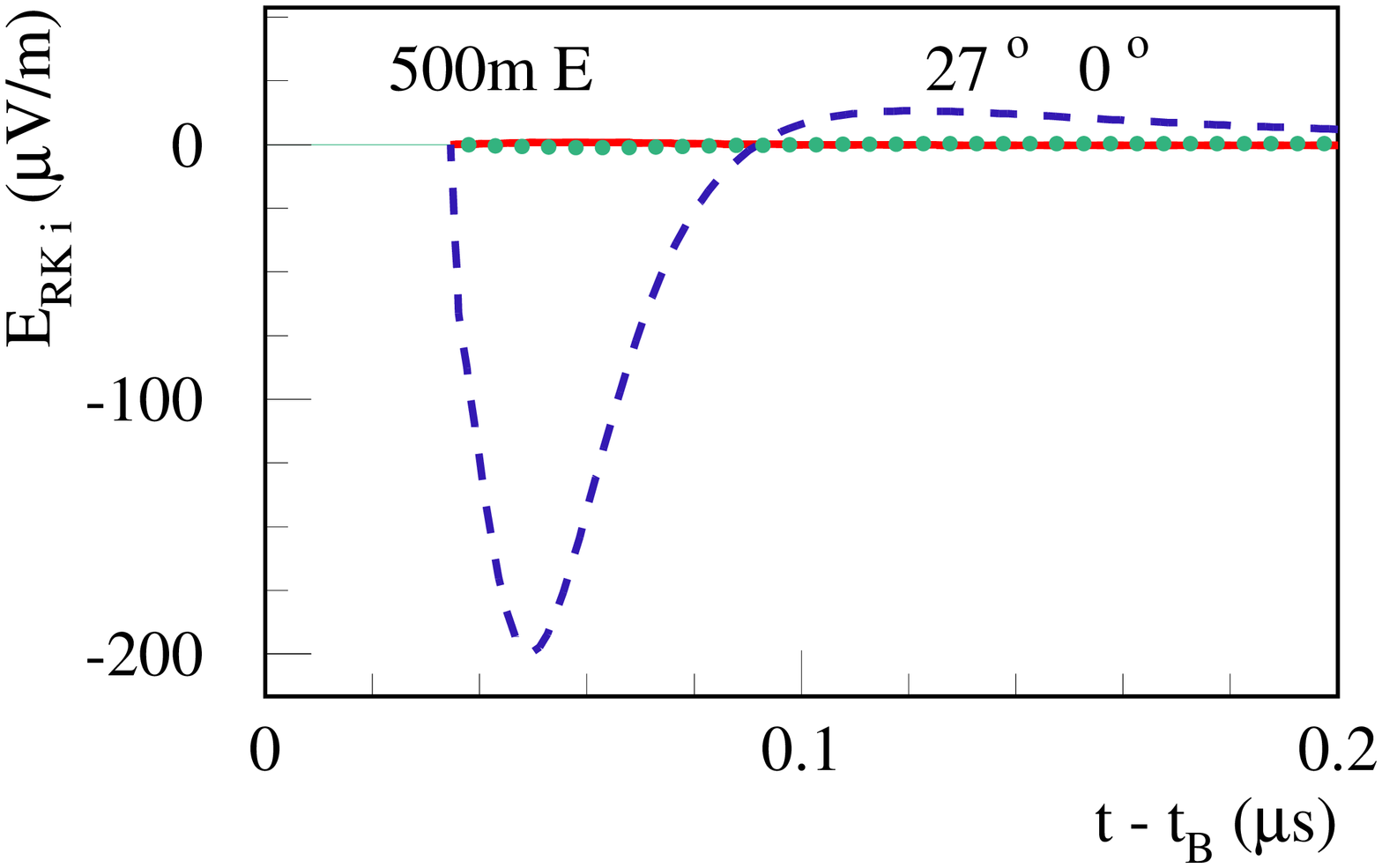}\end{center}
\vspace{-1.4cm}

\noindent \begin{flushleft}{\large (c)\hspace*{7.5cm}(d)}\end{flushleft}{\large \par}

\vspace{-2.3cm}
\begin{center}{\large \hspace*{-1cm}}\includegraphics[%
  scale=0.32,
  angle=270]{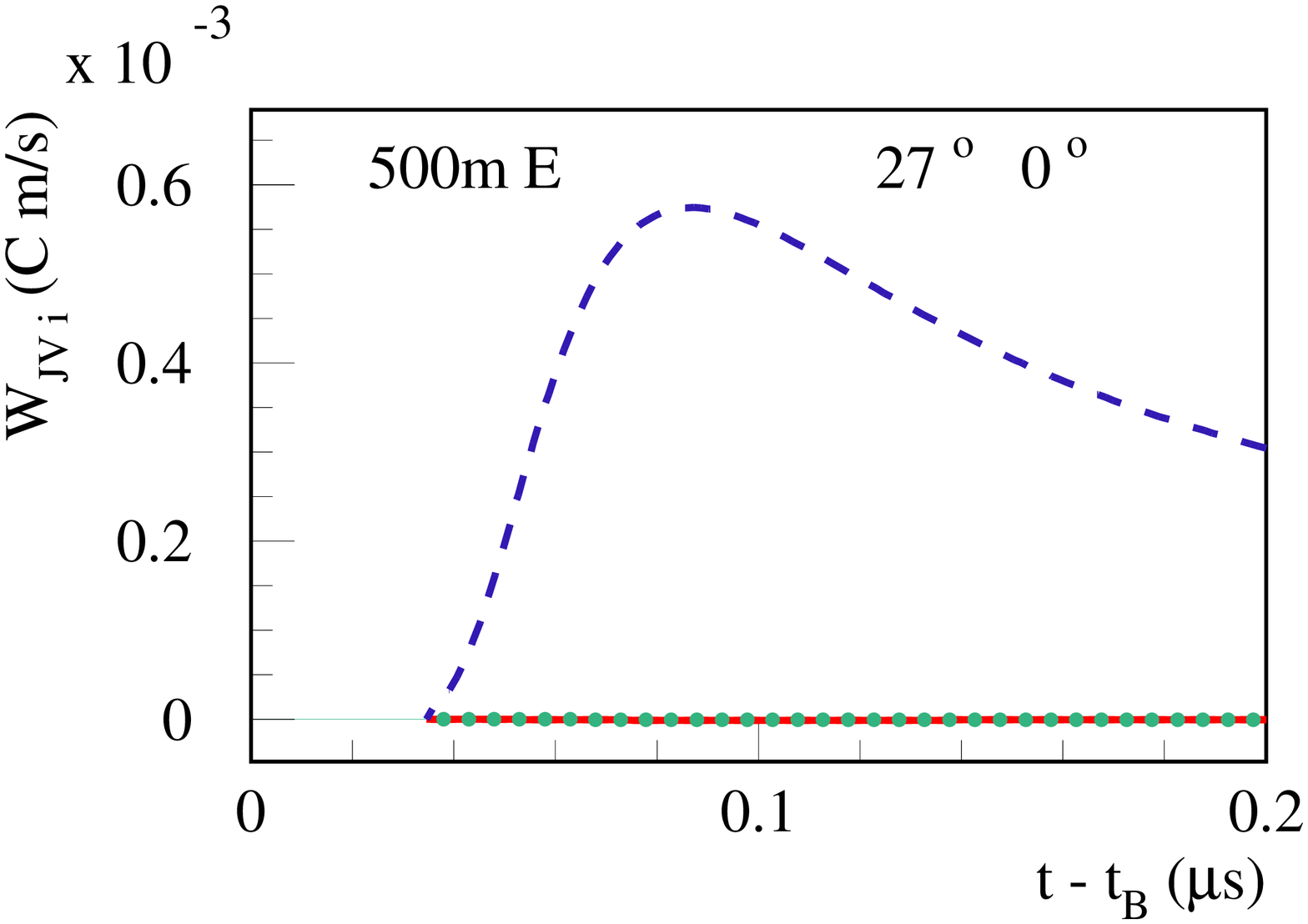}{\large \hspace*{-1.0cm}}\includegraphics[%
  scale=0.32,
  angle=270]{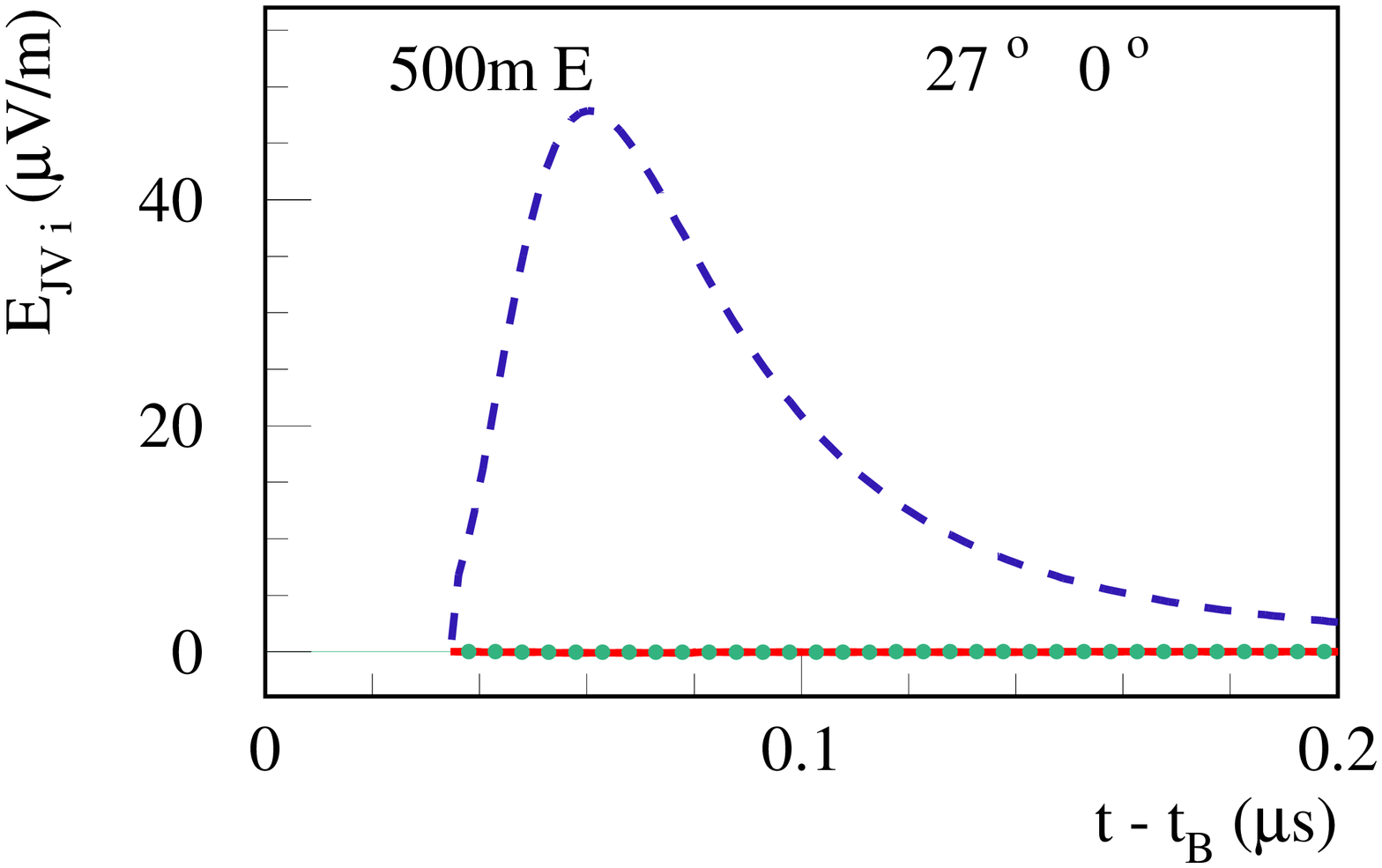}\end{center}
\vspace{-1.4cm}

\noindent \begin{flushleft}{\large (e)\hspace*{7.5cm}(f)}\end{flushleft}{\large \par}

\vspace{-2.5cm}
\begin{center}{\large \hspace*{-1cm}}\includegraphics[%
  scale=0.32,
  angle=270]{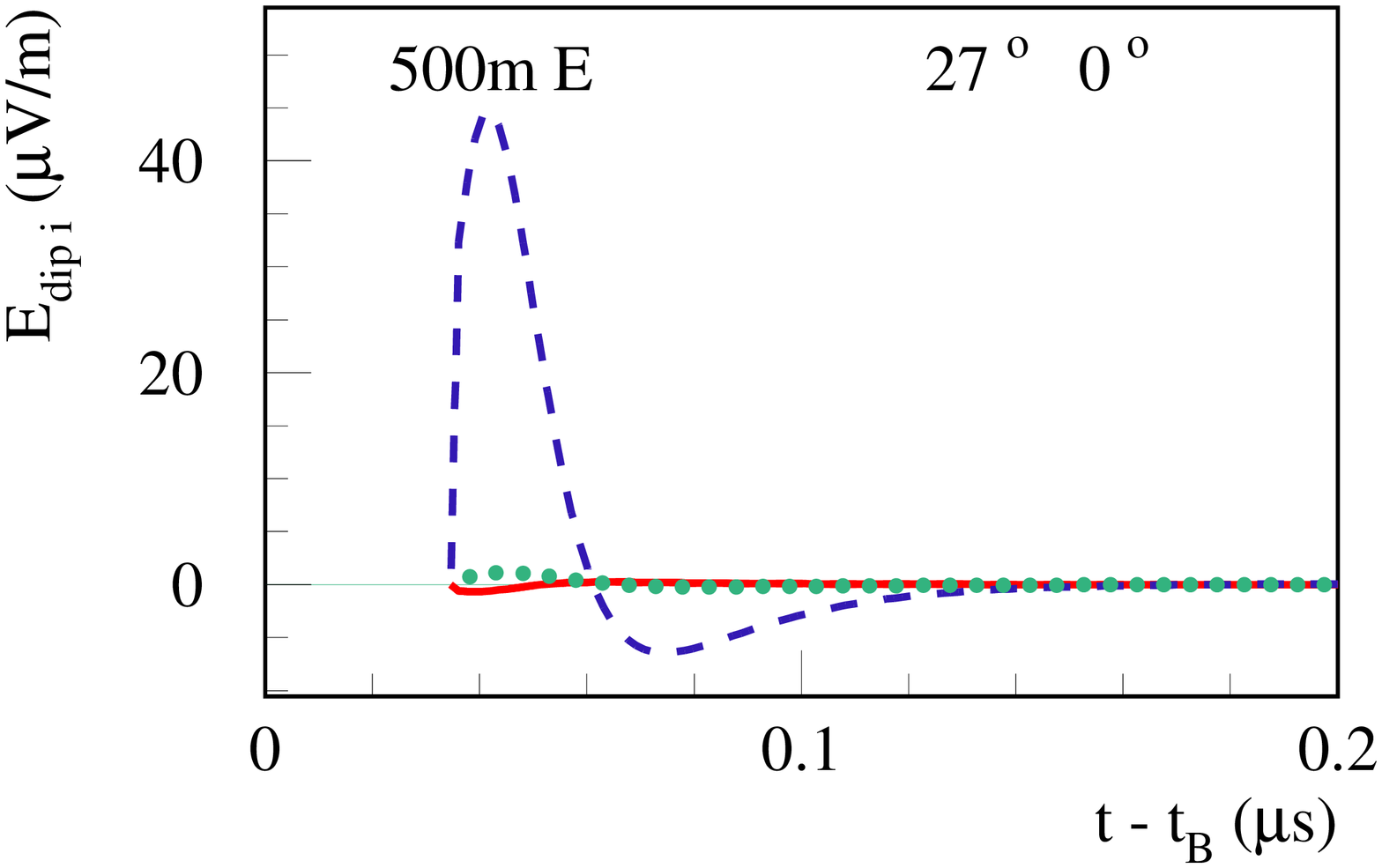}{\large \hspace*{-1.0cm}}\includegraphics[%
  scale=0.32,
  angle=270]{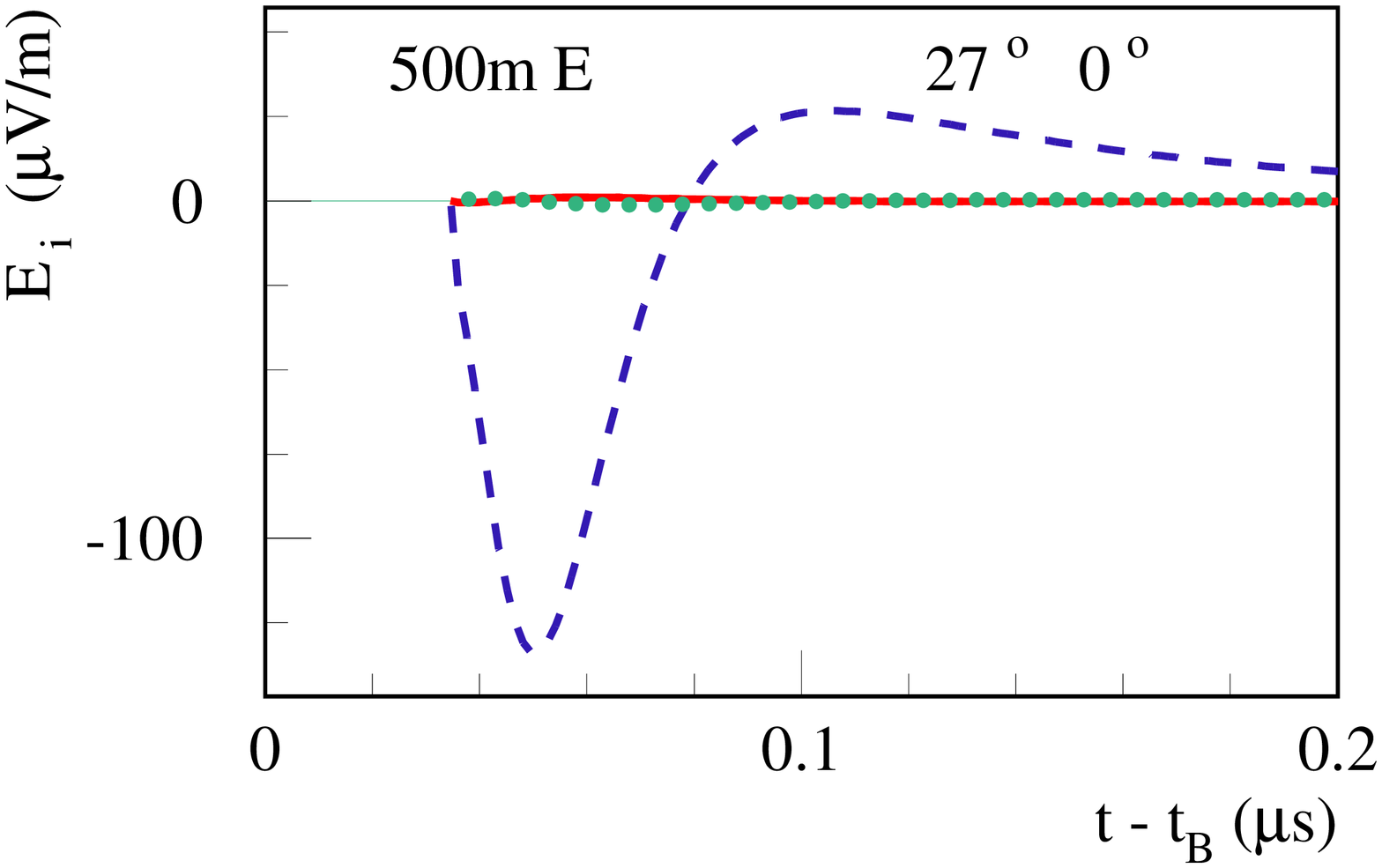}\end{center}
\vspace{-1.4cm}

\caption{The $x$- (full line), $y$- (dashed) and $z$-component (dotted)
of $\vec{W}_{RK}$ (a), $\vec{W}_{JV}$ (c), \label{cap:econtr} and
of the corresponding fields $\vec{E}_{RK}$ (b), $\vec{E}_{JV}$(d),
the dipole contribution $\vec{E}_{dip}$ (e), and the components of
the total electric field, with all contributions (f). The vector components
refer to the Earth frame.}
\end{figure*}

The angle $\psi$ refers to the origin of the shower, it is related
to the angle $\varphi$ from the preceding chapter as $\psi=180^{0}-\varphi$.
So $\psi=0$ means the shower moves from north to south.  We consider
again the magnetic field at the {\small CODALEMA} site, so the shower
makes an angle of roughly $54^{0}$ with the magnetic field. We consider
an observer at $x=0$, \textbf{$y=500\,$}m, $z=140\,$m (which means:
east of the impact pont). We suppose $a=z$ (so the shower hits the
ground at $x=y=0$). We plot again the components of the current,
divided by $Nec$, with $e$ being the electron charge, and $c$ the
velocity of light, see fig. \ref{cap:current0x}. The components $J^{x}/Nec$
and $J^{y}/Nec$ represent the average transverse drift velocity,
in units of $c$, caused by the magnetic field. Contrary to the case
studied in the preceding chapter, we have now a non-vanishing transverse
drift and $J^{x}/Nec$ is non-zero. The other component $J^{y}/Nec$
vanishes because both the shower trajectory and the magnetic field
are in the $y-z$ plane (in the shower frame). The component $J^{z}/Nec$
represents the charge excess. The dimensionless quantities $J^{i}/Nec$
vary only little with time, corresponding to a transverse drift velocity
 of 0.04 and a fractional charge excess of -0.2.   The small difference
compared to the result from the previous chapter is actually due to
event by event fluctuations.  The time variation of the currents is
mainly due to the time variation  of $N$. For the numerical calculations,
also the small time variations of $J^{i}/Nec$ are taken into account.
Not only the transverse current is non-vanishing, also the dipole
moment contributes, see fig. \ref{cap:current0x}.

Expressing the retarded time  in terms of the observation time, the
time dependence $J(t')$ of the current can be expressed as$J\big(t^{*}(t)\big)$
versus the observer time. In fig. \ref{cap:curr-earth}, we show the
components of the current and their derivatives in the Earth frame. 

We now discuss the electric fields, \begin{equation}
\vec{E}=\vec{E}_{\mathrm{dip}}+\vec{E}_{RK}+\vec{E}_{JV}=\vec{E}_{\mathrm{dip}}+\frac{\vec{W}_{RK}}{D}+\frac{\vec{W}_{JV}}{D}.\end{equation}
The three components of the vectors $\vec{W}$ and the corresponding
field contributions, the dipole field components, as well as the three
components of the complete field $\vec{E}$ are shown in fig. \ref{cap:econtr}.
The largest contribution is $\vec{E}_{RK}$, which has its origin
in the time variation of the currents. Also $\vec{E}_{JV}$ also contributes
significantly. This contribution is broader but smaller than $\vec{E}_{RK}$,
and of opposite sign, so the sum of the two gives finally a bipolar
field. Compared to the shower parallel to the magnetic field, studied
in the previous chapter, both contributions are much bigger, which
means that the dominant contribution is due to the transverse current
caused by the geomagnetic field. And here, the contribution due to
the time variation of this current is by far dominant. But we learn
as well that the charge excess contribution cannot be neglected. Also
the dipole field contributes significantly, with an opposite sign
relative to $\vec{E}_{RK}$.

\section{Different observer positions}

In this chapter, we will compare the fields seen by an observer at
four different positions: south, east, north, and west of the impact
point, always the same distance (500 m). %
\begin{figure}[htb]
\begin{center}\hspace*{-1cm}\includegraphics[%
  scale=0.35,
  angle=270]{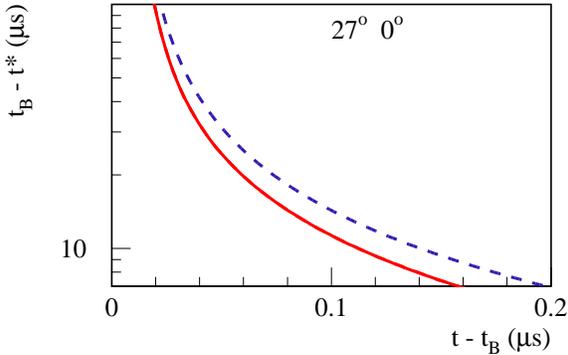}\end{center}
\vspace{-0.4cm}

\caption{The retarded time expressed via $t_{B}-t^{*}$ versus observer time
$t-t_{B}$, for a shower with $\theta=27^{0}$ , $\psi=0^{0}$, seen
by observers south (full ), or east (dashed) of the point of impact,
at a distance of $500\,\mathrm{m}$. \label{cap:tstar27b}}
\end{figure}
 We show in fig. \ref{cap:tstar27b} the retarded time versus observer
time $t-t_{B}$, for our shower with $\theta=27^{0}$ , $\psi=0^{0}$,
as seen by different observers. For a given $b$, the curves are different
for different observers. This fact will help to understand the different
widths of the signals seen from the different observers.

\begin{figure*}[htb]
~\vspace*{-2cm}

\begin{center}{\large \hspace*{-1cm}}\includegraphics[%
  scale=0.3,
  angle=270]{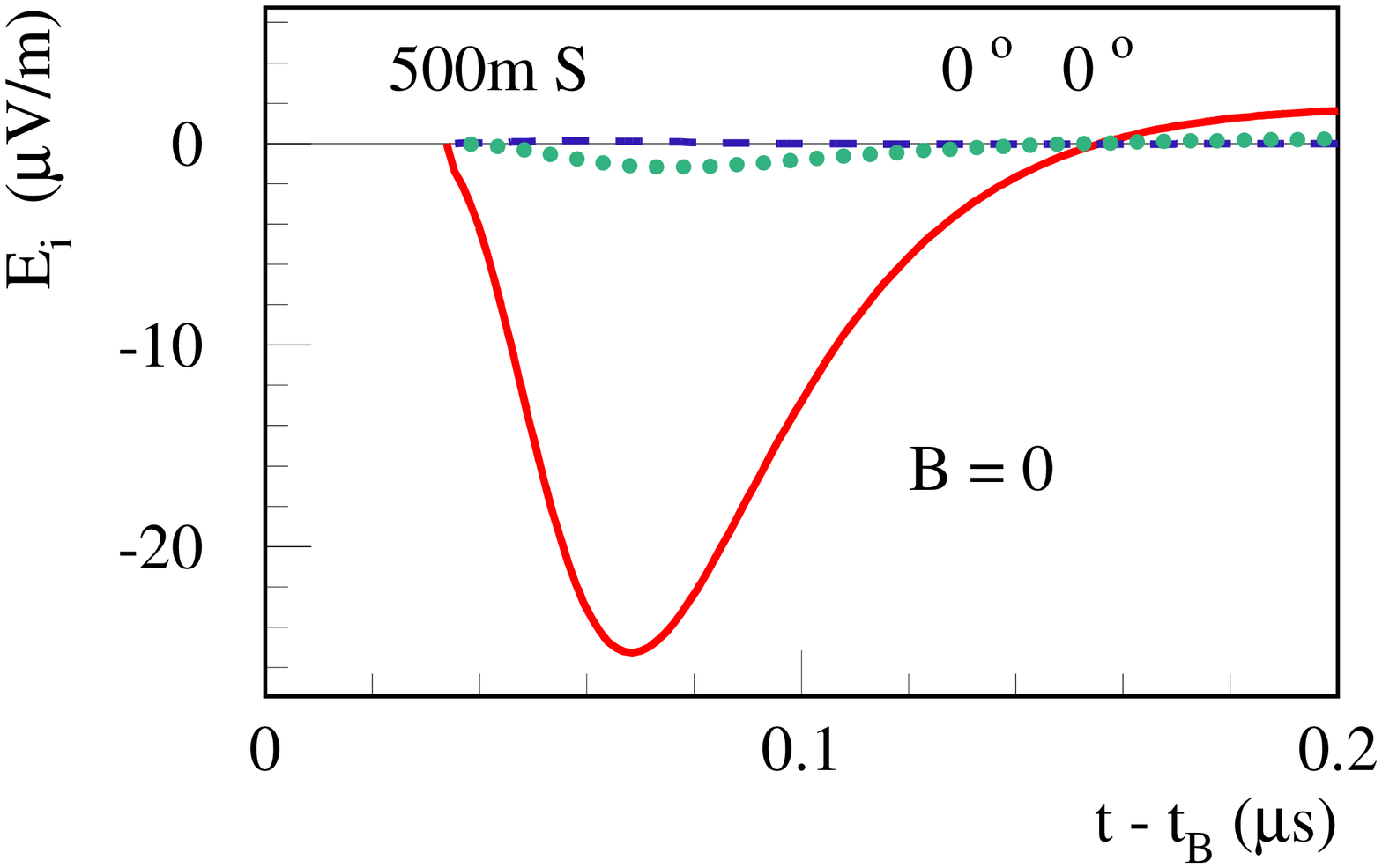}\hspace*{-1.5cm}\includegraphics[%
  scale=0.3,
  angle=270]{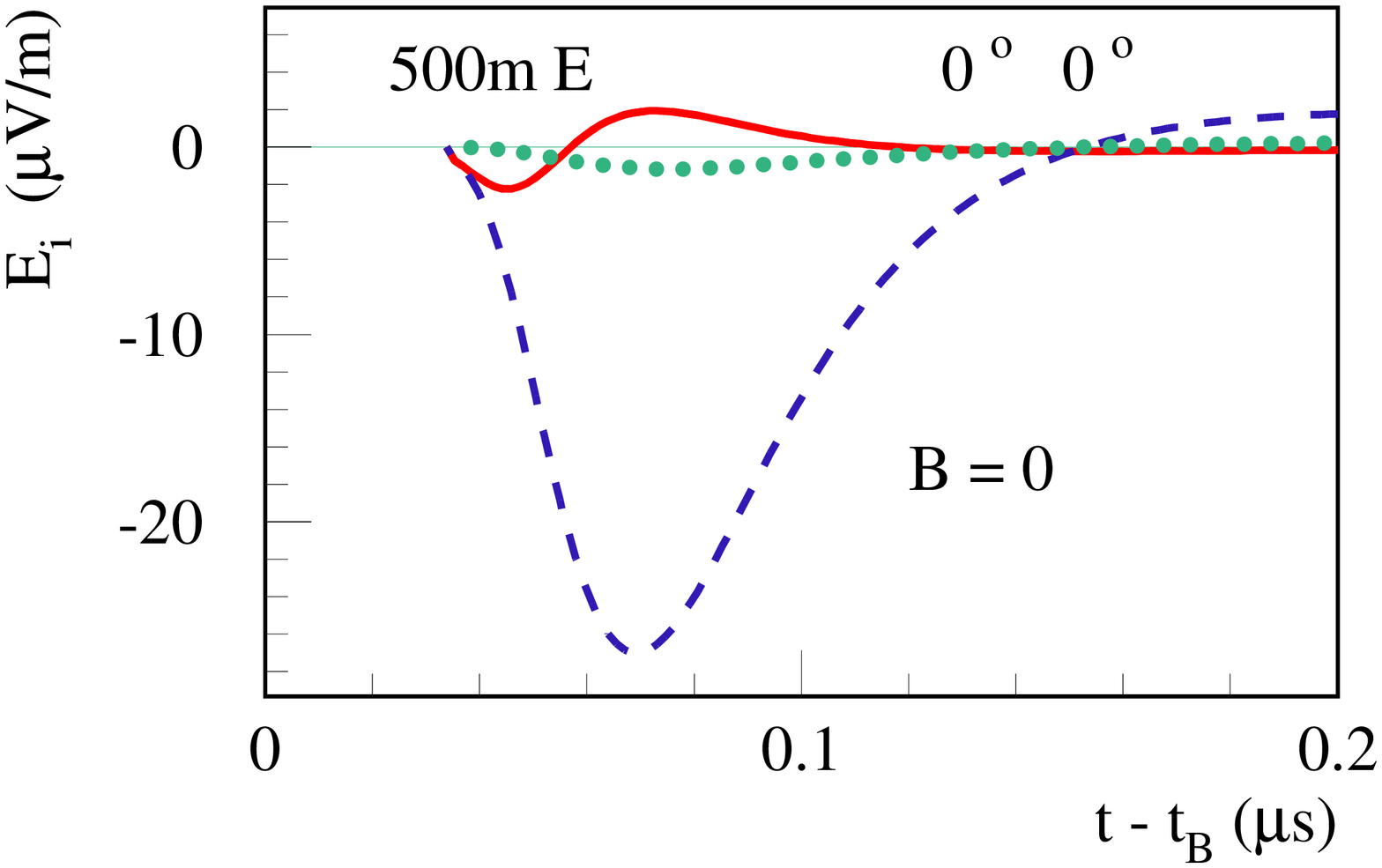}\end{center}

\vspace{-2.8cm}
\begin{center}{\large \hspace*{-1cm}}\includegraphics[%
  scale=0.3,
  angle=270]{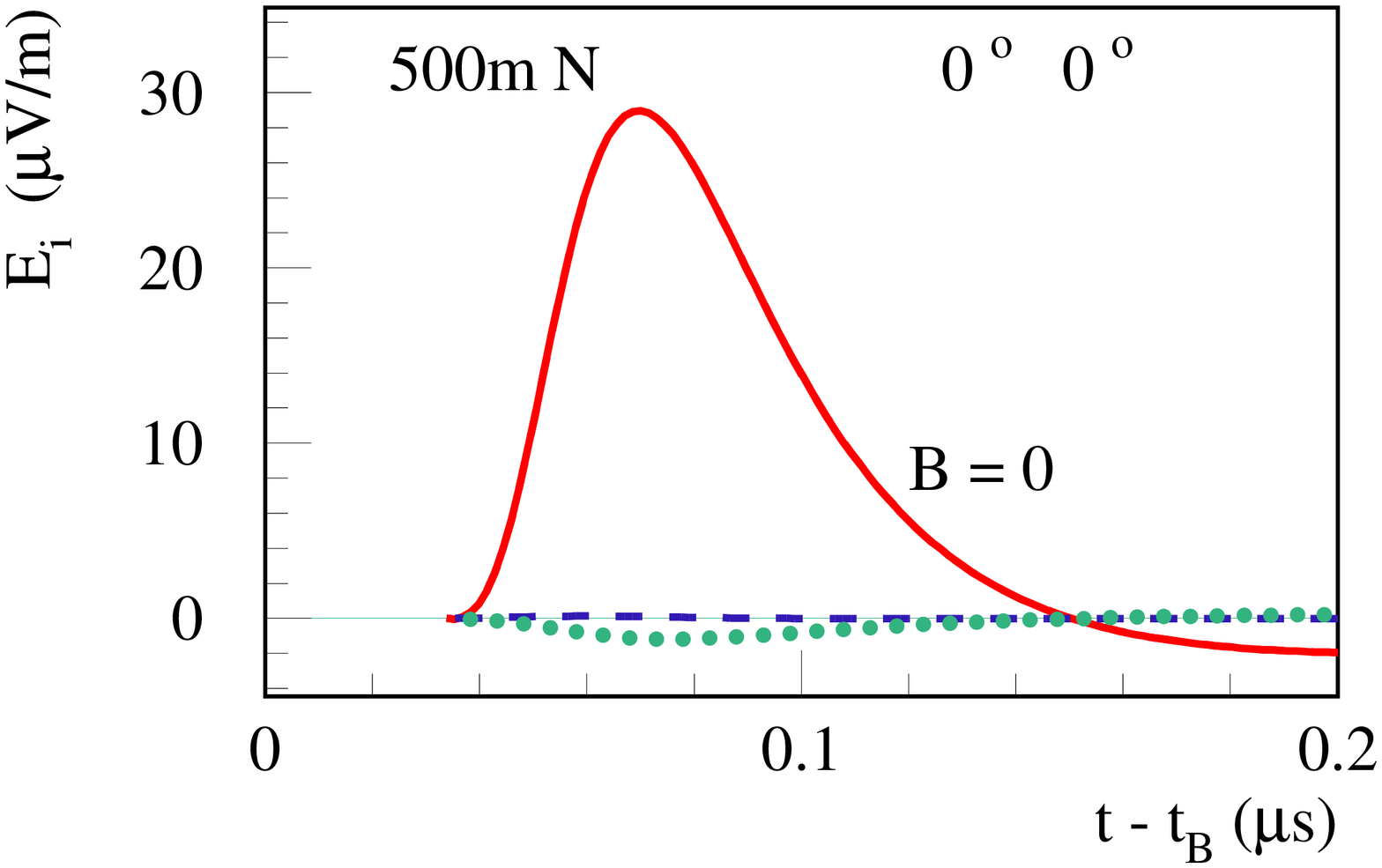}\hspace*{-1.5cm}\includegraphics[%
  scale=0.3,
  angle=270]{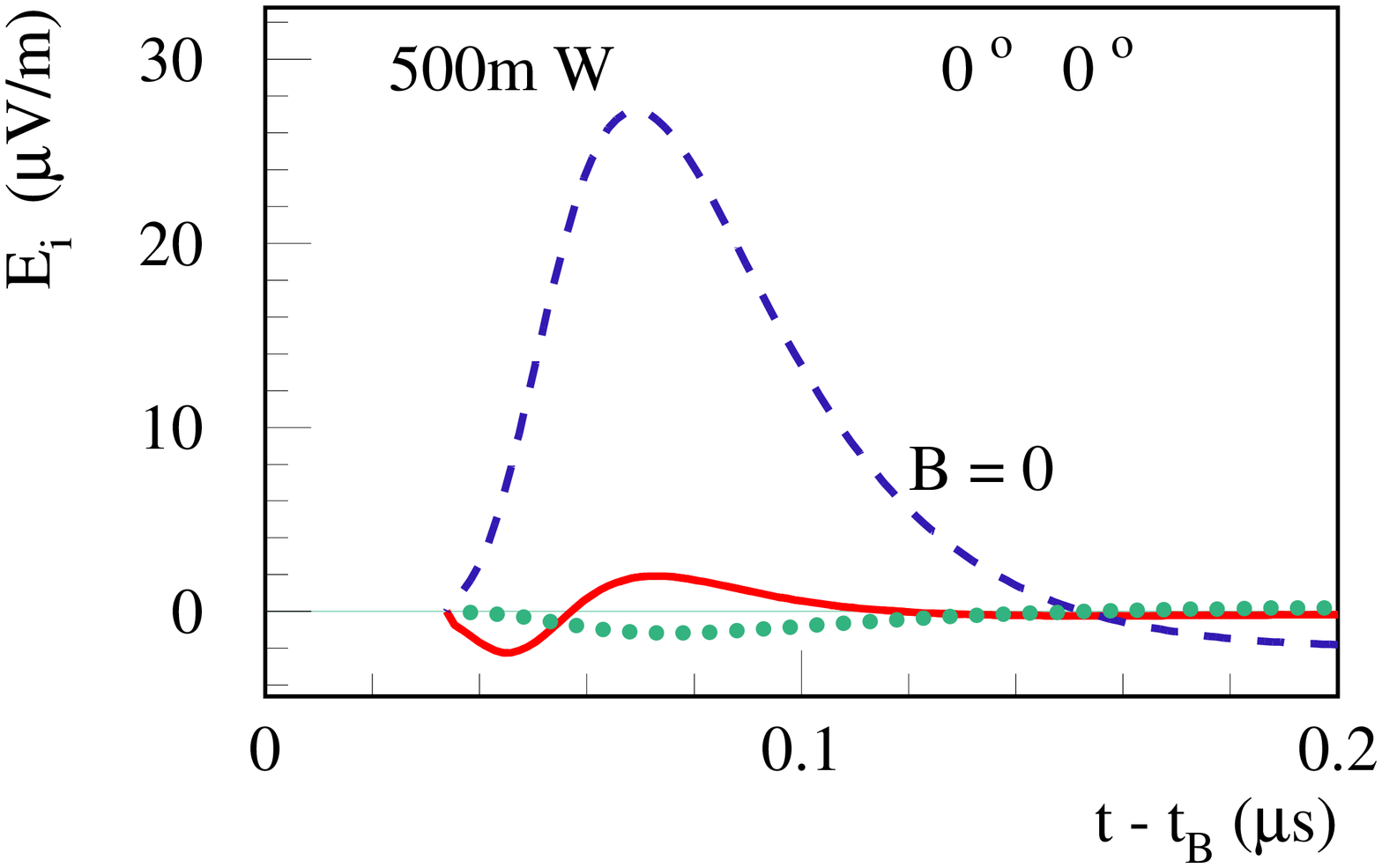}\end{center}
\vspace{-2.5cm}

{\large \hspace*{-1cm}}\includegraphics[%
  scale=0.3,
  angle=270]{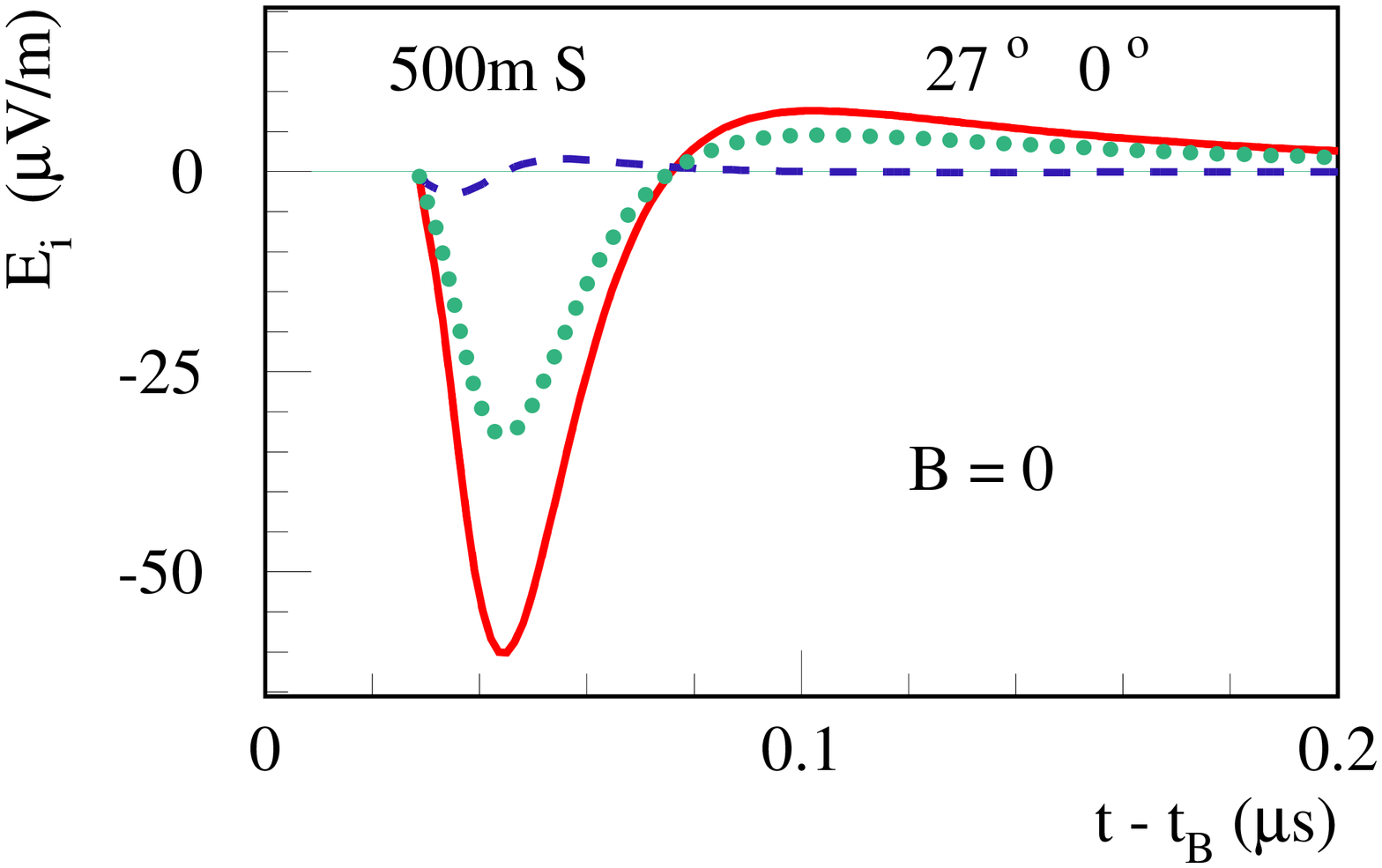}\hspace*{-1.5cm}\includegraphics[%
  scale=0.3,
  angle=270]{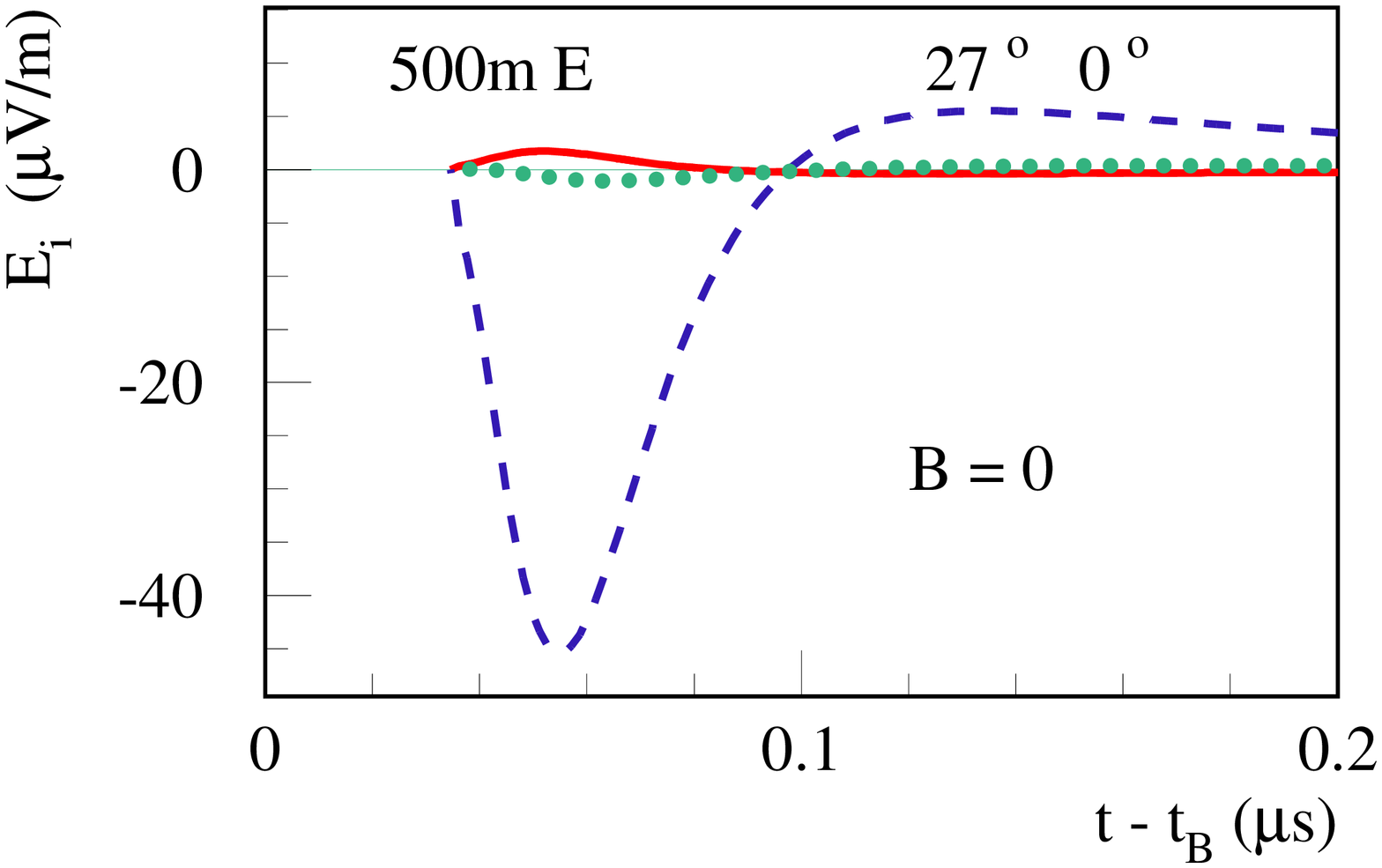}

\vspace{-2.5cm}
\begin{center}{\large \hspace*{-1cm}}\includegraphics[%
  scale=0.3,
  angle=270]{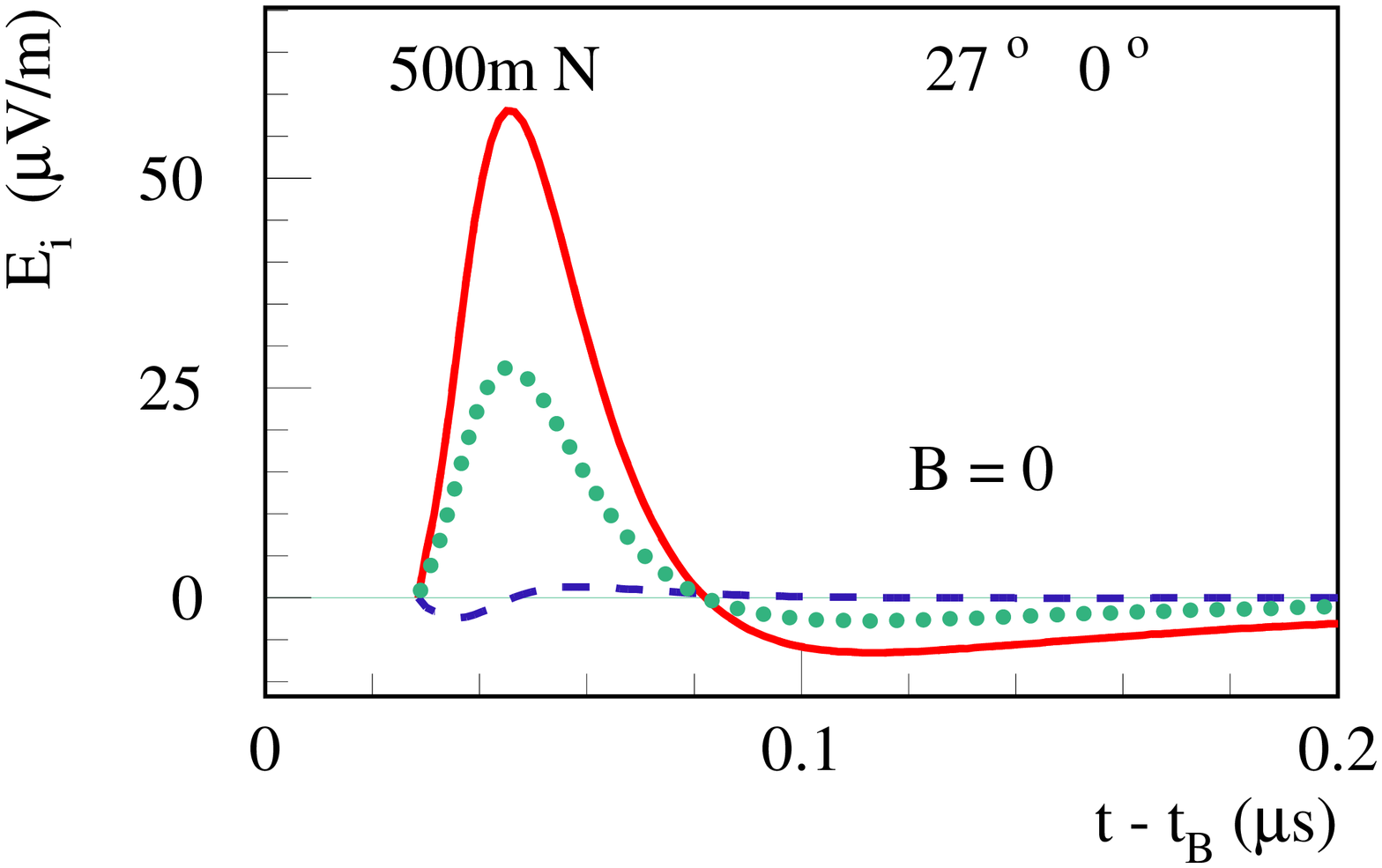}\hspace*{-1.5cm}\includegraphics[%
  scale=0.3,
  angle=270]{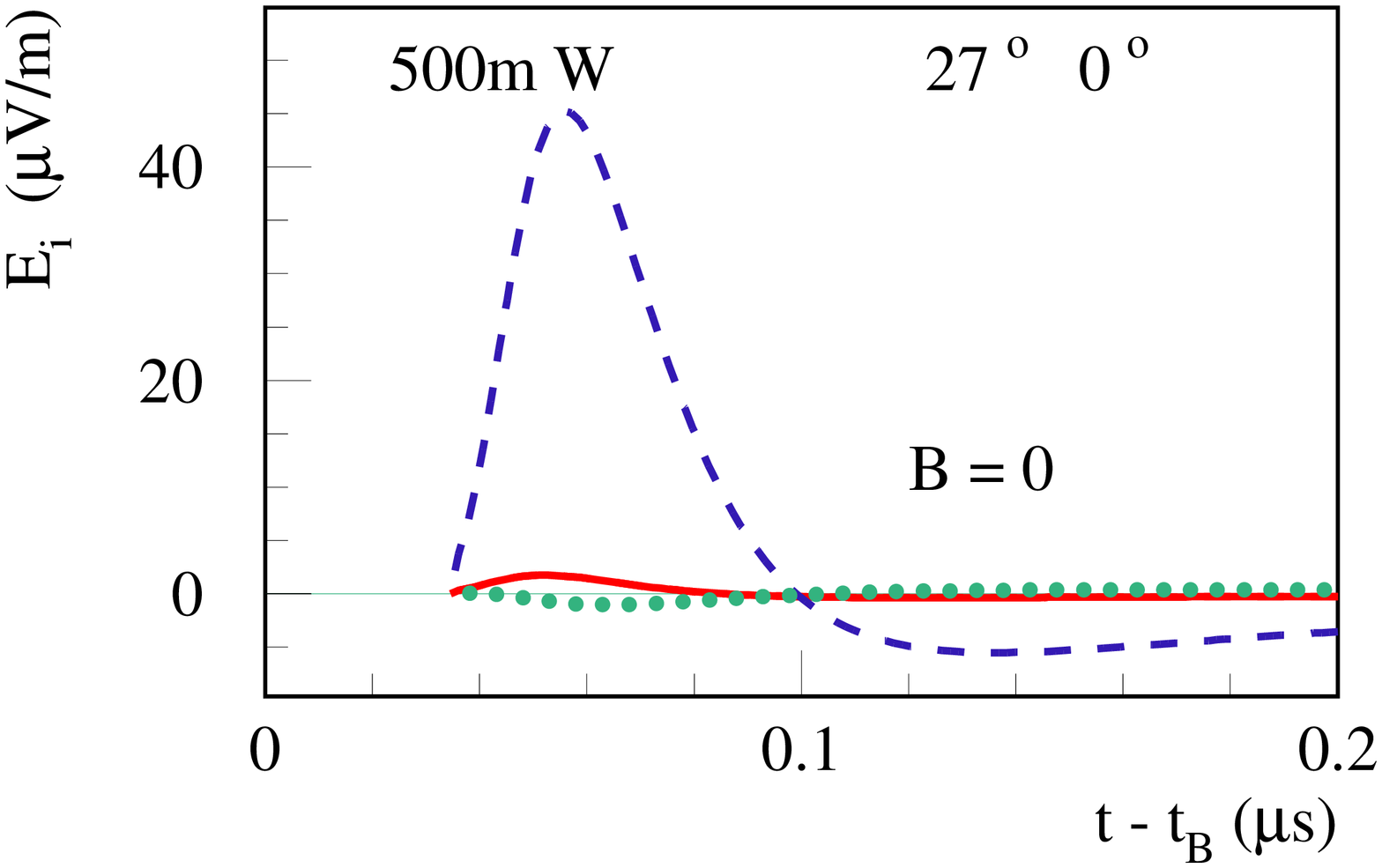}\end{center}
\vspace{-1.4cm}

\caption{The $x$- (full line), $y$- (dashed) and $z$-component (dotted)
of the electric field of a vertical shower (upper four plots) and
a shower with $\theta=27^{0}$, $\psi=0^{0}$ (lower four plots),
without geo-magnetic field, for four observers 500 m south, east,
north, and west of the point of impact.\label{cap:obs00000no} The
vector components refer to the Earth frame.}
\end{figure*}
\begin{figure*}[htb]
~\vspace*{-2cm}

\begin{center}\hspace*{-1cm}\includegraphics[%
  scale=0.3,
  angle=270]{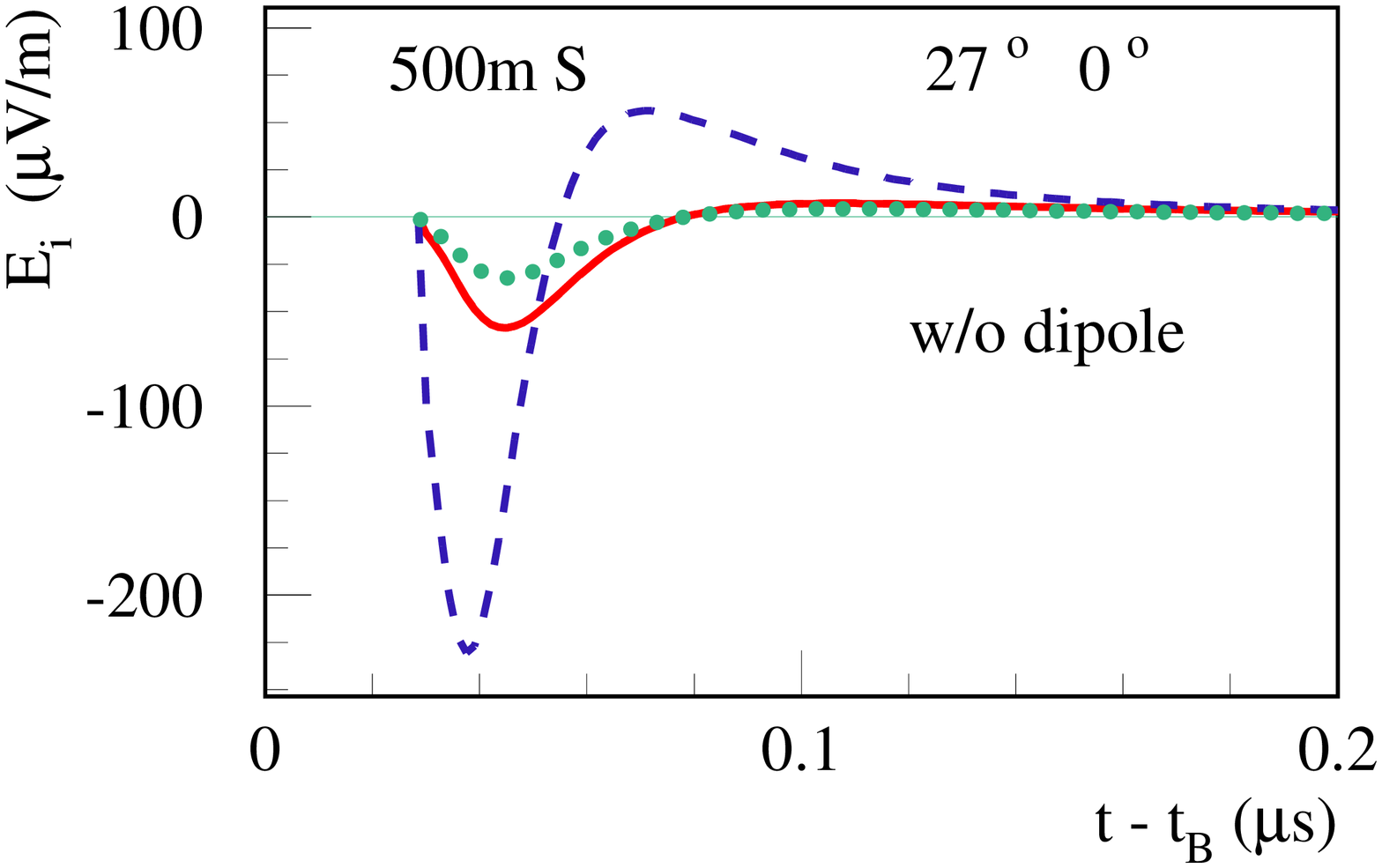}\hspace*{-1cm}\includegraphics[%
  scale=0.3,
  angle=270]{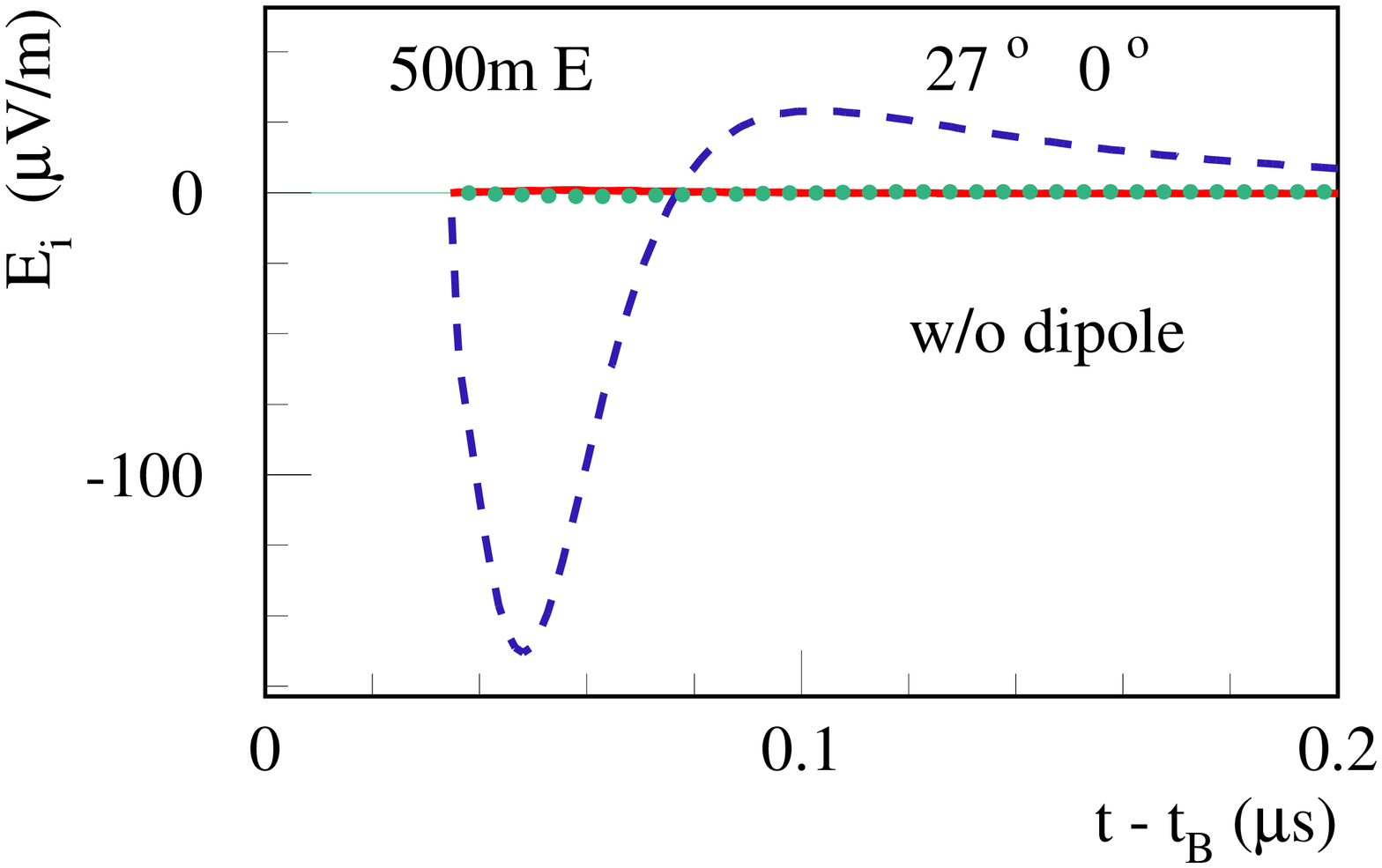}\end{center}

\vspace{-2.8cm}
\begin{center}\hspace*{-1cm}\includegraphics[%
  scale=0.3,
  angle=270]{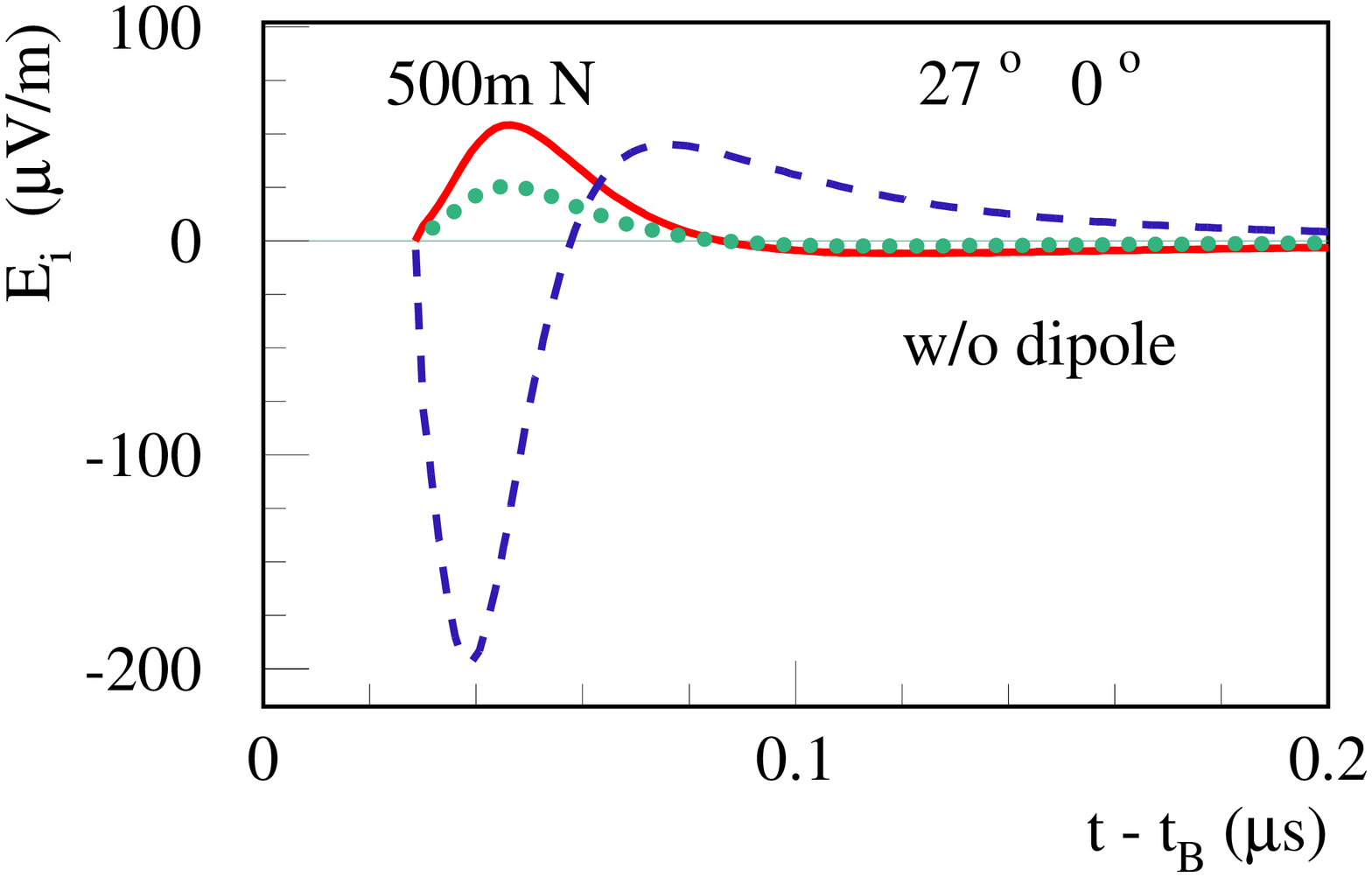}\hspace*{-1cm}\includegraphics[%
  scale=0.3,
  angle=270]{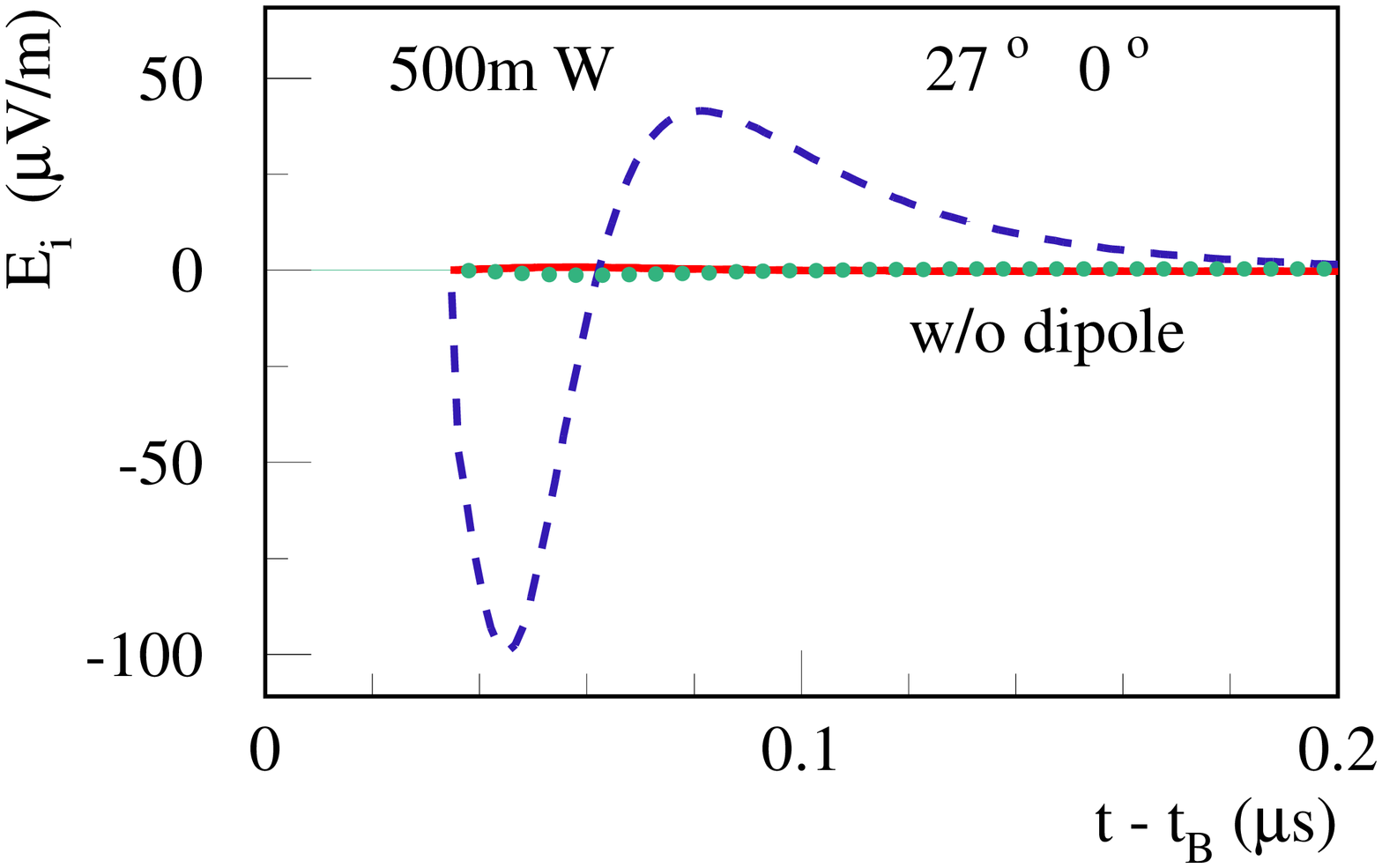}\end{center}
\vspace{-2.8cm}

\begin{center}\hspace*{-1cm}\includegraphics[%
  scale=0.3,
  angle=270]{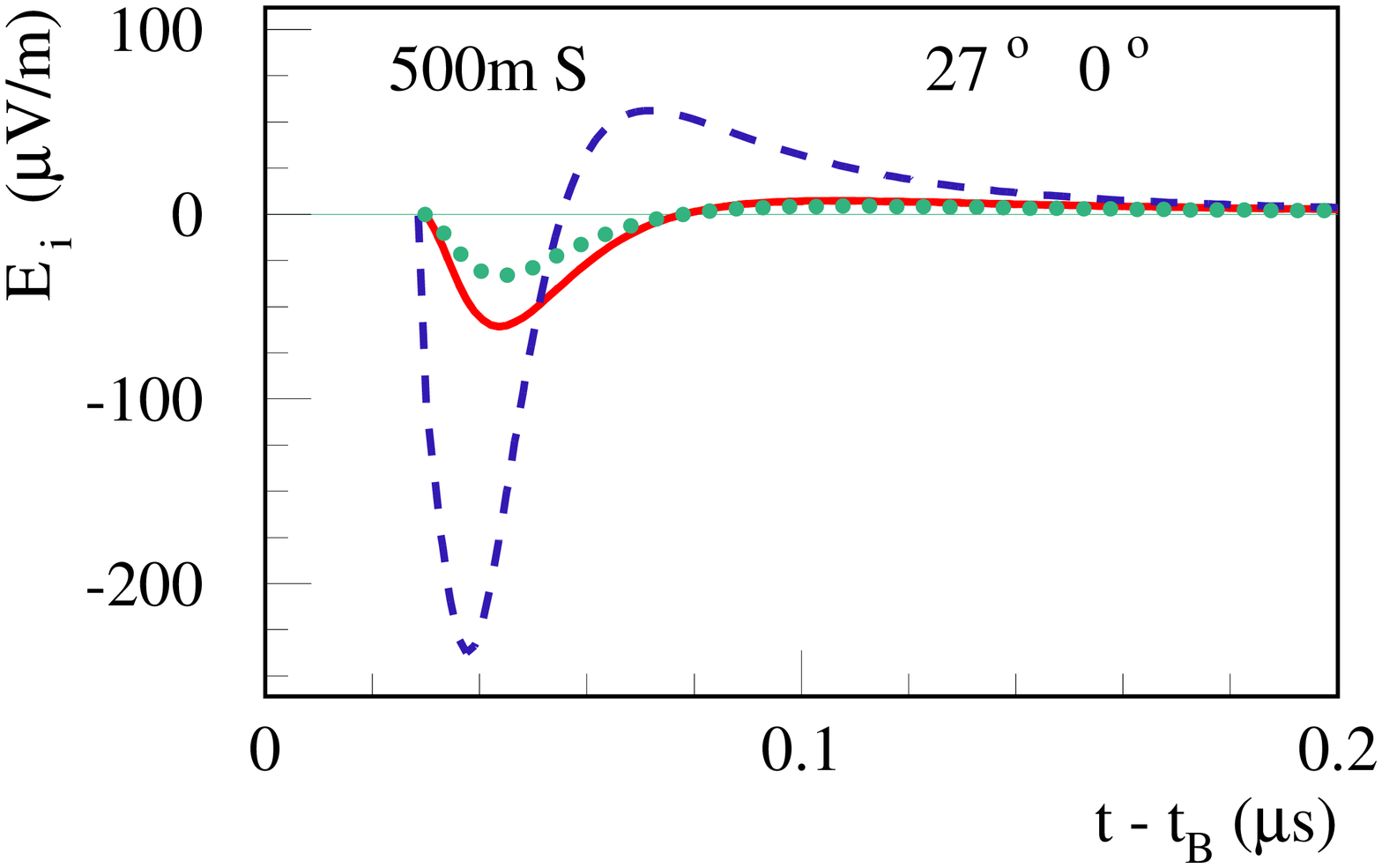}\hspace*{-1cm}\includegraphics[%
  scale=0.3,
  angle=270]{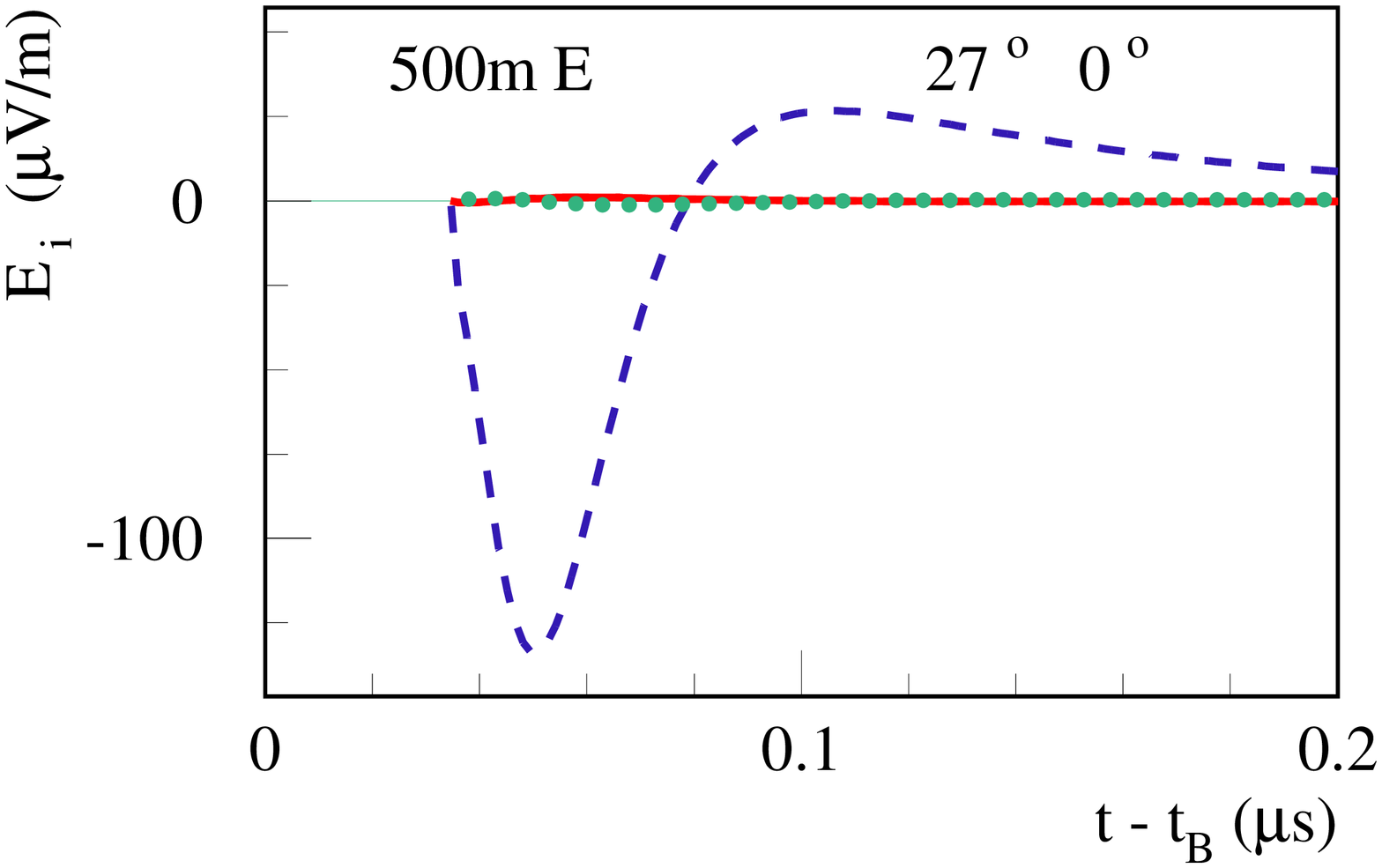}\end{center}

\vspace{-2.9cm}
\begin{center}\hspace*{-1cm}\includegraphics[%
  scale=0.3,
  angle=270]{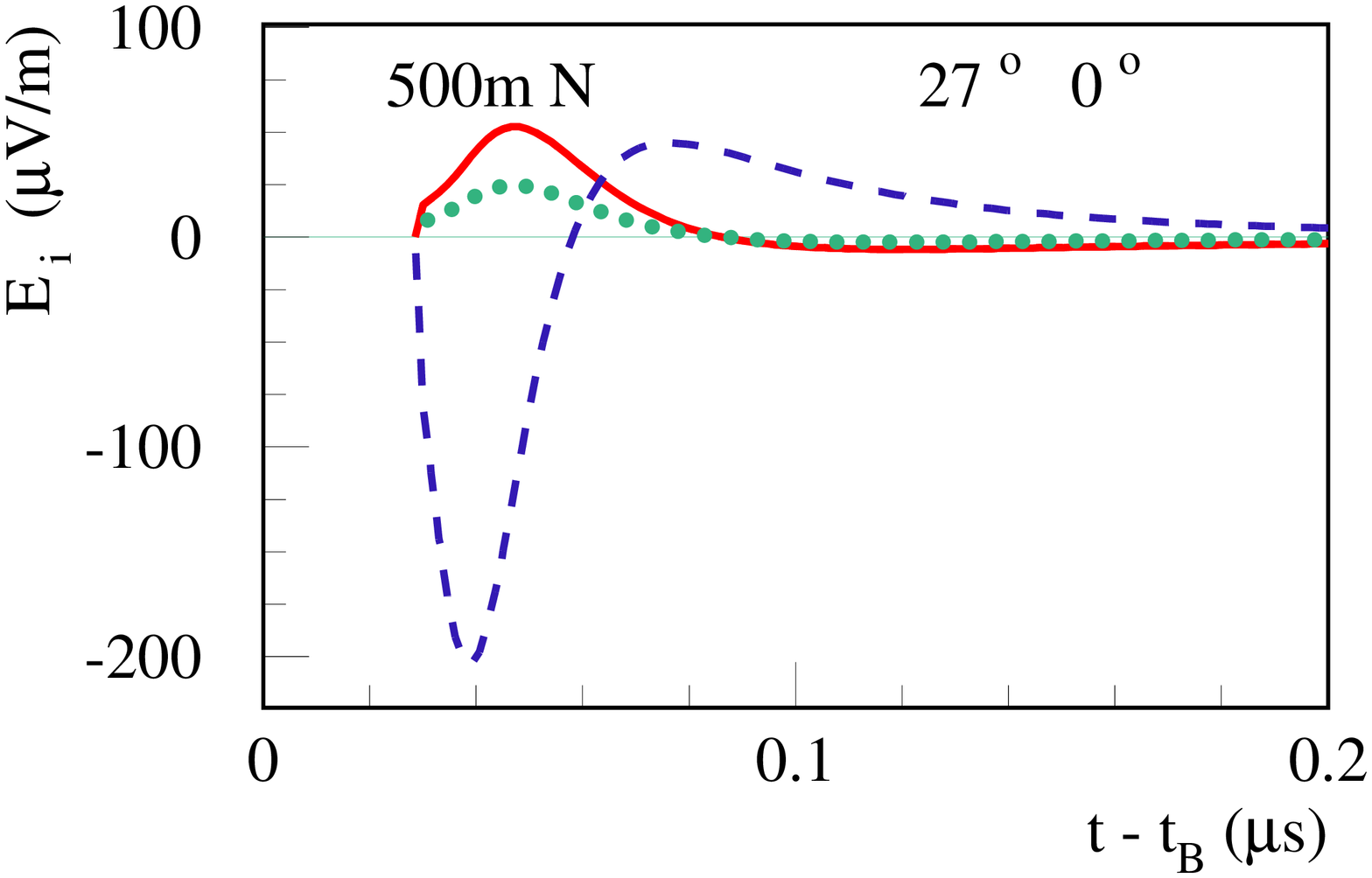}\hspace*{-1cm}\includegraphics[%
  scale=0.3,
  angle=270]{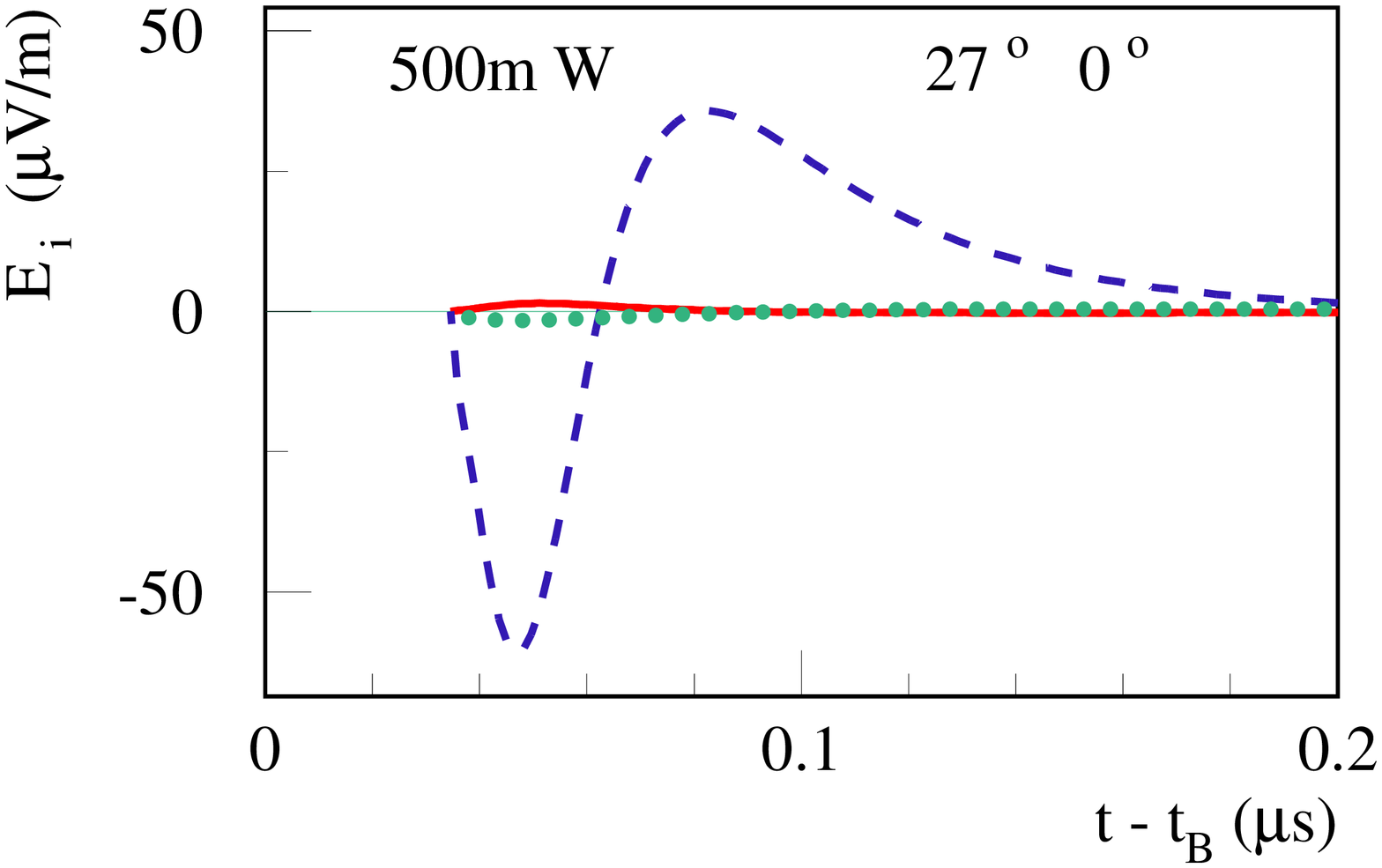}\end{center}
\vspace{-1.7cm}

\caption{The $x$- (full line), $y$- (dashed) and $z$-component (dotted)
of the electric field of a shower with $\theta=27^{0}$, $\psi=0^{0}$
, with the full geo-magnetic field, for four observers south, east,
north, and west of the point of impact, at a distance of $500\,\mathrm{m}$.
The upper four plots show the the fields without the dipole contribution,
the lower four plots represent all contributions. \label{cap:obs27000}
The vector components refer to the Earth frame.}
\end{figure*}
We consider first of all a vertical shower ($\theta=0^{0}$), with
the magnetic field artificially turned off. South corresponds to positive
$x$, east to positive $y$. The results are shown in fig. \ref{cap:obs00000no}.
We observe a radial electric field, as expected. The slight deviation
of a perfect axial symmetry is due to the fact that the calculation
is based on a Monte Carlo generated current, which is of course only
approximately symmetric. So even without geo-magnetic field, one observes
a sizable field of the order of $25\,\mu\mathrm{V/m}$. 

Also shown in fig. \ref{cap:obs00000no} are the fields for a slightly
inclined shower, with $\theta=27^{0}$, $\psi=0^{0}$, still without
geomagnetic field. The results are quite similar, we still observe
a radial electric field, with respect to the shower axis (with gives
a non-zero $z-$component in the Earth frame in case of a north or
south observer). The signal is also narrower, which is also easy to
understand, since in case of non-zero inclination the maximum of the
current corresponds to an earlier retarded time $t^{*}$ compared
to a vertical shower, and correspondingly the derivative $dt^{*}/dt$
is bigger. Therefore a current with a given width in $t^{*}$ will
be narrower in $t$, in case of an inclined shower, leading to a narrower
signal. 

In fig.\ref{cap:obs27000} (upper four plots), we show the results
for the same inclined shower, with a realistic geo-magnetic field,
but without the dipole contribution. We get in addition to the charge
excess contribution (seen in the case without magnetic field) a strong
additional contribution perpendicular to the plane defined by shower
axis and geo-magnetic field. As discussed earlier, the dominant contribution
is $\vec{E}_{RK}$, due to the time variation $K$ of the currents,
and here mainly due to the transverse current from the geo-magnetic
field. However, seen from the comparison with the $B=0$ case, the
charge excess is not negligible.

We finally show in fig. \ref{cap:obs27000} (lower four plots) the
complete fields for the four observers, including the dipole field.
Whereas for the north and the south observers including the dipole
field does not change the results, we find for the east and west observers
a considerable reduction of the strength due to the dipole contribution.

\section{Conclusion}

We introduced a macroscopic calculation of radio emission from air
showers initiated by ultra-high energy cosmic rays, based on currents
obtained from {\small CONEX} Monte Carlo simulations of air showers
in a realistic geo-magnetic field. We took into account a realistic
index of refraction $n$, which varies along the trajectory of the
shower. 

We demonstrated that in the case of a realistic index of refraction
$n(\vec{y}$) (depending on the space position $\vec{y}$) one can
derive expressions for the electric field which are formally equivalent
to those for $n=1$, however with a modified definition of the retarded
time. One has to take into account the fact that light moves in general
on curved trajectories, and that the velocity of light changes along
the trajectory. These curves are solutions of differential equations
containing $n(\vec{y})$. Finding the retarded time amounts to integrating
this differential equation, with given boundary conditions. An accurate
approximation can be obtained by assuming straight light trajectories.
 Another delicate point is denominator $D$ in the expressions for
the electric field, which is inversely proportional to the square
of the time derivative $\partial t^{*}/\partial t$ of the retarded
time. This derivative may become singular, and the position of the
singularities is strongly affected by the precise value of $n$. So
even though $n$ is very close to unity, the small deviations from
unity are important. 

We showed that the electric field can be expressed in terms of the
four-vector $R$ joining the observer's position and the shower, the
velocity four-vector $V$ of the shower, the four-current $J$, and
the derivative $K$ of the four-current, as $\vec{E}=\vec{E}_{\mathrm{dip}}+\vec{E}_{RK}+\vec{E}_{JV}$
$=\vec{E}_{\mathrm{dip}}+(\vec{W}_{RK}+\vec{W}_{JV})/D$ with $\vec{W}_{AB}$
defined as $\vec{A}B^{0}-A^{0}\vec{B}$. The term $\vec{E}_{JV}$
is zero in case of a longitudinally moving point charge, and non-zero
in case of transverse currents generated by the magnetic field. This
contribution therefore vanishes for shower trajectories parallel to
 the geo-magnetic field, and the contribution increases with increasing
 angle between shower axis and magnetic field.  The most important
contribution though is the contribution $\vec{E}_{RK}$ caused by
the time variation $K$ of the four-current. Here we can identify
actually two contributions: one due to time variation of the charge
excess, which is always present, and an second one due to the time
variation of the transverse current due to the geo-magnetic field.
This second contribution is only present in case of a non-zero angle
between the shower axis and the magnetic field. For the shower we
investigated ($\theta=27^{0}$, $\psi=0^{0}$), the second contribution
is four times larger than the first.  The field is thus clearly dominated
by the contributions due to the transverse current in the geo-magnetic
field, but the charge excess contribution can not be neglected. Also
the dipole field $\vec{E}_{\mathrm{dip}}$ contributes significantly.

We get in any case a characteristic bipolar signal, reflecting first
of all the time dependence of the variation $K$ of the currents.
In case of geo-magnetic contributions this bipolarity is even more
pronounced, since here we get an additional contribution $\vec{E}_{JV}$
whose time signal reflects the time dependence of the current $J$,
giving therefore a much broader signal, of opposite sign compared
to $\vec{E}_{RK}$. The sum is a bipolar signal with two strong pulses.

We emphasis that this bipolarity is in qualitative contrast to the
findings of \cite{Fal03,Sup03,Hue03,Hue05,Hue07}, where a microscopic
picture has been employed, leading to a single pulse. The fundamental
difference is that the time variation of the currents in the macroscopic
picture treats in an effective way the creation and disappearance
of charges. In a microscopic picture this has to be treated explicitly,
otherwise one misses an important part.

\appendix

\section{Potential and field from a pointlike current\label{sec:Potential-and-Field}}

We have\begin{equation}
j^{\beta}(x^{0},\vec{x})=\int dx'^{0}J^{\beta}(x'^{0})\,\delta^{(4)}(x-\xi(x'^{0})),\end{equation}
with $x^{0}=ct$ and $\xi_{0}(x'^{0})=x'^{0}$. The potential due
to the current $J$ is\begin{eqnarray}
A^{\beta}(x) & = & \mu_{0}\int d^{4}y\frac{1}{2\pi}\theta(x^{0}-y^{0})\delta((x-y)^{2})\\
 &  & \qquad\times\int dx'^{0}J^{\beta}(x'^{0})\,\delta^{(4)}(y-\xi(x'^{0})).\nonumber \end{eqnarray}
We change $\int d^{4}y...\int dx'^{0}...$ into \begin{equation}
\int dx'^{0}...\int d^{4}y\,\delta^{4}(y-\xi(x'^{0}))...,\end{equation}
which gives\begin{eqnarray}
A^{\beta}(x) & = & \frac{\mu_{0}}{2\pi}\int dx'^{0}\,\theta\big(x^{0}-\xi^{0}(x'^{0})\big)\,\\
 &  & \times\delta\big((x-\xi(x'^{0}))^{2}\big)J^{\beta}(x'^{0}).\nonumber \end{eqnarray}
Using $R=x-\xi(x'^{0})$, we have\begin{equation}
A^{\beta}(x)=\frac{\mu_{0}}{2\pi}\int dx'^{0}\,\theta(x^{0}-\xi^{0}(x'^{0}))\,\delta(R^{2})J^{\beta}(x'^{0}).\end{equation}
Using\begin{equation}
\delta(R^{2})=\frac{\delta(x'^{0}-\tilde{x}{}^{0})}{\mid\left(dR^{2}/dx'^{0}\right)_{x'^{0}=\tilde{x}{}^{0}}\mid},\end{equation}
with $\mathrm{x^{*}{}^{0}}$ being the root of $R^{2}$(supposing
there is only one), and\[
\frac{dR^{2}}{dx'^{0}}=2R\frac{dR}{dx'^{0}}=-2RU,\]
with $V=d\xi/dx'^{0}$, we get for $x_{0}\neq\xi_{0}(x'^{0})$\begin{equation}
A^{\beta}(x)=\frac{\mu_{0}}{4\pi}\,\frac{J^{\beta}}{|R\, V|},\end{equation}
with all quantities considered at the retarded time $x^{*}{}^{0}$.
The electric field is given as\begin{equation}
E^{i}=c\left(\partial^{i}A^{0}-\partial^{0}A^{i}\right).\end{equation}
We have\begin{equation}
\partial^{\alpha}A^{\beta}(x)=\frac{\mu_{0}}{4\pi}\,\partial^{\alpha}\frac{J^{\beta}}{|R\, V|}.\end{equation}
We have\begin{equation}
\partial^{\alpha}\frac{J^{\beta}}{|RV|}=\frac{dJ^{\beta}/d\tilde{x}{}^{0}}{|RV|}\partial^{\alpha}x^{*}{}^{0}-\frac{J^{\beta}}{RV\,|RV|}\partial^{\alpha}\{ RV\}.\end{equation}
The derivative of $RV$ is\begin{equation}
\partial^{\alpha}\{ RV\}=V^{\alpha}-VV\,\partial^{\alpha}x^{*}{}^{0}\,.\end{equation}
From $R^{2}=0$ we conclude\begin{equation}
\partial^{\alpha}R^{2}=2R^{\alpha}-2RV\,\partial^{\alpha}x^{*}{}^{0}=0,\end{equation}
so\begin{equation}
\partial^{\alpha}x^{*}{}^{0}=\frac{R^{\alpha}}{RV}\,.\end{equation}
Putting all together, we obtain\begin{equation}
\partial^{\alpha}A^{\beta}(x)=\frac{\mu_{0}}{4\pi}\,\frac{L^{\alpha\beta}}{|RV|^{3}}\,,\end{equation}
with \begin{equation}
L^{\alpha\beta}=RV\, R^{\alpha}\dot{J}^{\beta}-RV\, J^{\beta}V^{\alpha}+VV\, R^{\alpha}J^{\beta}.\end{equation}
where the {}``dot'' means derivative with respect to $x^{*}{}^{0}$.
The electric field is then\begin{equation}
E^{i}(x)=\frac{1}{4\pi\epsilon_{0}c}\,\frac{L^{i0}-L^{0i}}{|RV|^{3}}\,,\end{equation}
with $n$ being the index of refraction at the observer position.

\section{Currents from air shower simulations}

We compute the current by using the {\small CONEX} shower simulation
program, as\begin{equation}
\vec{J}=\sum_{\mathrm{positrons\, i}}e\vec{v}_{i}-\sum_{\mathrm{electrons\, j}}e\vec{v}_{j},\end{equation}
considering all electrons and positrons present at a given time. With
$t_{0}$ being the creation time of a particle, its velocity some
time later can be written as\begin{equation}
\vec{v}(t)=\vec{v}_{\Vert}(t_{0})+|\vec{v}_{\bot}(t_{0})|\left(\vec{e}_{1}\,\cos\omega t+\vec{e}_{2}\,\sin\omega t\right),\label{eq:velo}\end{equation}
where $\vec{v}_{\Vert}$ and $\vec{v}_{\bot}$are the velocity components
parallel and orthogonal to the magnetic field, $\omega=\pm eB/\gamma m$,
and where \{$\vec{e}_{i}$\} is a system of orthonormal vectors, defined
via\begin{equation}
\vec{e}_{3}=\vec{B}/|\vec{B}|,\quad\vec{e}_{1}=\vec{v}_{\bot}(t_{0})/|\vec{v}_{\bot}(t_{0})|,\quad\vec{e}_{2}=\vec{e}_{1}\times\vec{e}_{3}.\end{equation}
We define a {}``shower frame'' $\{ A$, $\vec{u}_{x}$, $\vec{u}_{y}$,
$\vec{u}_{z}\}$, where the vectors can be expressed with respect
to the Earth frame basis as \begin{equation}
\vec{u}_{x}=\left(\begin{array}{c}
0\\
1\\
0\end{array}\right),\;\vec{u}_{y}=\left(\begin{array}{c}
\cos\theta\\
0\\
\sin\theta\end{array}\right),\:\vec{u}_{z}=\left(\begin{array}{c}
\sin\theta\\
0\\
-\cos\theta\end{array}\right).\end{equation}
The vector $\vec{u}_{z}$is parallel to the shower axis ($z$ is the
longitudinal shower variable). Again with respect to the Earth frame
basis, the magnetic field may be written as\begin{equation}
\vec{B}=B\left(\begin{array}{c}
\sin\alpha\cos\varphi\\
\sin\alpha\sin\varphi\\
\cos\alpha\end{array}\right),\end{equation}
where $\alpha$ is the angle between $\vec{B}$ and the $\vec{w}_{z}$,
and $\varphi$ the angle between the shower plane and the plane containing
$\vec{B}$ and the vertical axis. Computing $\vec{B}\vec{u}_{x,y,z}$,
we obtain $\vec{B}$ in the shower coordinate system, and as well
$\vec{e}_{3}=\vec{B}/B$,\begin{equation}
\vec{e}_{3}=\left(\begin{array}{c}
\sin\alpha\sin\varphi\\
\sin\alpha\cos\varphi\cos\theta+\cos\alpha\sin\theta\\
\sin\alpha\cos\varphi\sin\theta-\cos\alpha\cos\theta\end{array}\right).\end{equation}
Then\begin{equation}
v_{\Vert}=\vec{v}\vec{e}_{3},\quad\vec{v}_{\Vert}=\, v_{\Vert}\vec{e}_{3},\quad,\quad\vec{v}_{\bot}=\vec{v}-\vec{v}_{\Vert},\quad v_{\bot}=|\vec{v}_{\bot}|,\end{equation}
and

\begin{equation}
\vec{e}_{1}=\vec{v}_{\bot}/v_{\bot},\quad\vec{e}_{2}=\vec{e}_{1}\times\vec{e}_{3}.\end{equation}
Now we know the components of the vectors $\vec{e}_{i}$ in the shower
frame, which allows us to evaluate eq. \ref{eq:velo} and determine
the components $J^{i}$ of the current in the shower frame. We finally
obtain easily the components of the currents with respect to the Earth
frame basis as \begin{equation}
\vec{J}=J^{x}\vec{u}_{x}+J^{y}\vec{u}_{y}+J^{z}\vec{u}_{z},\end{equation}
and its derivatives.

\end{document}